\documentclass[a4paper,aps,prd,showpacs,showkeys,amsmath,amssymb,amsfonts,nofootinbib,floats,floatfix,superscriptaddress,eqsecnum]{revtex4}

\usepackage{ifthen}
\newboolean{adp}
\setboolean{adp}{true}
\newboolean{arxiv}
\setboolean{arxiv}{false}
\newboolean{notadp}
\setboolean{notadp}{true}
\ifadp
\setboolean{notadp}{false}
\fi
\newboolean{notarxiv}
\setboolean{notarxiv}{true}
\ifarxiv
\setboolean{notarxiv}{false}
\fi

\ifarxiv
\usepackage{times}
\fi

\usepackage[american]{babel}

\ifnotadp
\usepackage{a4wide} % do not use with PRD
\input{fleqn.clo}
\PassOptionsToPackage{fleqn}{amsmath}
\PassOptionsToPackage{fleqn}{amstex}
\fi

\usepackage{amsmath}
\usepackage{amsfonts}
\usepackage{amssymb}

\allowdisplaybreaks
\numberwithin{equation}{section}

\ifnotarxiv
\usepackage{xcolor}
\xdefinecolor{mylinkcolor}{rgb}{0,0,0.5}
 \usepackage[
 	bookmarksnumbered, bookmarksopen, bookmarksopenlevel=2,
 	colorlinks=true, % breaklinks=true,
 	filecolor=mylinkcolor, citecolor=mylinkcolor, linkcolor=mylinkcolor,
 	urlcolor=mylinkcolor, menucolor=mylinkcolor,
 ]{hyperref}
\fi

\ifarxiv % breaklinks must be removed
\usepackage{xcolor}
\xdefinecolor{mylinkcolor}{rgb}{0,0,0.5}
 \usepackage[
 	bookmarksnumbered, bookmarksopen, bookmarksopenlevel=1,
 	colorlinks=true,
 	filecolor=mylinkcolor, citecolor=mylinkcolor, linkcolor=mylinkcolor,
 	urlcolor=mylinkcolor, menucolor=mylinkcolor,
 ]{hyperref}
\fi

\ifnotadp
 \usepackage[square,comma,numbers,sort&compress]{natbib}
\fi

\newcommand{\Sec}[1]{Sect.\ \ref{#1}}

\def\nl{\\ & \quad}

\def\vct#1{\mathbf{#1}}

\def\picSin{((\vct{p}_1 \times \vct{S}_1) \cdot \vct{n}_{1 2})}
\def\piicSin{((\vct{p}_2 \times \vct{S}_1) \cdot \vct{n}_{1 2})}
\def\piicSiin{((\vct{p}_2 \times \vct{S}_2) \cdot \vct{n}_{1 2})} 
\def\picSiin{((\vct{p}_1 \times \vct{S}_2) \cdot \vct{n}_{1 2})}
\def\picSipii{((\vct{p}_1 \times \vct{S}_1) \cdot \vct{p}_2)}
\def\SiSii{(\vct{S}_1 \!\cdot\! \vct{S}_2)}
\def\pipii{(\vct{p}_1 \!\cdot\! \vct{p}_2)}
\def\Sipi{(\vct{S}_1 \!\cdot\! \vct{p}_1)}
\def\Siipii{(\vct{S}_2 \!\cdot\! \vct{p}_2)}
\def\Sipii{(\vct{S}_1 \!\cdot\! \vct{p}_2)}
\def\Siipi{(\vct{S}_2 \!\cdot\! \vct{p}_1)}
\def\Sin{(\vct{S}_1 \!\cdot\! \vct{n}_{1 2})}
\def\Siin{(\vct{S}_2 \!\cdot\! \vct{n}_{1 2})}
\def\pin{(\vct{p}_1 \!\cdot\! \vct{n}_{1 2})}
\def\piin{(\vct{p}_2 \!\cdot\! \vct{n}_{1 2})}
\def\pipi{\vct{p}_1^2}
\def\piipii{\vct{p}_2^2}

\def\SiSi{\vct{S}_1^2}

\def\picSinNW{((\vct{p}_1 \times \hat{\vct{S}}_1) \cdot \hat{\vct{n}}_{1 2})}
\def\piicSinNW{((\vct{p}_2 \times \hat{\vct{S}}_1) \cdot \hat{\vct{n}}_{1 2})}
\def\piicSiinNW{((\vct{p}_2 \times \hat{\vct{S}}_2) \cdot \hat{\vct{n}}_{1 2})}
\def\picSiinNW{((\vct{p}_1 \times \hat{\vct{S}}_2) \cdot \hat{\vct{n}}_{1 2})}
\def\picSipiiNW{((\vct{p}_1 \times \hat{\vct{S}}_1) \cdot \vct{p}_2)}
\def\SiSiiNW{(\hat{\vct{S}}_1 \!\cdot\! \hat{\vct{S}}_2)}

\def\SipiNW{(\hat{\vct{S}}_1\! \cdot\! \vct{p}_1)}
\def\SiipiiNW{(\hat{\vct{S}}_2 \!\cdot\! \vct{p}_2)}
\def\SipiiNW{(\hat{\vct{S}}_1 \!\cdot\! \vct{p}_2)}
\def\SiipiNW{(\hat{\vct{S}}_2 \!\cdot\! \vct{p}_1)}
\def\SinNW{(\hat{\vct{S}}_1 \!\cdot\! \hat{\vct{n}}_{1 2})}
\def\SiinNW{(\hat{\vct{S}}_2 \!\cdot\! \hat{\vct{n}}_{1 2})}
\def\pinNW{(\vct{p}_1 \!\cdot\! \hat{\vct{n}}_{1 2})}
\def\piinNW{(\vct{p}_2 \!\cdot\! \hat{\vct{n}}_{1 2})}

\def\SiSiNW{\hat{\vct{S}}_1^2}

\def\pa{\partial}
\def\dd{\mathrm{d}}

\DeclareMathOperator{\Order}{\mathcal{O}}

\begin{document}

\def\tit{Elimination of the spin supplementary condition in the effective field
	theory approach to the post-Newtonian approximation}
\def\titleShort{Elimin. of the SSC in the EFT approach to the pN approx.}

\def\Fauthor{Steven Hergt}
\def\FauthorShort{S.\ Hergt}
\def\Sauthor{Jan Steinhoff}
\def\SauthorShort{J.\ Steinhoff}
\def\Tauthor{Gerhard Sch\"afer}
\def\TauthorShort{G.\ Sch\"afer}

\def\FauthorFootnote{Corresponding author\quad
	E-mail:~\textsf{Steven.Hergt@uni-jena.de},
	Phone: +49\,3641\,947\,107,
	Fax: +49\,3641\,947\,102}

\def\FauthorInst{1}
\def\SauthorInst{1,2}
\def\TauthorInst{1}

% uncomment this line if all authors have the same address information
% \renewcommand{\inst}[1]{}

\def\Fadr{Theoretisch--Physikalisches Institut, \\
	Friedrich--Schiller--Universit\"at Jena, \\
	Max--Wien--Platz 1, 07743 Jena, Germany, EU}

\def\Sadr{Centro Multidisciplinar de Astrof\'isica (CENTRA), Departamento de F\'isica, \\
	Instituto Superior T\'ecnico (IST), Universidade T\'ecnica de Lisboa, \\ %(UTL)
	Avenida Rovisco Pais 1, 1049-001 Lisboa, Portugal, EU}

\def\abs{The present paper addresses open questions regarding the handling of
the spin supplementary condition within the effective field theory approach to
the post-Newtonian approximation. In particular it is shown how the covariant spin
supplementary condition can be eliminated at
the level of the potential (which is subtle in various respects) and how
the dynamics can be cast into a fully reduced Hamiltonian form. Two different
methods are used and compared, one based on the well-known Dirac bracket and the
other based on an action principle. It is discussed how the latter approach can
be used to improve the Feynman rules by formulating them in terms of reduced
canonical spin variables.}

\def\pacsn{04.25.Nx, 04.20.Fy}
\def\keyw{Post-Newtonian approximation, classical spin, canonical formalism, Effective Field Theory.}

\def\ack{The authors wish to thank M.\ Levi for useful discussions on the EFT
formalism and very useful comments on the manuscript.
This work is supported by the Deutsche Forschungsgemeinschaft (DFG) through
SFB/TR7 ``Gravitational Wave Astronomy,'' project STE 2017/1-1, and GRK 1523,
and by the FCT (Portugal) through PTDC project CTEAST/098034/2008.}

\title{\tit}
 \author{Steven Hergt}
\email{steven.hergt@uni-jena.de}
\affiliation{Theoretisch--Physikalisches Institut, \\
	Friedrich--Schiller--Universit\"at Jena, \\
	Max--Wien--Platz 1, 07743 Jena, Germany, EU}

 \author{Jan Steinhoff}
\email{jan.steinhoff@uni-jena.de}
 \affiliation{Theoretisch--Physikalisches Institut, \\
	Friedrich--Schiller--Universit\"at Jena, \\
	Max--Wien--Platz 1, 07743 Jena, Germany, EU}
 \affiliation{Centro Multidisciplinar de Astrof\'isica (CENTRA), Departamento de F\'isica, \\
	Instituto Superior T\'ecnico (IST), Universidade T\'ecnica de Lisboa, \\ %(UTL)
	Avenida Rovisco Pais 1, 1049-001 Lisboa, Portugal, EU}

\author{Gerhard Sch\"afer}
\email{gos@tpi.uni-jena.de}
\affiliation{Theoretisch--Physikalisches Institut, \\
	Friedrich--Schiller--Universit\"at Jena, \\
	Max--Wien--Platz 1, 07743 Jena, Germany, EU}
\pacs{\pacsn}
\begin{abstract}
\abs
\end{abstract}

\date{\today}

\maketitle

% \ifadp
% \title[\titleShort]{\tit}
% \author[\FauthorShort]{\Fauthor\footnote{\FauthorFootnote}\inst{\FauthorInst}}
% \author[\SauthorShort]{\Sauthor\inst{\SauthorInst}}
% \author[\TauthorShort]{\Tauthor\inst{\TauthorInst}}
% \affiliation[\inst{1}]{\Fadr}
% \affiliation[\inst{2}]{\Sadr}
% \keywords{\keyw}
% \begin{abstract}
% \abs
% \end{abstract}
% \maketitle
% \fi
% 
% \ifnotarxiv
% \hypersetup{pdftitle={\tit}, pdfauthor={\Fauthor, \Sauthor, \Tauthor}}
% \fi

%document starts here

\section{Introduction\label{sec:intro}}
Recently important progress has been made in the analytic treatment of
self-gravitating \emph{spinning} compact objects in general relativity using
different methods. One of these methods is based on an effective field theory
(EFT) point of view, which was applied, e.g., within the post-Newtonian (pN)
approximation to non-spinning compact objects \cite{Goldberger:Rothstein:2006,
Gilmore:Ross:2008, Chu:2009, Foffa:Sturani:2011, Kol:Smolkin:2009} and to
spinning objects \cite{Porto:2006, Kol:Smolkin:2008, Porto:Rothstein:2008:1, Levi:2008,
Porto:Rothstein:2008:2, Porto:Rothstein:2008:2:err, Porto:2010, Levi:2010,
Perrodin:2010, Levi:2011}. An advantage of this approach is that some of the
very sophisticated and systematic techniques for perturbative calculations used
in high energy physics can be applied in a straightforward manner. The present
paper addresses certain open questions regarding the handling of the spin
supplementary condition (SSC) within this approach, though some aspects may be
applicable to other approaches as well. In particular it is shown how the covariant SSC
can be eliminated at
the level of the potential (which is subtle in various respects) and how the dynamics can be cast
into a fully reduced Hamiltonian form. For a review of spin in relativity and
the problem of the SSC see, e.g., \cite{Fleming:1965, Westpfahl:1967,
Westpfahl:1969:1}.

In classical Newtonian mechanics the spin of an object is described by a
3-dimensional antisymmetric tensor or by its dual vector. In some situations
(e.g., when the dynamics depends on the spin but not on the absolute orientation
of the objects) it is very convenient to associate a Poisson bracket
representation of the so(3) Lie algebra (i.e., the angular momentum algebra)
with the spin variables and describe the spin dynamics in terms of a function
generating the time evolution via these Poisson brackets. This function may be a
Hamiltonian or a Routhian \cite{Landau:Lifshitz:Vol1,
Goldstein:Poole:Safko:2000}. (The latter is a Hamiltonian for a part of the
variables only and a Lagrangian for the remaining variables.) Such a formulation
of spin is also desirable in the relativistic case. But in this case the spin is
given by a 4-dimensional antisymmetric tensor, so the best one can immediately
achieve is to relate it to the so$(1,3)$ Lie algebra of the SO$(1,3)$ Lorentz group.
However, it is well-known that the fixation of a representative worldline or
center of the spinning objects is equivalent to a supplementary condition on the
spin components. With this SSC the independent components of the 4-dimensional
antisymmetric spin tensor are given by its 3-dimensional spatial part. But the
reduction of the so$(1,3)$ algebra for the spin components to a so(3) algebra is
subtle and must be discussed within the framework of constrained Hamiltonian
dynamics. Using the Dirac bracket approach this was performed for
flat spacetimes \cite{Hanson:Regge:1974} and for test-spinning objects in curved
spacetime \cite{Barausse:Racine:Buonanno:2009}. For self-gravitating spinning
objects the reduction succeeded to linear order in spin with the help of an
action principle \cite{Steinhoff:Schafer:2009:2, Steinhoff:2011} and agrees with
a construction via generators of the global Poincar\'e algebra valid to
next-to-next-to-leading pN order \cite{Steinhoff:Schafer:Hergt:2008,
Steinhoff:Wang:2009, Steinhoff:2011}. But this derivation is focused on the
canonical formalism of Arnowitt, Deser, and Misner (ADM)
\cite{Arnowitt:Deser:Misner:1962, Arnowitt:Deser:Misner:2008} and is restricted
to the Schwinger time gauge family of vierbein gauges \cite{Schwinger:1963:1}
(which unfortunately does not include the gauge used in the EFT approach).

A consistent way to deal with the (covariant) SSC within the EFT method
is described in \cite{Porto:Rothstein:2008:1} (based on developments in
\cite{Porto:2006}, see also \cite{Levi:2010} and \cite{Kol:Smolkin:2008} for a sophisticated Kaluza-Klein-like reduction
of the components of the metric tensor for further construction of the Feynman rules).
There a Routhian generates the spin evolution via the
so$(1,3)$ algebra and the center of mass motion is given by Euler-Lagrange
equations (see also \cite{Yee:Bander:1993}). The covariant SSC is eliminated at the level of the equations
of motion. In the present paper we show how the covariant SSC can consistently be eliminated already at
the level of the potential using two different methods. One method is based on
the well-known Dirac bracket and the other on an action principle. It is then
straightforward to obtain a fully reduced Hamiltonian form of the dynamics. This
is very convenient, e.g., for deriving the fully reduced equations of motion
(where the length of the 3-dimensional spin vector is constant without further variable transformations) or
for an implementation into the effective one body formalism, see
\cite{Damour:Jaranowski:Schafer:2008:3, Barausse:Buonanno:2009, Nagar:2011} and
references therein. We treat the pN next-to-leading order (NLO) spin-orbit,
spin(1)-spin(2), and spin(1)-spin(1) level here. The relevant potentials were
derived in \cite{Porto:Rothstein:2008:1, Levi:2008, Porto:Rothstein:2008:2,
Porto:Rothstein:2008:2:err, Perrodin:2010, Porto:2010, Levi:2010} within the EFT
formalism. Recently also the corresponding source multipole moments were
calculated \cite{Porto:Ross:Rothstein:2010} (see also \cite{Kidder:1995,
Poisson:1998}). The NLO spin-orbit and spin(1)-spin(2) dynamics was obtained
earlier in \cite{Tagoshi:Ohashi:Owen:2001, Faye:Blanchet:Buonanno:2006,
Damour:Jaranowski:Schafer:2008:1, Steinhoff:Schafer:Hergt:2008,
Steinhoff:Hergt:Schafer:2008:2} and extended to arbitrary many objects in
\cite{Hartung:Steinhoff:2010}. The NLO spin(1)-spin(1) Hamiltonian for binary
black holes was given with correct center of mass motion in
\cite{Steinhoff:Hergt:Schafer:2008:1, Hergt:Schafer:2008}. In
\cite{Hergt:Steinhoff:Schafer:2010:1} the NLO spin(1)-spin(1) dynamics for
general compact objects (including neutron stars) was reproduced and put into
fully reduced Hamiltonian form. Higher orders in spin were treated in
\cite{Hergt:Schafer:2008:2, Hergt:Schafer:2008}. For radiation-reaction effects
on the motion of a binary due to spin see, e.g.,
\cite{Wang:Steinhoff:Zeng:Schafer:2011} and references therein. Very recently
even the next-to-next-to-leading order (NNLO) spin-orbit
\cite{Hartung:Steinhoff:2011:1} and spin(1)-spin(2) level was tackled, the
latter simultaneously by a potential within the EFT approach \cite{Levi:2011}
and by a fully reduced Hamiltonian \cite{Hartung:Steinhoff:2011:2}. An extension
of the results in the present article should be useful to relate these two
results at NNLO spin(1)-spin(2). Notice that above results were obtained only
very recently compared to the first treatments of self-gravitating spinning
objects within the post-Minkowskian
\cite{Goenner:Gralewski:Westpfahl:1967} and post-Newtonian
\cite{Barker:OConnell:1975, DEath:1975} approximations (see also
\cite{Bennewitz:Westpfahl:1971, Ibanez:Martin:Ruiz:1984, Bona:Fustero:Verdaguer:1983,
Barker:OConnell:1979, Thorne:Hartle:1985}).

It should be noted that the basic approach used in the present paper was already
described in \cite{Hergt:Steinhoff:Schafer:2010:1}. Therein also the result for
the transformation to canonical variables was used in advance, but the
presentation of the derivation was reserved for the present paper (some more
details can also be found in \cite{Hergt:2011}). In the meantime the basic
approach was also applied in \cite{Levi:2010, Levi:2008}, but still the
transformation to canonical variables was not derived from general principles.

The paper is organized as follows. In \Sec{sec:setup} an overview of the problem
addressed in the present article is given. In \Sec{sec:Ltrafo} the known
potentials are transformed into non-reduced canonical form, i.e., still with a
so$(1,3)$ Poisson bracket algebra for the spins. In \Sec{sec:db} the Dirac bracket for
the covariant SSC is calculated and transformed to standard canonical form, i.e., with a
so$(3)$ Poisson bracket algebra for the spins.
An alternative elimination procedure via an action principle is performed in
\Sec{sec:action}. At the end of \Sec{sec:action} it is discussed how this
alternative approach can be used to improve the Feynman rules of the EFT
formalism by formulating them in terms of reduced canonical spin variables.
In \Sec{sec:app} the non-reduced Hamiltonians obtained earlier
are transformed into fully reduced ones and compared with other results. Finally
conclusions and outlook are given in \Sec{sec:conclusions}.

Our units are such that $c=1$, where $c$ is the velocity of light and also 
the implicit inverse pN expansion parameter with the formal counting rule
$1/c^n\sim$ $\frac{n}{2}$pN order. Adapted to EFT convention we work in the 
spacetime signature $-2$ which is important to remember especially when working on the
action level, see \Sec{sec:action}. Three different frames are utilized in this article, denoted by different indices.
Greek indices $(\alpha,\mu,\ldots)$ refer to the coordinate frame, lower case Latin indices from the beginning
of the alphabet $(a,b,\ldots)$ belong to the local Lorentz frame, and upper case Latin indices from the beginning
of the alphabet $(A,B,\ldots)$ denote the so called body-fixed Lorentz frame. Lower case Latin indices
from the middle of the alphabet $(i,j,\ldots)$ are used for the spatial part of the mentioned frames and
are running through $(i=1,2,3)$. In order to distinguish the three frames when splitting them
into spatial and time part, we write $a=(0),(i)$ for Lorentz indices (or $a=(0),(1),(2),(3)$ in more detail),
$A=[0],[i]$ for the body-fixed frame, and $\mu=0,i$ for the coordinate frame.
Letters $I$ and $J$ are body labels, i.e.\ $I,J \in\{1,2\}$ and 
$\vct{z}\equiv(z^i)$ denotes a point in the 3-dimensional Euclidean
space $\mathbb{R}^3$ endowed with a standard Euclidean metric $(z^i=z_{i}=\delta^{ij}z_{j})$
and a scalar product (denoted by a dot), so $\vct{z}_{I}\in\mathbb{R}^3$ denotes the
position of the $I$th body. Indices appearing twice in a product
are implicitly summed over its range, except for label indices of the objects.
Round and square brackets are also used for index symmetrization and antisymmetrization,
respectively, e.g., $A^{(\mu\nu)}\equiv\frac{1}{2}\left(A^{\mu\nu}+A^{\nu\mu}\right)$.
We also define $\vct{r}_{I}:=\vct{z}-\vct{z}_{I}$, $r_{I}:=|\vct{r}_{I}|$,
 $\vct{n}_{I}:=\vct{r}_{I}/r_{I}$; and for $I\neq J$, $\vct{r}_{IJ}:=\vct{z}_{I}-\vct{z}_{J}$, which is the
distance vector of the two bodies. Likewise we define $r_{IJ}:=|\vct{r}_{IJ}|$, $\vct{n}_{IJ}:=\vct{r}_{IJ}/r_{IJ}$;
$|\,\cdotp|$ stands for the length of a vector. The linear momentum of the $I$th body is denoted
by $\vct{p}_{I}=(p_{Ii})$, and $m_{I}$ denotes its mass parameter. An overdot, as in $\dot{\vct{z}}$, means the
total time derivative. The spin vector of the $I$th body is denoted by $\vct{S}_{I}$ and is always supposed to be
referred to the local Lorentz frame in all potentials and Hamiltonians that are considered in the present article.
The connection between the antisymmetric spin tensor $S_{(i) (j)}$ and the spin vector is made by usage of the totally antisymmetric
$\epsilon$-symbol $\epsilon_{ijk}=\frac{1}{2}(i-j)(j-k)(k-i)$ as $S_{(i)(j)}=\epsilon_{ijk}S_{(k)}$.
Each body is within our approximation completely characterized by its three
parameters mass, momentum and spin. We associate to them the following relative pN order
\begin{equation}
 m_{I}\sim\mathcal{O}\left(\frac{1}{c^2}\right)\,,\quad \vct{p}_{I}\sim\mathcal{O}\left(\frac{1}{c^3}\right)\,,\quad \vct{S}_{I}\sim\mathcal{O}\left(\frac{1}{c^3}\right)\,,
\end{equation}
yielding the formal pN order for the Newtonian (N), leading-order (LO) and next-to-leading-order (NLO) Hamiltonians in question
\begin{equation}
 H_{\text{N}}\sim\mathcal{O}\left(\frac{1}{c^4}\right)\,,\quad H_{\text{LO}}\sim\mathcal{O}\left(\frac{1}{c^6}\right)\,,\quad H_{\text{NLO}}\sim\mathcal{O}\left(\frac{1}{c^8}\right)\,,
\end{equation}
starting with the non-relativistic Newtonian Hamiltonian for two interacting bodies in canonical conjugate variables:
\begin{equation}\label{HNewt}
 H_{\text{N}}=\frac{\vct{p}_{1}^2}{2m_{1}}+\frac{\vct{p}_{2}^2}{2m_{2}}-\frac{Gm_{1}m_{2}}{r_{12}}\,.
\end{equation}
Notice that $H_{\text{N}}$ defines the zeroth pN order for Hamiltonians, so the
pN orders of the LO and NLO Hamiltonians must be counted relative to $H_{\text{N}}$
(which is at $1/c^4$ in the formal counting given above).
% Notice that the correct pN order of $H_{\text{N}}$ is zero,
% so in order to arrive at the correct pN order of the LO and NLO Hamiltonians
% one has to cancel $1/c^4$ from their overall pN order.

\section{Setup of the Problem\label{sec:setup}}
In \cite{Porto:Rothstein:2008:1} an EFT approach to include spin within the pN
approximation is given, based on developments in \cite{Porto:2006} (see also \cite{Levi:2010}). This approach is able to deliver a
pN approximate description of the dynamics of spinning compact objects via a
Routhian, namely a function which is a Lagrangian for the objects 3-dimensional
position $\vct{z}$ and a Hamiltonian for the 4-dimensional antisymmetric spin
tensor $S_{ab}$ (in this section we may drop the object labels). Therefore this
Routhian generates the spin evolution via an so$(1,3)$ Poisson bracket algebra
and the center of mass motion is given by Euler-Lagrange equations (such an
approach was already used in \cite{Yee:Bander:1993}). In order to
allow a treatment of the covariant SSC via Dirac brackets one needs a canonical
description of all variables, including $\vct{z}$. Therefore we first perform a
standard Legendre transformation in order to replace the velocity
$\vct{v} = \dot{\vct{z}}$ by its generalized momentum $\vct{p}$, which is of
course the canonical conjugate of $\vct{z}$. (Eventually accelerations or even
higher time derivatives must be eliminated using the method in
\cite{Damour:Schafer:1991} first.)

It is important that $S_{ab}$ is actually the generalized momentum of the
4-dimensional angular velocity tensor defined by
\begin{equation}
\Omega^{ab} = \Lambda_{A}{}^{a} \dot{\Lambda}^{Ab} \,.
\end{equation}
The latter is build from a Lorentz matrix $\Lambda^{Aa} \in \text{SO}(1,3)$,
\begin{align}
% \Lambda_{A\mu}\Lambda^{A}_{\;\;\;\nu}=g_{\mu\nu}\,,\quad
\Lambda_{Aa}\Lambda^{A}_{\;\;\;b}=\eta_{ab} \,,
% \quad\text{with}\quad(A,B,..)\in\{[0],[i]\}\,.
\end{align}
which relates the body-fixed and the local Lorentz frames; to be more precise $\Lambda^{Aa}$ is a
representation of elements of $\text{SO}(1,3)$. Here
$\eta_{ab} = \text{diag}(1, -1, -1, -1)$ is the Lorentz metric with signature $-2$. Notice that this
property of the Lorentz matrix makes the angular velocity tensor antisymmetric.
The complete Poisson brackets read, see, e.g., \cite{Hanson:Regge:1974} or
\Sec{sec:action} below,
\begin{subequations}
\begin{align}
\{ z^{i}, p_{j} \}& = \delta_{ij} \,,\\
\{ \Lambda^A{}_b, S_{cd} \} &= \eta_{bc} \Lambda^A{}_d - \eta_{bd} \Lambda^A{}_c \,, \label{PB4dim1} \\
\{ S_{ab}, S_{cd} \} &= S_{ca} \eta_{bd} - S_{da} \eta_{bc} + S_{db} \eta_{ac} - S_{cb} \eta_{ad} \label{PB4dim2} \,,
\end{align}
\end{subequations}
and \emph{all other zero}. We will call these brackets canonical here, though
they are not canonical in a strict sense (e.g., the spin components are not
commuting). But it is possible to relate $\Lambda^{Ab}$ and $S_{cd}$ to
variables for which the phase space structure is manifest, see Sect.\ 3.A in
\cite{Hanson:Regge:1974}. For example, one can parametrize $\Lambda^{Aa}$ by
\emph{independent} angle-type variables. The time derivatives of these variables
are angular velocities and their generalized momenta are canonically conjugate
to the angle-type variables. (These angular velocities are also contained in
$\Omega^{ab}$, but with prefactors depending on the angle-type variables.)
Notice that \cite{Porto:Rothstein:2008:1} uses a different sign convention for
the spin part of the Poisson bracket shown above. Further the Hamiltonian does
actually not depend on $\Lambda^{Aa}$ in our case, so one can ignore
$\Lambda^{Aa}$ and its Poisson brackets for obtaining the equations of motion.
For the present article, however, $\Lambda^{Aa}$ is of crucial importance.

The Poisson brackets above are fully canonical, but the degrees of freedom are
not fully reduced. An SSC must be imposed, which corresponds to a choice for the
representative worldline of the compact object. A similar condition must be
given for $\Lambda^{Aa}$ \cite{Hanson:Regge:1974}. According to
\cite{Hanson:Regge:1974} the covariant supplementary conditions read (consistent with the
SSC used in \cite{Porto:Rothstein:2008:1})
% \begin{subequations} 
\begin{align}\label{supplcond}
S^{ab} p_{b}^{\text{D}} = 0 \,, \qquad \eta^{[0]A} = \Lambda^{Ab} \frac{p_{b}^{\text{D}}}{m_D} \,,
% \quad\Leftrightarrow\quad S_{(i)b}u^{b}=0\quad\Leftrightarrow\quad S_{(i)(0)}+S_{(i)(j)}\frac{u^{(j)}}{u^{(0)}}=0\,,\label{cssc1}\\
%  \Lambda^{[i]a}u_{a}&=0\quad\Leftrightarrow\quad\Lambda^{[i](0)}+\Lambda^{[i](j)}\frac{u_{(j)}}{u_{(0)}}=0\,,\label{lambdassc1}\\
%  \Lambda_{[0]a}&=\frac{u_{a}}{\sqrt{u_{b}u^{b}}}\,,\label{lambda0ssc1}
\end{align}
% \end{subequations}
where $p_{\mu}^{\text{D}}$ is Dixon's momentum of the compact object and $m_D$
the dynamical mass. It holds
$m^2_D = g^{\mu\nu} p_{\mu}^{\text{D}} p_{\nu}^{\text{D}}$, where $g_{\mu\nu}$
is the 4-dimensional metric. From (5.13) in \cite{Steinhoff:2011} we get
\begin{equation}\label{Dixp}
p_{\mu}^{\text{D}} = m \frac{u_{\mu}}{\sqrt{u_{\nu}u^{\nu}}} + \Order{(S^2)} \,, \qquad
m = m_D + \Order{(S^2)} \,,
\end{equation}
where $m$ is the constant mass parameter of the object. It should be noted that
the mass-shell constraint is already implicitly eliminated within the approach
in \cite{Porto:Rothstein:2008:1}, as a gauge-fixing for the worldline parameter
was performed. (Indeed, only a 3-dimensional canonical momentum $p_{i}$ is
defined by the Legendre transformation mentioned above, but not its time component $p_0$.) The
worldline parameter was chosen to be the coordinate time in \cite{Porto:Rothstein:2008:1}, $u^0 = 1$. Notice that
(\ref{supplcond}) guarantees that in the rest frame the spin tensor contains the
3-dimensional spin $S^{(i)(j)}$ only (i.e., the mass dipole part $S^{(0)(i)}$
vanishes) and that in the rest frame $\Lambda^{A b}$ describes a pure
3-dimensional rotation (no Lorentz boosts). This obviously reduces the degrees
of freedom to the physically relevant ones, which are given by $S^{(i)(j)}$ and
$\Lambda^{[i](j)}$.

The most prominent way to handle the constraints (\ref{supplcond}) on the phase
space described by (\ref{PB4dim1}, \ref{PB4dim2}) is provided by the Dirac
bracket, denoted by $\{\cdot,\cdot\}_{\text{D}}$ here. It is straightforward to
calculate the Dirac bracket for the current situation in a pN approximate way.
Notice that only three components of each condition in (\ref{supplcond}) are
independent, so one has six independent constraints for each particle. To the
considered approximation the derivation will turn out to be very similar to
\cite{Hanson:Regge:1974}, with the notable exception that in
\cite{Hanson:Regge:1974} the mass-shell constraint together with the
gauge-fixing of the worldline parameter was also treated using the Dirac
bracket. The Dirac bracket is essentially the Poisson brackets of the reduced
phase space, but the variables used above are not canonical any more with
respect to the Dirac bracket. Our next step is thus to transform $z^i$, $p_i$,
$S^{(i)(j)}$, and $\Lambda^{[i](j)}$ to new canonical variables denoted by a hat
such that
 \begin{subequations}
\begin{align}
\{ \hat{z}^{i}, \hat{p}_{j} \}_{\text{D}}& = \delta^i_j \,,\\
\{ \hat{\Lambda}^{[i](j)}, \hat{S}^{(m)(n)} \}_{\text{D}}& =
	- \delta_{jm} \hat{\Lambda}^{[i](n)} + \delta_{jn} \hat{\Lambda}^{[i](m)} \,, \label{DB1} \\
\{ \hat{S}^{(i)(j)}, \hat{S}^{(k)(l)} \}_{\text{D}}& =
	\delta_{jl} \hat{S}^{(i)(k)} - \delta_{jk} \hat{S}^{(i)(l)}
	+ \delta_{ik} \hat{S}^{(j)(l)} - \delta_{il} \hat{S}^{(j)(k)} \,, \label{DB2}
\end{align}
 \end{subequations}
and \emph{all other zero}. The algebra for the spin was reduced from so$(1,3)$
to so$(3)$ and the Lorentz matrix $\Lambda^{Aa} \in \text{SO}(1,3)$ was
transformed into $\hat{\Lambda}^{[i](j)} \in \text{SO}(3)$. These variables are
a suitable generalization of the Newton-Wigner
variables defined in flat spacetime \cite{Newton:Wigner:1949}, but they are not
canonically equivalent to the Newton-Wigner variables introduced in
\cite{Porto:Rothstein:2008:2} (for a discussion see
\cite{Steinhoff:Schafer:2009:1}). Notice that the transformation to canonical
variables is highly ambiguous, i.e., one may always perform a canonical
transformation. It turns out that we are able to choose $\hat{p}_i = p_i$ here.
Further we will actually not derive the transformation to
$\hat{\Lambda}^{[i](j)}$ via Dirac brackets, as the Hamiltonian does not depend
on it anyway. For the same reason the components of the spin tensor in the
body-fixed frame $\hat{S}_{[i][j]}$ and thus the spin length $s$ given by
$2 s^2 = \hat{S}_{[i][j]} \hat{S}^{[i][j]} = \hat{S}^{(i)(j)} \hat{S}^{(i)(j)}$
are conserved. It should be noted that in \cite{Barausse:Racine:Buonanno:2009}
a canonical Newton-Wigner SSC for test-bodies in curved spacetime was handled by
Dirac brackets directly (and consistently implemented into the action).
However, from a general point of view it is very convenient to start with a
covariant SSC, as this manifestly displays the covariance of the effective
theory, in particular when higher dimensional operators are included in the
worldline action. In the following the SSC is always assumed to be the covariant
one, if not otherwise stated.

Spin in relativity can also be treated by an action principle
\cite{Goenner:Westpfahl:1967, Romer:Westpfahl:1969, Westpfahl:1969:2}, see also
\cite{Hanson:Regge:1974, Bailey:Israel:1975, Porto:2006,
Steinhoff:Schafer:2009:2, Steinhoff:2011} and appendix A of \cite{DeWitt:2011}. It is indeed possible to derive the
transformation to \emph{reduced} canonical variables using an action approach.
The Poisson brackets (\ref{PB4dim1}, \ref{PB4dim2}) are essentially represented
by a term in the action of the form
$p_{i} \dot{z}^{i} + \frac{1}{2} S_{ab} \dot{\Lambda}^{Aa} \Lambda_A{}^b$.
After the supplementary conditions (\ref{supplcond}) are inserted, one must find
new variables such that this term takes on the form
$\hat{p}_{i} \dot{\hat{z}}^{i} + \frac{1}{2} \hat{S}_{(i)(j)} \dot{\hat{\Lambda}}^{[k](i)} \hat{\Lambda}_{[k]}{}^{(j)}$,
which precisely represents the reduced brackets (\ref{DB1}, \ref{DB2}). Here it
is important that $\hat{\Lambda}^{[k](i)} \in \text{SO}(3)$ must be a 3-dimensional rotation
matrix, $\hat{\Lambda}^{[k](i)}\hat{\Lambda}^{[k](j)}=\delta_{ij}$. This
approach is very similar to \cite{Steinhoff:Schafer:2009:2, Steinhoff:2011}.
Notice that one needs the transformation from $\Lambda^{[i](j)}$ to
$\hat{\Lambda}^{[i](j)}$ for the action approach, which is not necessary for the
Dirac bracket approach.

\section{Legendre transformation\label{sec:Ltrafo}}
The effective potential usually depends on velocities and positions when referred to a Lagrangian $L$ defined as the
difference between the non-relativistic Newtonian kinetic part $T_{\text{N}}$ and the effective potential $V_{\text{eff}}$:
\begin{equation}
 L_{\text{eff}}=T_{\text{N}}-V_{\text{eff}}=\frac{m_{1}}{2}\vct{v}_{1}^2+\frac{m_{2}}{2}\vct{v}_{2}^2-V_{\text{eff}}\,.
\end{equation}
The effective potential is the only part of the Lagrangian which
is pN expanded and is further decomposed into different spin contributions.
The first step to arrive at a reduced canonical Hamiltonian when starting with a non-reduced effective potential
$V_{\text{eff}}=V_{\text{eff}}(\vct{x}_{I},\vct{v}_{I},S_{I(i)(j)},S_{I(0)(i)})$ in pN approximation is to Legendre transform it
only with respect to the velocities/momenta (the spin variables are formally kept unchanged by this procedure)
to a non-reduced effective Hamiltonian $H_{\text{eff}}=H_{\text{eff}}(\vct{x}_{I},\vct{p}_{I},S_{I(i)(j)},S_{I(0)(i)})$.
In both expressions the SSC is not yet imposed, so that $S_{I(0)(i)}$ will be treated as an independent variable.
Furthermore the effective Hamiltonians are not supposed to contain any time
derivatives of the variables except for the one of the position variable being defined as the velocity. If the case arose
that a spin variable (including the constrained $S_{I(0)(i)}$ ones) would carry a time derivative one could either replace it
with its lower order equations of motion or one could shift it
onto positions and/or velocities in the same term by 
neglecting total time derivatives which serve as surface terms in the action. The last procedure ensures
to leave us only with time derivatives of variables which are not further subject to a constraint, 
because the mass-shell constraint is already eliminated, when performing Legendre transformation.
Those variables can
therefore be treated differently aside from the Dirac bracket formalism but rather with a fully
reduced Poisson bracket and a subleading potential/Hamiltonian, see the discussion at the end of \Sec{sec:action}.
But nevertheless one ends up with inserting lower order equations of motion for eliminating higher order
time derivatives as outlined in \cite{Schafer:1984,
Damour:Schafer:1991}. The Dirac bracket formalism
will be carried out below for the SSCs in order to find a canonical set of variables after using the SSCs.
The effective potential $V_{\text{eff}}$ for two interacting bodies is pN expanded up to NLO spin effects:
\begin{equation}
V_{\text{eff}} = V_{\text{pp}}
	+ V^{\text{LO}}_{\text{SO}} + V^{\text{LO}}_{S_1^2}
		+ V^{\text{LO}}_{S_2^2} + V^{\text{LO}}_{S_1S_2}
	+ V^{\text{NLO}}_{\text{SO}} + V^{\text{NLO}}_{S_1^2}
		+ V^{\text{NLO}}_{S_2^2}
		+ V^{\text{NLO}}_{S_1S_2}\,.
\end{equation}
$V_{\text{pp}}$ is the point particle interaction potential. This again is decomposed into
\begin{equation}
 V_{\text{pp}}=V_{\text{N}}+V_{\text{EIH/1pN}}+V_{\text{2pN}}\,,
\end{equation}
starting with the Newtonian potential
\begin{equation}
 V_{\text{N}}=-\frac{Gm_{1}m_{2}}{r_{12}}\,,
\end{equation}
and continuing with the Einstein-Infeld-Hoffmann (EIH) potential $(V_{\text{EIH/1pN}}=-L_{\text{EIH}})$ \cite{Landau:Lifshitz:Vol2:2,Goldberger:Rothstein:2006}
\begin{align}
 \begin{aligned}
  L_{\text{EIH}}&=\frac{1}{8}\sum_{a}m_{a}\vct{v}_{a}^4+\frac{Gm_{1}m_{2}}{2r_{12}}\left[3\left(\vct{v}_{1}^2+\vct{v}_{2}^2\right)-7\left(\vct{v}_{1}\cdot\vct{v}_{2}\right)-\left(\vct{v}_{1}\cdot\vct{n}_{12}\right)\left(\vct{v}_{2}\cdot\vct{n}_{12}\right)\right]\\
                &\quad-\frac{G^2m_{1}m_{2}(m_{1}+m_{2})}{2r_{12}^2}\,.
 \end{aligned}
\end{align}
The 2pN point particle potential is in fact not needed for the Legendre transformation, the interested reader can find it in \cite{Gilmore:Ross:2008}.
The leading order (LO) spin contributions are decomposed into the LO spin-orbit (SO) contribution in non-reduced form, see \cite{Porto:2010}:
\begin{align}\label{LOSO}
 \begin{aligned}
  V^{\text{LO}}_{\text{SO}}=\frac{Gm_{2}}{r_{12}^2}n_{12}^j\left[S_{1}^{(j)(0)}+S_{1}^{(j)(k)}\left(v_{1}^k-2v_{2}^k\right)\right]+(1\leftrightarrow 2)\,,
 \end{aligned}
\end{align}
the LO spin(1)-spin(2) contribution, e.g. from \cite{Porto:Rothstein:2008:1}
\begin{equation}
V^{\text{LO}}_{S_1S_2}=\frac{G}{r_{12}^3} \left[ 3(\vct{S}_{1}\cdot\vct{n}_{12})(\vct{S}_{2}\cdot\vct{n}_{12})-(\vct{S}_{1}\cdot\vct{S}_{2}) \right]
\end{equation}
and the LO spin(I)-spin(I) (finite size) contributions from \cite{Poisson:1998,Porto:Rothstein:2008:2}
\begin{equation}\label{LOS12pot}
 V^{\text{LO}}_{S_I^2}=C_{Q_{I}}\frac{Gm_{J}}{2m_{I}r_{IJ}^3}\left(3(\vct{S}_{I}\cdot\vct{n}_{IJ})^2-\vct{S}_{I}^2\right)
\end{equation}
with the spin quadrupole constant $C_{Q_{I}}$ which is chosen such that it is equal to one for black holes and correspondingly bigger for
white dwarfs or neutron stars.
The Legendre transformation with respect to the velocities/momenta is done by using the formula
\begin{align}
\begin{aligned}
 H_{\text{eff}}&=\vct{v}_{1}\cdot\vct{p}_{1}+\vct{v}_{2}\cdot\vct{p}_{2}-L_{\text{eff}}\\
  &=\vct{v}_{1}\cdot\vct{p}_{1}+\vct{v}_{2}\cdot\vct{p}_{2}-\frac{1}{2}m_{1}\vct{v}_{1}^2-\frac{1}{2}m_{2}\vct{v}_{2}^2+V_{\text{eff}}\\
  &=\frac{\vct{p}_{1}^2}{2m_{1}}+\frac{\vct{p}_{2}^2}{2m_{2}}-\frac{1}{2m_{1}}\left(\vct{p}_{1}-m_{1}\vct{v}_{1}\right)^2-\frac{1}{2m_{2}}\left(\vct{p}_{2}-m_{2}\vct{v}_{2}\right)^2+V _{\text{eff}}
\end{aligned}
\end{align}
with
\begin{equation}
 \vct{p}_{I}=\frac{\partial L_{\text{eff}}}{\partial\vct{v}_{I}}=m_{I}\vct{v}_{I}-\frac{\partial V_{\text{eff}}}{\partial\vct{v}_{I}}\,.
\end{equation}
This leaves us with the expression
\begin{align}\label{Heff}
 H_{\text{eff}}&=\frac{\vct{p}_{1}^2}{2m_{1}}+\frac{\vct{p}_{2}^2}{2m_{2}}-\frac{1}{2m_{1}}\left(\frac{\partial V_{\text{eff}}}{\partial\vct{v}_{1}}\right)^2-\frac{1}{2m_{2}}\left(\frac{\partial V_{\text{eff}}}{\partial\vct{v}_{2}}\right)^2+V_{\text{eff}}\\
               &=\frac{\vct{p}_{1}^2}{2m_{1}}+\frac{\vct{p}_{2}^2}{2m_{2}}+H_{(\partial V)^2}+V_{\text{eff}}\,,\\
 H_{(\partial V)^2}&\equiv-\frac{1}{2m_{1}}\left(\frac{\partial V_{\text{eff}}}{\partial\vct{v}_{1}}\right)^2-\frac{1}{2m_{2}}\left(\frac{\partial V_{\text{eff}}}{\partial\vct{v}_{2}}\right)^2\label{Hp}
\end{align}
with $\vct{v}_{I}=\vct{v}_{I}\left(\vct{p}_{I},\vct{p}_{J}\right)_{\text{1pN}}$, so it is sufficient to know the momentum up to 1pN order, because at 2pN level correction terms to the Hamiltonian will be induced only by the square of the velocity derivative of the potential. The momentum is given by
\begin{subequations}
\begin{align}
\vct{p}_{I} &= m_I \vct{v}_I - \frac{\partial( V_{\text{EIH/1pN}}+V^{\text{LO}}_{\text{SO}})}{\partial \vct{v}_I}+\mathcal{O}(c^{-7}) \\
\begin{split}
	&= \left( 1 + \frac{1}{2} \vct{v}_I^2 \right) m_I \vct{v}_I
	+ \frac{G m_I m_J}{2 r_{IJ}} [ 6 \vct{v}_I - 7 \vct{v}_J
		- (\vct{n}_{IJ} \cdot \vct{v}_J) \vct{n}_{IJ} ] \nl
	+ \frac{G}{r_{IJ}^2} [ m_J (\vct{n}_{IJ} \times \vct{S}_I)
		+ 2 m_I (\vct{n}_{IJ} \times \vct{S}_J) ]+\mathcal{O}(c^{-7})\,. \label{momentum}
\end{split}
\end{align}
\end{subequations}
The derivative of the potential is given by
\begin{align}\label{potder}
\begin{aligned}
 \frac{\partial V_{\text{eff}}}{\partial\vct{v}_{I}}&=-\Bigg[ \frac{m_{I}}{2} \vct{v}_I^2 \vct{v}_I
	+ \frac{G m_I m_J}{2 r_{IJ}} [ 6 \vct{v}_I - 7 \vct{v}_J
		- (\vct{n}_{IJ} \cdot \vct{v}_J) \vct{n}_{IJ} ] \nl
	+ \frac{G}{r_{IJ}^2} [ m_J (\vct{n}_{IJ} \times \vct{S}_I)
		+ 2 m_I (\vct{n}_{IJ} \times \vct{S}_J) ]\Bigg]+\mathcal{O}(c^{-7})\,.
\end{aligned}
\end{align}
After evaluating its square and replacing the velocity by inverting Eq.\ (\ref{momentum})
\begin{align}\label{geschw}
 \begin{aligned}
  \vct{v}_{I}&=\left( 1 - \frac{1}{2} \frac{\vct{p}_I^2}{m_{I}^2} \right) \frac{\vct{p}_{I}}{m_{I}}
	-\frac{Gm_J}{2 r_{IJ}} \left[ 6\frac{\vct{p}_{I}}{m_{I}} - 7 \frac{\vct{p}_J}{m_{J}}
		- \frac{(\vct{n}_{IJ} \cdot \vct{p}_J)}{m_{J}} \vct{n}_{IJ} \right] \nl
	- \frac{G}{r_{IJ}^2} \left[ \frac{m_J}{m_{I}} (\vct{n}_{IJ} \times \vct{S}_I)
		+ 2 (\vct{n}_{IJ} \times \vct{S}_J) \right]+\mathcal{O}(c^{-5})\,,
 \end{aligned}
\end{align}
the Legendre transformation can be performed. The contribution $H_{(\partial V)^2}$ from (\ref{Hp}) then reads
\begin{align}\label{Hpcon}
 \begin{aligned}
  H_{(\partial V)^2}&=\frac{G}{r_{12}^2}\left(\frac{1}{m_{1}^2}\vct{p}_{1}^2\,\vct{n}_{12}\cdot\left(\vct{p}_{1}\times\vct{S}_{2}\right)+\frac{m_{2}}{2m_{1}^3}\vct{p}_{1}^2\,\vct{n}_{12}\cdot\left(\vct{p}_{1}\times\vct{S}_{1}\right)\right)\\
                              &\quad+\frac{G^2}{r_{12}^3}\bigg(\frac{3m_{2}^2}{m_{1}}\vct{n}_{12}\cdot\left(\vct{p}_{1}\times\vct{S}_{1}\right)-\frac{7m_{2}}{2}\vct{n}_{12}\cdot\left(\vct{p}_{2}\times\vct{S}_{1}\right)\\
                              &\quad\quad-7m_{1}\vct{n}_{12}\cdot\left(\vct{p}_{2}\times\vct{S}_{2}\right)+6m_{2}\vct{n}_{12}\cdot\left(\vct{p}_{1}\times\vct{S}_{2}\right)\bigg)\\
                              &\quad+\frac{G^2}{r_{12}^4}\Bigg[\frac{m_{2}^2}{2m_{1}}\bigg(-\vct{S}_{1}^2+\left(\vct{S}_{1}\cdot\vct{n}_{12}\right)^2\bigg)+2m_{1}\bigg(-\vct{S}_{2}^2+\left(\vct{S}_{2}\cdot\vct{n}_{12}\right)^2\bigg)\\
                              &\quad\quad+2m_{2}\bigg(-\left(\vct{S}_{1}\cdot\vct{S}_{2}\right)+\left(\vct{S}_{1}\cdot\vct{n}_{12}\right)\left(\vct{S}_{2}\cdot\vct{n}_{12}\right)\bigg)\Bigg]+(1\leftrightarrow 2)+\mathcal{O}(c^{-10})\,.
 \end{aligned}
\end{align}
Certainly the replacement of the velocities through (\ref{geschw}) has also to be done in the LO spin-orbit and the EIH potential to arrive at
the fully correct NLO Hamiltonians. The LO spin-orbit potential (\ref{LOSO}) yields the contribution $H_{\text{LOSO}}^{v\rightarrow p}$ excluding the SSC,
which remains still untouched:
\begin{align}\label{HLOSOcon}
 \begin{aligned}
  H_{\text{LOSO}}^{v\rightarrow p}&=\frac{G}{r_{12}^2}\vct{S}_{1}\cdot\Bigg[\frac{m_{2}}{m_{1}}\left(1-\frac{\vct{p}_{1}^2}{2m_{1}^2}\right)(\vct{n}_{12}\times\vct{p}_{1})-\left(1-\frac{\vct{p}_{2}^2}{m_{2}^2}\right)(\vct{n}_{12}\times\vct{p}_{2})\Bigg]\\
    &+\frac{G^2}{r_{12}^3}\Bigg[-m_{2}\left(7+\frac{3m_{2}}{m_{1}}\right)\vct{S}_{1}\cdot(\vct{n}_{12}\times\vct{p}_{1})+\left(6m_{1}+7m_{2}\right)\vct{S}_{1}\cdot(\vct{n}_{12}\times\vct{p}_{2})\\
    &-m_{2}\left(4+\frac{m_{2}}{m_{1}}\right)\left((\vct{S}_{1}\cdot\vct{n}_{12})^2-\vct{S}_{1}^2\right)-2(m_{1}+m_{2})\bigg((\vct{S}_{1}\cdot\vct{n}_{12})(\vct{S}_{2}\cdot\vct{n}_{12})\\
    &-(\vct{S}_{1}\cdot\vct{S}_{2})\bigg)\Bigg]+(1\leftrightarrow 2)+\mathcal{O}(c^{-10})\,.
 \end{aligned}
\end{align}
Likewise the EIH potential gives rise to the contribution $H_{\text{EIH}}^{v\rightarrow p}$:
\begin{align}\label{HEIHcon}
 \begin{aligned}
  H_{\text{EIH}}^{v\rightarrow p}&=\frac{G}{r_{12}^2}\left(\frac{1}{m_{2}^2}\vct{p}_{2}^2\piicSin-\frac{m_{2}}{2m_{1}^3}\vct{p}_{1}^2\picSin\right)\\
                              &\quad\frac{G^2}{r_{12}^3}\left(\left(-7m_{2}-\frac{3m_{2}^2}{m_{1}}\right)\picSin+\left(6m_{1}+\frac{7m_{2}}{2}\right)\piicSin\right)\\
                              &\quad+(1\leftrightarrow 2)+\mathcal{O}(c^{-10})\,.
 \end{aligned}
\end{align}
Now we are able to evaluate from the effective potentials for the NLO spin-orbit, spin(1)-spin(2) and spin(1)-spin(1)
case their effective Hamiltonian counterparts $H_{\text{eff}}=H_{\text{eff}}(\vct{x}_{I},\vct{p}_{I},S_{I(i)(j)},S_{I(0)(i)})$,
which result from Legendre transformation with respect to the velocities/momenta only. So these Hamiltonians still remain non-reduced 
in phase space as long as the SSC is not imposed.
\subsection{The non-reduced NLO spin-orbit Hamiltonians}
We consider two effective potentials from literature, one from Levi \cite{Levi:2010}, and the other one from Porto \cite{Porto:2010}. Both will be
subject to a Legendre transformation to arrive at $H_{\text{eff}}(\vct{x}_{I},\vct{p}_{I},S_{I(i)(j)},S_{I(0)(i)})$. These Hamiltonians still depend on
the SSC, so they are far from being canonical and deserve only formally to be called Hamiltonians, in the sense that they are the result of a Legendre
transformation of the potentials but only to a subset of variables, likewise the reduction of unphysical degrees of freedom is only achieved
on a subspace of phase space. Both Levi and Porto include in their NLO potentials the SSC term arising from the LO SO potential, because the
elimination of the SSC is itself subject to a pN expanded expression and will therefore lift LO expressions in the potentials to NLO ones and so one.
For this reason it is important to keep track of the SSC in all terms where it is present.
\subsubsection{The non-reduced NLO spin-orbit Hamiltonian of the potential of Levi}
We start with the effective NLO SO potential of Levi as given in \cite{Levi:2010} Eq.\ (109) $(V^{\text{NLO}}_{\text{SO(L)}}=-L^{\text{NLO}}_{\text{SO(L)}})$:
\begin{subequations}
\begin{align}\label{Levipo}
 \begin{aligned}
  &L^{\text{NLO}}_{\text{SO(L)}}=\frac{Gm_{2}}{r_{12}^2}\vct{S}_{1}\cdot\bigg[\vct{v}_{1}\times\vct{n}_{12}\left(\frac{1}{2}\vct{v}_{1}\cdot\vct{v}_{2}-\frac{1}{2}v_{2}^2-\frac{3}{2}(\vct{v}_{1}\cdot\vct{n}_{12})(\vct{v}_{2}\cdot\vct{n}_{12})\right)\\
&+\vct{v}_{2}\times\vct{n}_{12}\left(\vct{v}_{1}\cdot\vct{v}_{2}-v_{2}^2+3(\vct{v}_{1}\cdot\vct{n}_{12})(\vct{v}_{2}\cdot\vct{n}_{12})\right)+\vct{v}_{1}\times\vct{v}_{2}\left(\frac{1}{2}\vct{v}_{1}\cdot\vct{n}_{12}+\vct{v}_{2}\cdot\vct{n}_{12}\right)\bigg]\\
                               &+\frac{G^2m_{2}}{r_{12}^3}\vct{S}_{1}\cdot\left[\vct{v}_{1}\times\vct{n}_{12}\left(2m_{1}-\frac{1}{2}m_{2}\right)+\vct{v}_{2}\times\vct{n}_{12}(2m_{2})\right]\\
                               &+\frac{Gm_{2}}{r_{12}^2}\Bigg[S_{1}^{(0)(i)}n_{12}^i\left(1-\frac{3}{2}\vct{v}_{1}\cdot\vct{v}_{2}+\frac{3}{2}v_{2}^2-\frac{3}{2}(\vct{v}_{1}\cdot\vct{n}_{12})(\vct{v}_{2}\cdot\vct{n}_{12})\right)\\
                               &+S_{1}^{(0)(i)}v_{2}^i\left(-\frac{3}{2}\vct{v}_{1}\cdot\vct{n}_{12}\right)\Bigg]+\frac{Gm_{2}}{r_{12}}\left[\frac{3}{2}\dot{S}_{1}^{(0)(i)}v_{2}^i\right]-\frac{G^2m_{2}}{r_{12}^3}S_{1}^{(0)(i)}n_{12}^i\left[m_{1}+2m_{2}\right]\,.
 \end{aligned}
\end{align}
This potential owns a term with a time derivative of ${S}_{1}^{(0)(i)}$. According to our agreed rule we shift it onto positions and velocities in the same term yielding
\begin{align}\label{Levipo2}
 \begin{aligned}
  &L^{\text{NLO}}_{\text{SO(L)}}=\frac{Gm_{2}}{r_{12}^2}\vct{S}_{1}\cdot\bigg[\vct{v}_{1}\times\vct{n}_{12}\left(\frac{1}{2}\vct{v}_{1}\cdot\vct{v}_{2}-\frac{1}{2}v_{2}^2-\frac{3}{2}(\vct{v}_{1}\cdot\vct{n}_{12})(\vct{v}_{2}\cdot\vct{n}_{12})\right)\\
&+\vct{v}_{2}\times\vct{n}_{12}\left(\vct{v}_{1}\cdot\vct{v}_{2}-v_{2}^2+3(\vct{v}_{1}\cdot\vct{n}_{12})(\vct{v}_{2}\cdot\vct{n}_{12})\right)+\vct{v}_{1}\times\vct{v}_{2}\left(\frac{1}{2}\vct{v}_{1}\cdot\vct{n}_{12}+\vct{v}_{2}\cdot\vct{n}_{12}\right)\bigg]\\
                               &+\frac{G^2m_{2}}{r_{12}^3}\vct{S}_{1}\cdot\left[\vct{v}_{1}\times\vct{n}_{12}\left(2m_{1}-\frac{1}{2}m_{2}\right)+\vct{v}_{2}\times\vct{n}_{12}(2m_{2})\right]+\frac{Gm_{2}}{r_{12}^2}\Bigg[S_{1}^{(0)(i)}v_{2}^i\left(-\frac{3}{2}\vct{v}_{1}\cdot\vct{n}_{12}\right)\\
                               &+S_{1}^{(0)(i)}n_{12}^i\left(1-\frac{3}{2}\vct{v}_{1}\cdot\vct{v}_{2}+\frac{3}{2}v_{2}^2-\frac{3}{2}(\vct{v}_{1}\cdot\vct{n}_{12})(\vct{v}_{2}\cdot\vct{n}_{12})\right)\Bigg]\\
                               &-\frac{3}{2}Gm_{2}{S}_{1}^{(0)(i)}\left[\frac{ {a}_{2}^i}{r_{12}}-\frac{(\vct{v}_{1}\cdot\vct{n}_{12})v_{2}^i}{r_{12}^2}+\frac{(\vct{v}_{2}\cdot\vct{n}_{12})v_{2}^i}{r_{12}^2}\right]-\frac{G^2m_{2}}{r_{12}^3}S_{1}^{(0)(i)}n_{12}^i\left[m_{1}+2m_{2}\right]\,.
 \end{aligned}
\end{align}
\end{subequations}
Next we insert lower order equations of motion to eliminate the accelaration term. By counting the pN order
of this term in the potential it is clear that only the Newtonian equation of motion is needed as replacement which reads
\begin{equation}
 \vct{a}_{2}=\frac{Gm_{1}\vct{n}_{12}}{r_{12}^2}\,.
\end{equation}
The resulting potential will then be subject to a Legendre transformation meaning we replace velocities in (\ref{Levipo2}) by momenta indicated by $(v\rightarrow p)$ in $V^{\text{NLO}(v\rightarrow p)}_{\text{SO(L)}}$
and add the specific NLO spin-orbit contributions from (\ref{Hpcon}), (\ref{HLOSOcon}) and (\ref{HEIHcon}), indicated by $\simeq$, to arrive at
the effective Hamiltonian $H^{\text{NLO(eff)}}_{\text{SO(L)}}(\vct{x}_{I},\vct{p}_{I},S_{Iab},S_{I(0)(i)})$:
\begin{align}
 \begin{aligned}
  H^{\text{NLO(eff)}}_{\text{SO(L)}}\simeq H_{(\partial V)^2}+H_{\text{LOSO}}^{v\rightarrow p}+H_{\text{EIH}}^{v\rightarrow p}+V^{\text{NLO}(v\rightarrow p)}_{\text{SO(L)}}
\end{aligned}
\end{align}
leading to the result
\begin{align}\label{LevieffH2}
 \begin{aligned}
  &H^{\text{NLO(eff)}}_{\text{SO(L)}}=\frac{G}{r_{12}^2}\Bigg(-\frac{m_{2}}{2m_{1}^3}\vct{p}_{1}^2\picSin-\frac{3}{2m_{1}^2}\pin\piin\picSin\\
                 &+\frac{1}{2m_{1}^2}\pipii\picSin-\frac{1}{2m_{1}m_{2}}\piipii\picSin\\
                 &+\frac{3}{m_{1}m_{2}}\pin\piin\piicSin+\frac{1}{m_{1}m_{2}}\pipii\piicSin\\
                 &+\frac{1}{2m_{1}^2}\pin\picSipii+\frac{1}{m_{1}m_{2}}\piin\picSipii\\
                 &+\frac{G^2}{r_{12}^3}\Bigg(-5m_{2}\picSin-\frac{7m_{2}^2}{2m_{1}}\picSin+6m_{1}\piicSin\\
                 &+\frac{11m_{2}}{2}\piicSin\Bigg)-\frac{G}{m_{1}r_{12}^2}\bigg[S_{1}^{(0)(i)}n_{12}^i\bigg(1-\frac{3}{2}\vct{p}_{1}\cdot\vct{p}_{2}+\frac{3}{2}p_{2}^2\\
                 &-\frac{3}{2}(\vct{p}_{1}\cdot\vct{n}_{12})(\vct{p}_{2}\cdot\vct{n}_{12})\bigg)+S_{1}^{(0)(i)}p_{2}^i\left(-\frac{3}{2}\vct{p}_{1}\cdot\vct{n}_{12}\right)\bigg]+\frac{G^2m_{2}}{r_{12}^3}S_{1}^{(0)(i)}n_{12}^i\left[m_{1}+2m_{2}\right]\\
                 &-\frac{3}{2}G{S}_{1}^{(0)(i)}\left[{\frac{Gm_{1}m_{2}n^i_{12}}{r_{12}^3}}-\frac{(\vct{p}_{1}\cdot\vct{n}_{12})p_{2}^i}{m_{1}r_{12}^2}+\frac{(\vct{p}_{2}\cdot\vct{n}_{12})p_{2}^i}{m_{2}r_{12}^2}\right]\,.
 \end{aligned}
\end{align}

\subsubsection{The non-reduced NLO spin-orbit Hamiltonian of the potential of Porto}
The second alternative potential we find in Porto \cite{Porto:2010} Eq.\ (53)
 \begin{align}
 \begin{aligned}
    V^{\text{NLO}}_{\text{SO(P)}}&= \frac{Gm_2}{r_{12}^2}\biggl[\bigg\{S^{(i)(0)}_1\left(1+2{\bf v}_2^2-2 {\bf v}_1\cdot {\bf v}_2-\frac{3}{2} ({\bf v}_2\cdot \vct{n}_{12})^2-\frac{G}{r} (3m_1+2m_2)\right) \\
   &\quad+ \left(1-\frac{3}{2} ({\bf v}_2\cdot \vct{n}_{12})^2+\frac{G}{2r}\left(4m_1-m_2\right)\right) S_1^{(i)(j)}{\bf v}_1^j\\
   &\quad-\left(2-2 {\bf v}_1\cdot {\bf v}_2-3({\bf v}_2\cdot \vct{n}_{12})^2+2{\bf v}_2^2-\frac{G}{2r}\left(2m_1+5m_2\right)\right)S_1^{(i)(j)}{\bf v}_2^j \bigg\}\vct{n}_{12}^i \\
   &\quad +S^{(i)(0)}_1({\bf v}_1-{\bf v}_2)^i {\bf v}_2\cdot \vct{n}_{12}+S_1^{(i)(j)}{\bf v}_1^j{\bf v}_2^i{\bf v}_2\cdot\vct{n}_{12}\biggr] + (1 \leftrightarrow 2)\,.
 \end{aligned}
\end{align}
Notice that accelerations were already eliminated by inserting equations of motion.
Legendre transformation yields like in the case of Levi's potential:
\begin{align}\label{VNLOSOPO}
 \begin{aligned}
   &H^{\text{NLO(eff)}}_{\text{SO(P)}}\simeq H_{(\partial V)^2}+H_{\text{LOSO}}^{v\rightarrow p}+H_{\text{EIH}}^{v\rightarrow p}+V^{\text{NLO}(v\rightarrow p)}_{\text{SO(P)}}\\
  &H^{\text{NLO(eff)}}_{\text{SO(P)}}=\frac{G}{r_{12}^2}\Bigg[-\frac{m_{2}}{2m_{1}^3}\vct{p}_{1}^2\picSin-\frac{3}{2m_{1}m_{2}}\piin^2\picSin\\
                 &+\frac{3}{m_{2}^2}\piin^2\piicSin+\frac{2}{m_{1}m_{2}}\pipii\piicSin\\
                 &-\frac{1}{m_{2}^2}\piipii\piicSin+\frac{1}{m_{1}m_{2}}\piin\picSipii\\
                 &+\left(-m_{2}+\frac{3}{2m_{2}}\piin^2+\frac{2}{m_{1}}\pipii-\frac{2}{m_{2}}\piipii\right)S_{1}^{(0)(i)}n_{12}^i\\
                 &-\frac{1}{m_{1}}\piin S_{1}^{(0)(i)}p_{1}^i+\frac{1}{m_{2}}\piin S_{1}^{(0)(i)}p_{2}^i\Bigg)\\
                 &+\frac{G^2}{r_{12}^3}\Bigg(-5m_{2}\picSin-\frac{7m_{2}^2}{2m_{1}}\picSin+7m_{1}\piicSin\\
                 &\qquad+6m_{2}\piicSin+3m_{1}m_{2}S_{1}^{(0)(i)}n_{12}^{i}+2m_{2}^2S_{1}^{(0)(i)}n_{12}^{i}\Bigg]\,.
 \end{aligned}
\end{align}
\subsection{The non-reduced NLO spin(1)-spin(2) Hamiltonian}
We take the NLO spin(1)-spin(2) potential $V^{\text{NLO}}_{S_{1}S_{2}\text{(P)}}$ of Porto/Rothstein from \cite{Porto:Rothstein:2008:1} Eq.\ (56) (modulo the non-SSC-dependent LO spin-orbit terms while keeping the important SSC dependent LO spin-orbit term)
\begin{align}
\begin{aligned}
 &V^{\text{NLO}}_{S_{1}S_{2}\text{(P)}}=-\frac{G}{r_{12}^3}\Bigg[(\delta^{ij}-3n_{12}^in_{12}^j)\bigg(S_{1}^{(i)(0)}S_{2}^{(j)(0)}+\frac{1}{2}\vct{v}_{1}\cdot\vct{v}_{2}S_{1}^{(i)(k)}S_{2}^{(j)(k)}\\
                                      &+v_{1}^{m}v_{2}^kS_{1}^{(i)(k)}S_{2}^{(j)(m)}-v_{1}^kv_{2}^mS_{1}^{(i)(k)}S_{2}^{(j)(m)}+S_{1}^{(i)(0)}S_{2}^{(j)(k)}(v_{2}^k-v_{1}^k)+S_{1}^{(i)(k)}S_{2}^{(j)(0)}(v_{1}^k-v_{2}^k)\bigg)\\
                                      &+\frac{1}{2}S_{1}^{(k)(i)}S_{2}^{(k)(j)}\bigg(3\vct{v}_{1}\cdot\vct{n}_{12}\vct{v}_{2}\cdot\vct{n}_{12}(\delta^{ij}-5n_{12}^in_{12}^j)\\
                                      &+3\vct{v}_{1}\cdot\vct{n}_{12}(v_{2}^jn_{12}^i+v_{2}^in_{12}^j)+3\vct{v}_{2}\cdot\vct{n}_{12}(v_{1}^jn_{12}^i+v_{1}^i n_{12}^{j})-v_{1}^i v_{2}^j-v_{2}^i v_{1}^j\bigg)\\
                                      &+(3n_{12}^l\vct{v}_{2}\cdot\vct{n}_{12}-v_{2}^l)S_{1}^{(0)(k)}S_{2}^{(k)(l)}+(3n_{12}^l\vct{v}_{1}\cdot\vct{n}_{12}-v_{1}^l)S_{2}^{(0)(k)}S_{1}^{(k)(l)}\Bigg]\\
                                      &+\left(\frac{G}{r_{12}^3}-\frac{3(m_{1}+m_{2})G^2}{r_{12}^4}\right)S_{1}^{(j)(k)}S_{2}^{(j)(i)}(\delta^{ki}-3n_{12}^kn_{12}^i)+\frac{Gm_{2}}{r_{12}^2}n_{12}^jS_{1}^{(j)(0)}-\frac{Gm_{1}}{r_{12}^2}n_{12}^jS_{2}^{(j)(0)}\,.
\end{aligned}
\end{align}
and apply to it the same Legendre transformation procedure as in the preceding section yielding
\begin{align}\label{NLOS1S2}
 \begin{aligned}
&H^{\text{NLO(eff)}}_{S_{1}S_{2}\text{(P)}}\simeq H_{(\partial V)^2}+H_{\text{LOSO}}^{v\rightarrow p}+H_{\text{EIH}}^{v\rightarrow p}+V^{\text{NLO}(v\rightarrow p)}_{S_{1}S_{2}\text{(P)}}\\
  &H^{\text{NLO(eff)}}_{S_{1}S_{2}\text{(P)}}=\frac{G}{m_{1}m_{2}r_{12}^3}\Bigg(-\frac{15}{2}\pin\piin\Sin\Siin\\
                                        &-\frac{3}{2}\pipii\Sin\Siin+\frac{9}{2}\piin\Sipi\Siin\\
                                        &-\frac{3}{2}\piin\Sin\Siipi-\frac{3}{2}\pin\Sipii\Siin+\frac{3}{2}\Sipii\Siipi\\
                                        &+\frac{9}{2}\pin\Sin\Siipii-\frac{5}{2}\Sipi\Siipii\\
                                        &-\frac{3}{2}\pin\piin\SiSii+\frac{3}{2}\pipii\SiSii\Bigg)\\
                                        &+\frac{G}{r_{12}^3}\Bigg[\left(\frac{3}{m_{1}}\picSiin-\frac{3}{m_{2}}\piicSiin\right) S_{1}^{(0)(i)}n_{12}^i\\
                                        &-\frac{3}{m_{2}}\piin\epsilon_{ijk}n_{12}^iS_{2}^{(j)}S_{1}^{(0)(k)}-\frac{1}{m_{1}}\epsilon_{ijk}p_{1i}S_{2(j)}S_{1}^{(0)(k)}+\frac{2}{m_{2}}\epsilon_{ijk}p_{2i}S_{2(j)}S_{1}^{(0)(k)}\\
                                        &+\left(\frac{3}{m_{2}}\piicSin-\frac{3}{m_{1}}\picSin\right) S_{2}^{(0)(i)}n_{12}^i\\
                                        &+3S_{1}^{(0)(i)}n_{12}^iS_{2}^{(0)(j)}n_{12}^j-S_{1}^{(0)(i)}S_{2}^{(0)(i)}\\
                                        &-\frac{3}{m_{1}}\pin\epsilon_{ijk}n_{12}^iS_{1}^{(j)}S_{2}^{(0)(k)}+\frac{2}{m_{1}}\epsilon_{ijk}p_{1}^iS_{1}^{(j)}S_{2}^{(0)(k)}-\frac{\epsilon_{ijk}p_{2}^iS_{1}^{(j)}S_{2}^{(0)(k)}}{m_{2}}\Bigg]\\
                                        &+\frac{Gm_{2}}{r_{12}^2}n_{12}^jS_{1}^{(j)(0)}-\frac{Gm_{1}}{r_{12}^2}n_{12}^jS_{2}^{(j)(0)}\\
                                        &+\frac{G^2}{r_{12}^4}\bigg(7(m_{1}+m_{2})\SiSii-13(m_{1}+m_{2})\Sin\Siin\bigg)\,.
 \end{aligned}
\end{align}
where $\simeq$ indicates here focusing only on NLO spin(1)-spin(2) terms with the SSC untouched.
\subsection{The non-reduced NLO spin(1)-spin(1) Hamiltonian}
We adopt the NLO spin(1)-spin(1) potential of Porto and Rothstein that was calculated in \cite{Porto:Rothstein:2008:2} up to a missing contribution stemming from
an acceleration term, which was corrected in \cite{Porto:Rothstein:2008:2:err}. The potential reads according to Eq.\ (49) of the arXiv version with the LO spin-orbit SSC term included:
\begin{align}
\begin{aligned}\label{fullspin}
&V^{\text{NLO}}_{S_{1}^2\text{(P)}} = C_{Q_{1}}\frac{Gm_2}{2m_1r^3}\bigg[ S_1^{(j)(0)}S_1^{(i)(0)}(3n^in^j-\delta^{ij}) 
\\
&-2 S_1^{(k)(0)} \left( ({\bf v}_1\times{\bf S}_1)^{k} - 3 ({\bf n}\cdot {\bf v}_1)({\bf n}\times{\bf S}_1)^k\right)\bigg] \\
&+ C_{Q_{1}}\frac{Gm_2}{2m_1r^3}\Bigg[ {\bf S}_1^2 \left( 6 ({\bf n}\cdot{\bf v}_1)^2 - \frac{15}{2} {\bf n}\cdot{\bf v}_1{\bf n}\cdot{\bf v}_2 + \frac{13}{2}{\bf v}_1\cdot{\bf v}_2 - \frac{3}{2}{\bf v}_2^2 - \frac{7}{2}{\bf v}_1^2-2{\bf a}_1\cdot {\bf r}\right)\\
&+   ({\bf S}_1\cdot {\bf n})^2 \left ( \frac{9}{2}({\bf v}_1^2+{\bf v}_2^2)-\frac{21}{2}{\bf v}_1\cdot{\bf v}_2 - \frac{15}{2} {\bf n}\cdot{\bf v}_1 {\bf n}\cdot{\bf v}_2\right)+ 2{\bf v}_1\cdot{\bf S}_1{\bf v}_1\cdot{\bf S}_1\\
& - 3{\bf v}_1\cdot{\bf S}_1{\bf v}_2\cdot{\bf S}_1 
-6 {\bf n}\cdot{\bf v}_1{\bf n}\cdot{\bf S}_1{\bf v}_1\cdot{\bf S}_1+ 9 {\bf n}\cdot{\bf v}_2{\bf n}\cdot{\bf S}_1{\bf v}_1\cdot{\bf S}_1\\
&+ 3 {\bf n}\cdot{\bf v}_1{\bf n}\cdot{\bf S}_1{\bf v}_2\cdot{\bf S}_1\Bigg] +  C_{Q_{1}}\frac{m_2G^2}{2r^4}\left(1+\frac{4m_2}{m_1}\right) \left( {\bf S}_1^2 - 3({\bf S}_1\cdot{\bf n})^2\right)\\
&-\frac{G^2m_2}{r^4}\left({\bf S}_1\cdot {\bf n}\right)^2 + \left({\bf \tilde a}^{so}_{1(1)}\right)^l S_1^{(0)(l)} + {\bf v}_1\times{\bf S}_1\cdot {\bf \tilde a}^{so}_{1(1)}+\frac{Gm_{2}}{r_{12}^2}n_{12}^jS_{1}^{(j)(0)}\,,
\end{aligned}
\end{align}
with the $\vct{S}_{1}$-dependent part of the acceleration of the local frame
\begin{equation}
\label{covasoloc}
{\bf \tilde a}^{so}_{1(1)} =  \frac{m_2G}{m_1r^3} \left[ -3{\bf v} \times {\bf S}_1 + 6{\bf n} ({\bf v} \times {\bf  S}_1)\cdot {\bf n} + 3 {\bf n}\cdot{\bf v} ({\bf n} \times {\bf  S}_1) \right]\,.
\end{equation}
The acceleration term with $\vct{a}_{1}$ appearing is eliminated by using the Newtonian EOMs for two bodies:
\begin{equation}
 \vct{a}_{1}=-\frac{Gm_{2}\vct{n}_{12}}{r_{12}^2}.
\end{equation}
This potential will be Legendre transformed like the other potentials above resulting in the effective SSC-dependent Hamiltonian ($\simeq$ indicates here
sole focus on spin(1)-spin(1) terms with the SSC-dependent terms untouched)
\begin{align}
 \begin{aligned}
  &H^{\text{NLO(eff)}}_{S_{1}^2\text{(P)}}\simeq H_{(\partial V)^2}+H_{\text{LOSO}}^{v\rightarrow p}+H_{\text{EIH}}^{v\rightarrow p}+V^{\text{NLO}(v\rightarrow p)}_{S_{1}^2\text{(P)}}\,,
\end{aligned}
\end{align}
which results in
\begin{align}\label{NLOS12}
\begin{aligned}
  &H^{\text{NLO(eff)}}_{S_{1}^2\text{(P)}}=\frac{G^2m_{2}}{r_{12}^4}\Bigg[\left(\left(2+\frac{C_{Q_{1}}}{2}\right)+\left(\frac{1}{2}+3C_{Q_{1}}\right)\frac{m_{2}}{m_{1}}\right)\SiSi\\
                                    &-\left(\left(3+\frac{3C_{Q_{1}}}{2}\right)+\left(\frac{1}{2}+6C_{Q_{1}}\right)\frac{m_{2}}{m_{1}}\right)\Sin^2\Bigg]+\frac{G}{r_{12}^3}\Bigg[-\frac{C_{Q_{1}}}{2}\frac{m_{2}}{m_{1}}S_{1}^{(0)(i)}S_{1}^{(0)(i)}\\
                                    &+\frac{6m_{2}}{m_{1}^2}\picSin S_{1}^{(0)(i)}n_{12}^i-\frac{6}{m_{1}}\piicSin S_{1}^{(0)(i)}n_{12}^i+\frac{3C_{Q_{1}}m_{2}}{2m_{1}}\left(S_{1}^{(0)(i)}n_{12}^i\right)^2\\
                                    &+\left(\frac{3(1-C_{Q_{1}})m_{2}}{m_{1}^2}\pin-\frac{3}{m_{1}}\piin\right)\epsilon_{ijk}n_{12}^iS_{1(j)}S_{1}^{(0)(k)}\\
                                    &-\frac{3(1-C_{Q_{1}})m_{2}}{m_{1}^2}\epsilon_{ijk}n_{12}^iS_{1}^{(j)}S_{1}^{(0)(k)}+\frac{3}{m_{1}}\epsilon_{ijk}p_{2i}S_{1}^{(j)}S_{1}^{(0)(k)}\Bigg]+\frac{Gm_{2}}{r_{12}^2}n_{12}^jS_{1}^{(j)(0)}\\
                                    &+\frac{G}{r_{12}^3}\frac{m_{2}}{m_{1}^3}\Bigg(\left(-6+\frac{9}{4}C_{Q_{1}}\right)\pipi\Sin^2+\left(9-3C_{Q_{1}}\right)\pin\Sin\Sipi\\
                                    &+\left(-3+C_{Q_{1}}\right)\Sipi^2+\left(-3+3C_{Q_{1}}\right)\pin^2\SiSi+\left(3-\frac{7C_{Q_{1}}}{4}\right)\pipi\SiSi\Bigg)\\
                                    &+\frac{G}{m_{1}^2r_{12}^3}\Bigg(-\frac{15}{4}C_{Q_{1}}\pin\piin\Sin^2+\left(6+\frac{21}{4}C_{Q_{1}}\right)\pipii\Sin\\
                                    &-\left(3-\frac{9}{2}C_{Q_{1}}\right)\piin\Sin\Sipi-\left(6-\frac{3}{2}C_{Q_{1}}\right)\pin\Sin\Sipii\\
                                    &+\left(3-\frac{3}{2}C_{Q_{1}}\right)\Sipi\Sipii+\left(3-\frac{15}{4}C_{Q_{1}}\right)\pin\piin\SiSi\\
                                    &+\left(-3+\frac{13}{4}C_{Q_{1}}\right)\pipii\SiSi\Bigg)+\frac{G\,C_{Q_{1}}}{m_{1}m_{2}r_{12}^3}\piipii\Bigg(\frac{9}{4}\Sin^2-\frac{3}{4}\SiSi\Bigg)\,.
 \end{aligned}
\end{align}
Notice that while the LO potential (\ref{LOS12pot}) is purely spin quadrupole dependent (via the constant $C_{Q_{1}}$) the corresponding NLO potential
(\ref{fullspin}) is not, likewise in the case of the Hamiltonian (\ref{NLOS12}).

These non-reduced Hamiltonians are now fit for further phase-space reduction procedures,
either by Dirac brackets or by the action principle in Eq.\ (\ref{effaction}).

\section{Reduction via Dirac brackets\label{sec:db}}
Canonical formalisms in the presence of constraints were analyzed in a very
general way by Dirac \cite{Dirac:1950, Dirac:1951, Dirac:1958:1, Dirac:1964},
for reviews see also \cite{Hanson:Regge:1974, Hanson:Regge:Teitelboim:1976,
Henneaux:Teitelboim:1992, Pons:2005}. A very important tool developed in this
area is nowadays called the Dirac bracket and is further explained in the
following. Other important contributions to constrained Hamiltonian dynamics
were made by Bergmann and his collaborators, e.g., the notion of primary
constraints and the understanding of gauge transformations
\cite{Anderson:Bergmann:1951}. Early contributions were already made by
Rosenfeld \cite{Rosenfeld:1930}, e.g., the discovery of what is nowadays called
the Dirac or total Hamiltonian. For a historical review see
\cite{Salisbury:2007}.

\subsection{Construction of the Dirac bracket}
The Hamiltonian formulation of a dynamical system needs the construction of a Poisson bracket type structure of the dynamics.
In the case of a constraint dynamical systems this is not an easy task, but the needed formalism is available as developed by 
Bergmann and Dirac. In the following we shall present the formalism to the extent we will need it.
At the beginning let us treat the following variables $\left(z_I^i,p_{Jj},S^{ab}_K,\Lambda_{L}^{Ab}\right)$ as unconstrained.
The time derivative of any function $Q$ of these variables reads
\begin{equation}\label{QEq}
\dot Q = \frac{\pa Q}{\pa z^{i}_{I}}\dot z^{i}_{I} + \frac{\pa Q}{\pa p_{Ii}}\dot p_{Ii} +  \frac{\pa Q}{\pa S_{Iab}}\dot S_{Iab}+\frac{\pa Q}{\pa \Lambda^{Ab}_{I}}\dot \Lambda^{Ab}_{I}\,.
\end{equation}
If one wants to restrict the time derivative to the independent degrees of freedom in all variables, one still has to reduce to the six degrees of freedom in the
Lorentz matrices, which can be achieved by writing (\ref{QEq}) in the following way
\begin{equation}
\dot Q = \frac{\pa Q}{\pa z^{i}_{I}}\dot z^{i}_{I} + \frac{\pa Q}{\pa p_{Ii}}\dot p_{Ii} +  \frac{\pa Q}{\pa S_{Iab}}\dot S_{Iab}-\frac{\partial Q}{\partial\Lambda^{A[a}_{I}}\Lambda_{I\,c]}^{A}\Omega_{I}^{ac}\quad\text{with}\quad\Omega^{ac}_{I}=\Lambda^{Aa}_{I}\dot{\Lambda}_{IA}^{\;\;\;\;\,c}\,.
\end{equation}
Taking into account, see \Sec{sec:action},
\begin{align}
\dot z^{i}_{I} &= \frac{\pa H_{\text{eff}}}{\pa p_{Ii}} \,, \qquad
\dot p_{Ii} = - \frac{\pa H_{\text{eff}}}{\pa z^{i}_{I}} \,, \qquad
\dot S_{Iab} = 4 S_{Ic[a} \eta_{b]d} \frac{\pa H_{\text{eff}}}{\pa S_{Icd}}-2\eta_{c[a}\Lambda^{A}_{I\,b]}\frac{\partial H_{\text{eff}}}{\partial\Lambda_{I\,c}^{A}}\,,\\
\dot \Lambda_A^{~\,\,a} &= \Lambda_{Ab}\Omega^{ab},\qquad \Lambda_{Ab}\Lambda^{Bb} = \delta^B_A \,,\qquad
\Omega^{ab}_{I} = - 2 \frac{\partial H_{\text{eff}}}{\partial S_{Iab}}
\end{align}
where, as seen from the previous equations, generally $H_{\text{eff}}=H_{\text{eff}}(\vct{z}_{I},\vct{p}_{J},S_{K}^{ab},\Lambda^{Ab}_{L})$ will hold, we get:
\begin{equation}
\dot Q = \{Q, H_{\text{eff}}\}\,. 
\end{equation}
The Poisson bracket $\{\cdot,\cdot\}$ as defined here has its standard properties which comprise bilinearity, fulfillment of the Leibniz rule and of the Jacobi identity when performed with standard canonical variables having vanishing Poisson brackets among themselves. Hereof one can derive a chain rule analogon for the Poisson bracket when applied to a continuous function $H$ depending on a canonical variable $\xi$ which reads
\begin{equation}
 \{Q,H(\xi)\}=\{Q,\xi\}\frac{\partial H}{\partial\xi}\,.
\end{equation}
Applying this formula one can read off all the Poisson brackets, see \Sec{sec:action} for a thorough derivation of them.
The results are (\ref{PBdim4a})-(\ref{PBdim4c}) and read
\begin{subequations}\label{nrbrackets}
\begin{align}
\{ z^{i}_{I}, p_{Jj} \} &= \delta_{ij}\delta_{IJ}\label{PBa} \,,\\
\{ S_{Iab}, S_{Icd} \} &= S_{Ica} \eta_{bd} - S_{Ida} \eta_{bc} + S_{Idb} \eta_{ac} - S_{Icb} \eta_{ad}\label{PBb} \,,\\
\{ S_{Iab},\Lambda^{A}_{I\,c}\}&=\eta_{bc}\Lambda^{A}_{I\,a}-\eta_{ac}\Lambda^{A}_{I\,b}\label{PBc}\,,
\end{align}
\end{subequations}
all other zero.
The physical evolution of the variables needs a SSC. We use the covariant one by Tulczyjew written in the local frame, Eq.\ (\ref{supplcond}), as this SSC is the one implemented in the potentials shown in the last section (i.e., this SSC is conserved by the time evolution given by the potentials). Referring to (\ref{supplcond}) we may replace Dixon's 4-momentum with the 4-velocity in the SSCs due to their equivalence to our approximation, cf.\ (\ref{Dixp}), while avoiding
unnecessary confusion when dealing with Dixon's 4-momentum whose 3-components $p^{\text{D}}_{i}=\frac{m u_{i}}{\sqrt{u_{\nu}u^{\nu}}}$ are essentially 
different from the canonical ones defined by Legendre transformation $p_{i}^{\text{L}}=\frac{\partial L}{\partial v^i}$.
This means that for the considered approximation the Tulczyjew SSC is equivalent to the Mathisson-Pirani one, $S_{ab}u^{b}=0$.
With the 4-velocity we get
\begin{subequations}\label{DiracSSC} 
\begin{align}
 S_{ab}u^{b}&=0\quad\Leftrightarrow\quad S_{(i)b}u^{b}=0\quad\Leftrightarrow\quad S_{(i)(0)}+S_{(i)(j)}\frac{u^{(j)}}{u^{(0)}}=0\,,\label{cssc3}\\
 \Lambda^{[i]a}u_{a}&=0\quad\Leftrightarrow\quad\Lambda^{[i](0)}+\Lambda^{[i](j)}\frac{u_{(j)}}{u_{(0)}}=0\,,\label{lambdassc3}\\
 \Lambda_{[0]a}&=\frac{u_{a}}{\sqrt{u_{b}u^{b}}}\,.\label{lambda0ssc3}
\end{align}
\end{subequations}
 %where $p_{(0)} = p^{(0)}= (p_{(i)}p_{(i)} + m^2)^{1/2}$ by using the mass-shell constraint $p_{b}p^b=m^2$.
Let us call the SSCs, including the $\Lambda$-relations, $\Phi_A =0, (A = 1,2, ... ,12)$. Then the Poisson brackets
\begin{equation}\label{cmatrix1}
\{\Phi_A, \Phi_B\} \equiv C_{AB}
\end{equation}
with the inverse matrix $C^{AB}$, i.e. $C^{AD}C_{DB} = \delta^A_B$,
are most important. Notice, the assumed non-degeneracy of $C_{AB}$ even for $\Phi_A =0$
makes the SSCs to be second class constraints. 
The Dirac brackets are defined in the form 
\begin{equation}
\{ Q_1, Q_2 \}_\text{D} = \{ Q_1, Q_2 \} - \{ Q_1, \Phi_A \}C^{AB}\{\Phi_B , Q_2 \}\,.
\end{equation}
The constrained evolution for the independent variables reads 
\begin{equation}
\dot z^{i}_{I} = \{  z^{i}_{I}, H_{\text{eff}}\}_\text{D}  \,, \qquad
\dot p_{Ii} = \{ p_{Ii}, H_{\text{eff}}\}_\text{D} \,, \qquad
\dot S_{I(i)(j)} = \{ S_{I(i)(j)}, H_{\text{eff}}\}_\text{D}      \,.
\end{equation}
The Dirac bracket therefore satisfies all the laws known from the Poisson bracket, which turns it also into a Lie bracket and it leads to
the correct equations of motion together with the Hamiltonian $H_{\text{eff}}$.
The Dirac bracket can thus be used as substitute for the Poisson bracket. However, whereas one may use the constraints only after all Poisson brackets were calculated,
the second class constraints $\Phi_{A}=0$ can be used \textit{before} an application of the Dirac bracket without changing the result, e.g. one has $\{Q,\Phi_{a}\}_{\text{D}}=0$ for all $Q$'s and $\Phi_{a}$ thus preserving the constraints in time when using the Hamiltonian for $Q$. Notice if one restricts to use the Dirac bracket instead of the Poisson bracket, the second class constraints $\Phi_{a}=0$ can be used off-shell to solve for certain phase space variables and eliminate them from all quantities, thus reducing the actual degrees of freedom.
The transition to new variables $(\hat{z}_I^i,\hat{p}_{Jj},\hat{S}^{(i)(j)}_K)$
which fulfill
\begin{subequations} 
\begin{align}
\{ \hat{z}^{i}_{I}, \hat{p}_{Jj} \}_{\text{D}}& = \delta^i_j\delta_{IJ} \,,\label{DB11}\\
\{ \hat{\Lambda}^{[i](j)}_{I}, \hat{S}^{(m)(n)}_{I} \}_{\text{D}}& =
	- \delta_{jm} \hat{\Lambda}^{[i](n)}_{I} + \delta_{jn} \hat{\Lambda}^{[i](m)}_{I} \,, \label{DB12} \\
\{ \hat{S}^{(i)(j)}_{I}, \hat{S}^{(k)(l)}_{I} \}_{\text{D}}& =
	\delta_{jl} \hat{S}^{(i)(k)}_{I} - \delta_{jk} \hat{S}^{(i)(l)}_{I}
	+ \delta_{ik} \hat{S}^{(j)(l)}_{I} - \delta_{il} \hat{S}^{(j)(k)}_{I} \,, \label{DB23}
\end{align}
\end{subequations}
all other brackets being zero, results in a standard canonical representation. They will be called Newton-Wigner variables, because it
can be shown that the ones proposed by those two in \cite{Newton:Wigner:1949} represent the only possible standard canonical set of variables at least in
Special Relativity. For an extension of definition of those variables to General Relativity see the comment \cite{Steinhoff:Schafer:2009:1}. The `unhatted' variables
corresponding to the covariant SSC (\ref{cssc3})-(\ref{lambda0ssc3}) are from now on shortly dubbed `covariant' variables.
More details are given in the application later on. 
The careful reader should have noticed that the space and momentum variables $(z_I^i,\,p_{Jj})$ were assumed non-constrained
throughout. This was done for simplicity reasons and because the pure spinless case is well known. On the other side,
the presented form is just the one delivered by researchers using the EFT approach, see later on.
% The supplementary conditions read
% \begin{subequations} 
% \begin{align}
%  S_{ab}u^{b}&=0\quad\Leftrightarrow\quad S_{(i)b}u^{b}=0\quad\Leftrightarrow\quad S_{(i)(0)}+S_{(i)(j)}\frac{u^{(j)}}{u^{(0)}}=0\,,\label{cssc1}\\
%  \Lambda^{[i]a}u_{a}&=0\quad\Leftrightarrow\quad\Lambda^{[i](0)}+\Lambda^{[i](j)}\frac{u_{(j)}}{u_{(0)}}=0\,,\label{lambdassc1}\\
%  \Lambda_{[0]a}&=\frac{u_{a}}{\sqrt{u_{b}u^{b}}}\,,\label{lambda0ssc1}
% \end{align}
% \end{subequations}

Now the supplementary conditions (\ref{cssc3})-(\ref{lambda0ssc3}) need to be explicitly known in dependence
of positions and canonical momenta in coordinate space when eliminating them in the Hamiltonians.
Upon agreement the spin is always defined in the local Lorentz frame,
so what we need are the 4-velocities in the coordinate frame mapped into the local frame,
which is linked to the coordinate one by a vierbein transformation matrix. By decomposing the curved spacetime metric $g_{\mu\nu}$ according to
\begin{equation}
g_{\mu\nu}=\eta_{\mu\nu}+h_{\mu\nu} 
\end{equation}
into the Minkowski background field part $\eta_{\mu\nu}$ and a perturbation part $h_{\mu\nu}$, whose indices are raised and lowered with $\eta_{\mu\nu}$, we can derive for $u_{I}^a$ the expanded expression
\begin{equation}\label{ulocal}
 u_{I}^a=e^a_{\,\,\mu}(\vct{z}_{I})u_{I}^{\mu}=\eta^{a\mu}\left(\eta_{\mu\nu}+\frac{1}{2}h_{\mu\nu}-\frac{1}{8}h_{\mu\rho}\eta^{\rho\sigma}h_{\sigma\nu}+\ldots\right)u_{I}^{\nu}
\end{equation}
up to a certain pN order. (Notice that a vierbein is not uniquely fixed by the
metric, but the one used here is entering the derivation of the Feynman rules.)
Expressed in variables defined in an Euclidean flat space coordinate frame the vierbein components are given up to all terms necessary for our declared approximation according to \cite{Steinhoff:Schafer:2009:1} by
(see also \cite{Porto:Rothstein:2008:1:err})
\begin{subequations}
\begin{align}
e_{(0)0} &= 1 + \frac{1}{2} h_{00} = 1 - G \frac{m_1}{r_1} - G \frac{m_2}{r_2}+\mathcal{O}(c^{-4})\label{e0c}\,, \\
% e_{(0)0} &= 1 + \frac{1}{2} \delta g_{00} = 1 - G \frac{m_1}{r_1}
% 	- 3 G v^i_1 S_1^{ij} \frac{n_1^j}{r_1^2} + (1 \leftrightarrow 2) \\
e_{(0)i} = e_{(i)0} &= \frac{1}{2} h_{0i} = 2G v^i_1 \frac{m_1}{r_1} - G S_1^{(i)(j)} \frac{n_1^j}{r_1^2} + (1 \leftrightarrow 2)+\mathcal{O}(c^{-5})\label{eic}\,,  \\
e_{(i)j} &= - \delta_{ij} + \frac{1}{2}  h_{ij} = - \delta_{ij} \left( 1
	+ G \frac{m_1}{r_1} + G \frac{m_2}{r_2} \right)+\mathcal{O}(c^{-4})\label{eijc} \,.
% e_{(i)j} &= - \delta_{ij} + \frac{1}{2} \delta g_{ij} = - \delta_{ij} \left( 1 + G \frac{m_1}{r_1}
% 	+ G v^i_1 S_1^{ij} \frac{n_1^j}{r_1^2} + (1 \leftrightarrow 2) \right)
\end{align}
The velocities depending on the momenta follow from Legendre transformation using the pN expanded Lagrangian for two spinning compact body interaction leading to Eq.\ (\ref{geschw}). Notice the vierbein is chosen to be symmetric so that local Lorentz indices and coordinate indices get indistinguishable yielding the 4-velocities in Euclidean flat space coordinates and depending on canonical momenta which replace
the coordinate velocities via Eq.\ (\ref{geschw}) indicated by $\simeq$
\end{subequations}
\begin{subequations}
\begin{align}
  u^{(0)}_{1}&=e^{(0)}_{\;\;\;\mu}(\vct{z}_1)u^{\mu}_{1}=e_{(0)0}(\vct{z}_1)u_{1}^{0}+e_{(0)i}(\vct{z}_1)u_{1}^{i}= 1 - G \frac{m_2}{r_{12}}+\mathcal{O}(c^{-4})\label{u0red}\\
  u^{(i)}_{1}&=e^{(i)}_{\;\;\;\mu}(\vct{z}_1)u^{\mu}_{1}=-e_{(i)0}(\vct{z}_1)u_{1}^{0}-e_{(i)j}(\vct{z}_1)u_{1}^{j}\,,\\
             &\simeq\frac{p_{1i}}{m_{1}}\left(1-\frac{\vct{p}_{1}^2}{2m_{1}^2}\right)-\frac{2Gm_{2}p_{1i}}{m_{1}r_{12}}+\frac{3Gp_{2i}}{2r_{12}}+\frac{Gn_{12}^{i}\piin}{2r_{12}}\nonumber\\
             &\quad-\frac{Gm_{2}n_{12}^kS_{1}^{(i)(k)}}{m_{1}r_{12}^2}-\frac{Gn_{12}^kS_{2}^{(i)(k)}}{r_{12}^2}+\mathcal{O}(c^{-5}) \label{uired}\,.
\end{align}
\end{subequations}
The covariant SSC (\ref{cssc3}) is then pN expanded to yield
\begin{subequations}
\begin{align}
0 &= S^{(0)(i)}_1 + S^{(i)(j)}_1 \frac{u^{(j)}_1}{u^{(0)}_1} \\
&= S^{(0)(i)}_1 + S^{(i)(j)}_1 \left[ v^j_1 + 2 G (v^j_1 - v^j_2) \frac{m_2}{r_{12}} + G S_2^{jk} \frac{n_{12}^k}{r_{12}^2} \right]+\mathcal{O}(c^{-8}) \\
&\simeq S^{(0)(i)}_1 + S^{(i)(j)}_1 \Bigg[ \frac{p_{1j}}{m_{1}}\left(1-\frac{\vct{p}_{1}^2}{2m_{1}^2}\right)
	+ \frac{G}{2 r_{12}} \left( 3 p_{2j} - 2 \frac{m_2}{m_1} p_{1j}
		+ (\vct{n}_{12} \cdot \vct{p}_2) n_{12}^j \right)\nonumber\\
	&\qquad\qquad- \frac{G n_{12}^k}{r_{12}^2} \left( S_2^{(j)(k)} + \frac{m_2}{m_1} S_1^{(j)(k)} \right) \Bigg]+\mathcal{O}(c^{-8})\,.
	\label{SSSC}
\end{align}
\end{subequations}
Likewise the Lorentz matrix constraint (\ref{lambda0ssc3}) is pN expanded to give
 \begin{subequations}
 \begin{align}
 0 &= \Lambda^{[0](i)}_1 - \frac{u^{(i)}_1}{\sqrt{\left( u^{(0)}_1 \right)^2 -\left(u^{(i)}_1\right)^2}} \\
 &= \Lambda^{[0](i)}_1 - ( 1 - \vct{v}_1^2 )^{-\frac{1}{2}} v^i_1
 	- 2G (v^i_1 - v^i_2) \frac{m_2}{r_{12}} - G S_2^{ij} \frac{n_{12}^j}{r_{12}^2}+\mathcal{O}(c^{-5}) \\
 &\simeq \Lambda^{[0](i)}_1 - \frac{p_{1i}}{m_1}
 	- \frac{G}{2 r_{12}} \left( 3 p_{2i} - 2 \frac{m_2}{m_1} p_{1i}
 		+ (\vct{n}_{12} \cdot \vct{p}_2) n_{12}^i \right)\nonumber\\
 	&\quad+ \frac{G n_{12}^j}{r_{12}^2} \left( S_2^{(i)(j)} + \frac{m_2}{m_1} S_1^{(i)(j)} \right)+\mathcal{O}(c^{-5})\,.
 	\label{LSSC}
 \end{align}
 \end{subequations}
% \begin{subequations}
% \begin{align}
% 0 &= \Lambda^{[0](i)}_1 - \frac{p^{(i)}_1}{m_{1}} \\
% %&= \Lambda^{[0](i)}_1 - ( 1 - \vct{v}_1^2 )^{-\frac{1}{2}} v^i_1
% 	%- 2G (v^i_1 - v^i_2) \frac{m_2}{r_{12}} - G S_2^{ij} \frac{n_{12}^j}{r_{12}^2}++\mathcal{O}(c^{-5}) \\
% &\simeq \Lambda^{[0](i)}_1 - \frac{p_{1i}}{m_1}
% 	- \frac{G}{2 r_{12}} \left( 3 p_{2i} - 2 \frac{m_2}{m_1} p_{1i}
% 		+ (\vct{n}_{12} \cdot \vct{p}_2) n_{12}^i \right)\nonumber\\
% 	&\quad+ \frac{G n_{12}^j}{r_{12}^2} \left( S_2^{(i)(j)} + \frac{m_2}{m_1} S_1^{(i)(j)} \right)+\mathcal{O}(c^{-5})\,.
% 	\label{LSSC}
% \end{align}
% \end{subequations}
These 6 constraints per body (12 in total) are comprised into a set of 12 elements $\Phi_A=0$, $A=1, \dots, 12$, which enables us to
construct the Dirac bracket out of them according to the standard rule, if we know the Poisson brackets between the various
quantities which enter the constraints and the Dirac bracket, which are given by (\ref{PBa})-(\ref{PBc}) e.g., we
have the following Poisson brackets between the constrained variables:
% For the spin variables we use the so(1,3) Poisson bracket (notice the signature of spacetime
%is still -2):
%\begin{equation}
% \{ S_{Iab}, S_{Icd} \} = S_{Ica} \eta_{bd} - S_{Ida} \eta_{bc} + S_{Idb} \eta_{ac} - S_{Icb} \eta_{ad} \,.
% \end{equation}
% For momentum and position variable we have the standard canonical Poisson bracket
% \begin{equation}
% \{ z^{i}_{I}, p_{Ij} \} = \delta_{ij}
% \end{equation}
\begin{subequations}
 \begin{align}
  \{\Lambda^{A(j)}_{I},S^{(m)(n)}_{I}\}&=-\delta_{mj}\Lambda^{A(n)}_{I}+\delta_{nj}\Lambda^{A(m)}_{I}\,,\\
  \{\Lambda_{I}^{A(0)},S_{I}^{(0)(k)}\}&=\Lambda_{I}^{A(k)}\,,\\
  \{\Lambda_{I}^{A(j)},S_{I}^{(0)(k)}\}&=\Lambda_{I}^{A(0)}\delta_{jk}\,.
 \end{align}
\end{subequations}
After applying all Poisson brackets, the constraints can be used in the results of the Dirac brackets.
The constraints $\Phi_A=0$ are given above by (\ref{SSSC}) for $A = 1, 2, 3$ for object 1
and $A = 7, 8, 9$ for object 2 and by (\ref{LSSC}) for $A = 4, 5, 6$  for object 1
and $A = 10, 11, 12$ for object 2. The matrix $C_{AB}$ from the definition (\ref{cmatrix1}) is decomposed in a Minkowski part $C_{(0)AB}$
and a curvature correction part linear in $G$, $C_{(1)AB}$, approximated by the needed pN order:
\begin{subequations}
\begin{align}
C_{AB} &= C_{(0)AB} + C_{(1)AB}\, \\
C^{AB} &= C_{(0)}^{AB} - C_{(0)}^{AC} C_{(1)CD} C_{(0)}^{DB}\,.
\end{align}
\end{subequations}
Referring to \cite{Hanson:Regge:1974} the Minkowski parts $C_{(0)AB}$ and $C_{(0)}^{AC}$ can be obtained
analytically exact with the %mass-shell constraint $p_{I\mu}p^{\mu}_{I}=m_{I}^2$ inserted as 
formal definition $p^0_I = \sqrt{m_I^2 + \vct{p}_I^2}$ yielding:
\begin{subequations}
\begin{align}
C_{(0)AB} &= \left( \begin{array}{cccc}
- \frac{m_1^2}{(p_1^0)^2} S_1^{(i)(j)} & - \frac{m_1}{p_1^0} \mathcal{P}_{1ij}^{-1} & 0 & 0 \\
\frac{m_1}{p_1^0} \mathcal{P}_{1ij}^{-1} & 0 & 0 & 0 \\
0 & 0 & - \frac{m_2^2}{(p_2^0)^2} S_2^{(i)(j)} & - \frac{m_2}{p_2^0} \mathcal{P}_{2ij}^{-1} \\
0 & 0 & \frac{m_2}{p_2^0} \mathcal{P}_{2ij}^{-1} & 0
\end{array} \right)\,, \\
C_{(0)}^{AB} &= \left( \begin{array}{cccc}
0 & \frac{p_1^0}{m_1} \mathcal{P}_{1ij} & 0 & 0 \\
- \frac{p_1^0}{m_1} \mathcal{P}_{1ij} &
	- \mathcal{P}_{1ik} \mathcal{P}_{1jl} S_1^{(k)(l)} & 0 & 0 \\
0 & 0 & 0 & \frac{p_2^0}{m_2} \mathcal{P}_{2ij} \\
0 & 0 & - \frac{p_2^0}{m_2} \mathcal{P}_{2ij}^{-1} &
	- \mathcal{P}_{2ik} \mathcal{P}_{2jl} S_2^{(k)(l)}
\end{array} \right)\,, \\
\mathcal{P}_{Iij} &= \delta_{ij} - \frac{p_{Ii}p_{Ij}}{(p_I^0)^2} \,, \quad
\mathcal{P}_{Iij}^{-1} = \delta_{ij} + \frac{p_{Ii}p_{Ij}}{m_I^2}\,.
\end{align}
\end{subequations}
The notation is adapted to the one used in \cite{Hanson:Regge:1974}, whence the same calculation was done in the Minkowski case which should be included here as a limiting case.
As a matter of fact we will only need the Minkowski part $C_{(0)AB}$ for our calculation, because all terms up to the next-to-leading pN order $c^{-8}$
of the Dirac bracket are produced by the vectors to the left and to the right of the inverse matrix $C^{AB}$,
which can be checked for all possible Dirac brackets, so no curvature terms of $C^{AB}$ will contribute to the Dirac bracket to NLO.
Interestingly the object $\mathcal{P}_{Iij}$ also appears in surface terms of the stress-energy tensor algebra in  Minkowski spacetime,
see Eq.\ (A7) in \cite{Steinhoff:Schafer:Hergt:2008}.

\subsection{Transition to Newton-Wigner variables}

The Dirac bracket reads in reduced approximation to linear order in G (indicated by $\simeq$)
\begin{equation}
 \{ Q_{1}, Q_{2} \}_{\text{D}} \simeq \{ Q_{1}, Q_{2} \} - \{ Q_{1}, \Phi_A \} C_{(0)}^{AB} \{ \Phi_B, Q_{2} \}\,.
\end{equation}
A list of all possible combinations of quantities which may enter the Dirac bracket and their results is given in the appendix \ref{sec:DBA}.
Assuming that we have calculated the Dirac brackets between all the variables entering the Hamiltonian the transition to Newton-Wigner variables
$\vct{\hat{z}}_1$, $\vct{\hat{p}}_1$, $\vct{\hat{S}}_1$,
$\vct{\hat{z}}_2$, $\vct{\hat{p}}_2$, and $\vct{\hat{S}}_2$,
can be performed. The Dirac brackets of these new variables should be standard canonical fulfilling their natural standard commutation relations
\begin{subequations}
\begin{align}
\{ \hat{z}_I^{i}, \hat{p}_{Jj} \}_{\text{D}} &= \delta^i_j\delta_{IJ}\label{relz} \,,\\
\{\hat{S}_{I}^{(i)(j)},\hat{S}_{I}^{(k)(l)}\}_{\text{D}}&=\delta_{jl}\hat{S}_{I}^{(i)(k)}-\delta_{jk}\hat{S}_{I}^{(i)(l)}+\delta_{ik}\hat{S}_{I}^{(j)(l)}-\delta_{il}\hat{S}_{I}^{(j)(k)}\label{relS}\,,\,%\text{\bf and}
\end{align}
\end{subequations}
and \emph{all other possible brackets should vanish}. For the sake of completion we also list the standard Poisson bracket relation for the standard canonical Lorentz matrix $\hat{\Lambda}^{[i](j)}$, which happens to be a pure 3-dimensional rotation matrix fulfilling the standard Poisson bracket relation
\begin{equation*}
 \{\hat{\Lambda}^{[i](j)}_{I},\hat{S}_{I(k)(l)}\}_{\text{D}}=\hat{\Lambda}^{[i](k)}_{I}\delta_{lj}-\hat{\Lambda}^{[i](l)}_{I}\delta_{kj}\,,
\end{equation*}
while all other possible brackets with $\hat{\Lambda}^{[i](j)}$ being zero.
The Newton-Wigner variables that are assumed to exist up to the considered order are fixed up to canonical transformations which opens up the possibility to choose a representation that leaves the momenta unchanged, $\vct{\hat{p}}_I=\vct{p}_I$. In order to find the position and spin variable transformation we make a general ansatz
with undetermined coefficients, e.g., for the position variable the general ansatz (excluding the Minkowski term which is known exactly) reads to the pN order of $c^{-5}$
\begin{align}\label{zansatz}
\begin{aligned}
 z_{1}^{i}=&~\hat{z}_{1}^{i}+G\bigg[\zeta_{1}\frac{p_{1k}\hat{S}_{1(i)(k)}}{\hat{r}_{12}}+\zeta_{2}\frac{p_{2k}\hat{S}_{1(i)(k)}}{\hat{r}_{12}}+\zeta_{3}\frac{n^k_{12}\pinNW\hat{S}_{1(i)(k)}}{\hat{r}_{12}}+\zeta_{4}\frac{n^k_{12}\piinNW\hat{S}_{1(i)(k)}}{\hat{r}_{12}}\\
&+\zeta_{5}\frac{n^k_{12}\hat{S}_{1(i)(k)}}{\hat{r}_{12}}+\zeta_{6}\frac{\hat{S}_{1(k)(l)}\hat{S}_{1(i)(l)}\hat{n}_{12}^{k}}{\hat{r}_{12}^2}+\zeta_{7}\frac{\hat{S}_{1(k)(l)}\hat{S}_{1(i)(l)}p_{1k}}{\hat{r}_{12}^2}+\zeta_{8}\frac{\hat{S}_{1(k)(l)}\hat{S}_{1(i)(l)}p_{2k}}{\hat{r}_{12}^2}\\
&+\zeta_{9}\frac{\hat{S}_{1(k)(l)}\hat{S}_{1(i)(l)}\pinNW\hat{n}_{12}^{k}}{\hat{r}_{12}^2}+\zeta_{10}\frac{\hat{S}_{1(k)(l)}\hat{S}_{1(i)(l)}\piinNW\hat{n}_{12}^{k}}{\hat{r}_{12}^2}\\
&+\zeta_{11}\frac{\hat{S}_{2(k)(l)}\hat{S}_{1(i)(l)}\hat{n}_{12}^{k}}{\hat{r}_{12}^2}+\zeta_{12}\frac{\hat{S}_{2(k)(l)}\hat{S}_{1(i)(l)}p_{1k}}{\hat{r}_{12}^2}+\zeta_{13}\frac{\hat{S}_{2(k)(l)}\hat{S}_{1(i)(l)}p_{2k}}{\hat{r}_{12}^2}\\
&+\zeta_{14}\frac{\hat{S}_{2(k)(l)}\hat{S}_{1(i)(l)}\pinNW\hat{n}_{12}^{k}}{\hat{r}_{12}^2}+\zeta_{15}\frac{\hat{S}_{2(k)(l)}\hat{S}_{1(i)(l)}\piinNW\hat{n}_{12}^{k}}{\hat{r}_{12}^2}\\
&+(1\leftrightarrow 2)\left(\zeta_{16},...,\zeta_{30}\right)\Bigg]+\mathcal{O}(c^{-5})\,.
\end{aligned}
\end{align}
It is worth mentioning that to first order `covariant' and Newton-Wigner variables agree,
so that the hat on the variables of the $G$-terms can also be thought of as erased when trying
to cancel similar terms in the Dirac bracket relations. The ansatz (\ref{zansatz}) contains 30
coefficients to be determined which in general fulfill no symmetries among themselves and are independent
of each other, although most of them will be set to zero. Notice the coefficients also depend on mass factors
and are therefore not dimensionless. In actuality it is more practical to choose a shorter ansatz which focuses
on terms which are present in the Dirac brackets, because only those terms have to be cancelled in order to arrive
at canonical Dirac brackets, all other terms in the general ansatz are a priori zero. The ansatz (\ref{zansatz})
is therefore to be inserted into the Dirac brackets which contain the position variable, e.g., (\ref{relzD}).
It turns out that the fulfillment of the crucial Poisson/Dirac bracket relation (\ref{relz}) is already enough
to uniquely fix all coefficients in (\ref{zansatz}), all other Dirac bracket relations that would make $\hat{z}$
canonical are then automatically fulfilled and serve merely as consistency checks. The coefficients for (\ref{zansatz}) are then given as
\begin{align}
 \zeta_{1}=\frac{m_{2}}{m_{1}^2}\,,\quad\zeta_{2}=-\frac{3}{2m_{1}}\,,\quad\zeta_{4}=-\frac{1}{2m_{1}}\,,\quad\zeta_{6}=-\frac{m_{2}}{m_{1}^2}\,,\quad\zeta_{11}=-\frac{1}{m_{1}}\,,
\end{align}
and all other coefficients are zero. The complete transformation formula (including the pN expanded Minkowski term) reads
\begin{align}
\begin{aligned}\label{xNW}
 z_{1}^{i}=&~\hat{z}_{1}^{i}-\bigg[\frac{1}{2m_{1}^2}p_{1k}\hat{S}_{1(i)(k)}\left(1-\frac{\vct{p}_{1}^2}{4m_{1}^2}\right)-G\frac{m_{2}}{m_{1}^2}\frac{p_{1k}\hat{S}_{1(i)(k)}}{\hat{r}_{12}}+\frac{3}{2}G\frac{p_{2k}\hat{S}_{1(i)(k)}}{m_{1}\hat{r}_{12}}\\
&+\frac{G}{2}\frac{\hat{n}_{12}^{k}(\vct{\hat{n}}_{12}\cdot\vct{p}_{2})\hat{S}_{1(i)(k)}}{m_{1}\hat{r}_{12}}+G\frac{m_{2}}{m_{1}^2}\frac{\hat{S}_{1(k)(l)}\hat{S}_{1(i)(l)}\hat{n}_{12}^{k}}{\hat{r}_{12}^2}+G\frac{\hat{n}_{12}^{k}\hat{S}_{1(i)(l)}\hat{S}_{2(k)(l)}}{m_{1}\hat{r}_{12}^2}\bigg]+\mathcal{O}(c^{-5})\,.
\end{aligned}
\end{align}
For the spin variable we make a similar ansatz, which we insert into (\ref{relSD}) and demand fulfillment of (\ref{relS}) which again uniquely determines the spin transformation reading
\begin{align}
\begin{aligned}\label{SNW}
 S_{1(i)(j)}=&~\hat{S}_{1(i)(j)}-\bigg[\frac{p_{1[i}\hat{S}_{1(j)](k)}p_{1k}}{m_{1}^2}\left(1-\frac{\vct{p}_{1}^2}{4m_{1}^2}\right)-\frac{2Gm_{2}}{m_{1}^2\hat{r}_{12}}p_{1[i}\hat{S}_{1(j)](k)}p_{1k}\\
 &+\frac{3G}{m_{1}\hat{r}_{12}}p_{1[i}\hat{S}_{1(j)](k)}p_{2k}+\frac{G}{m_{1}\hat{r}_{12}}p_{1[i}\hat{S}_{1(j)](k)}\hat{n}_{12}^{k}(\vct{\hat{n}}_{12}\cdot\vct{p}_{2})\\
 &+\frac{2Gm_{2}}{m_{1}^2\hat{r}_{12}^2}p_{1[i}\hat{S}_{1(j)](l)}\hat{S}_{1(k)(l)}\hat{n}_{12}^{k}+\frac{2G}{m_{1}\hat{r}_{12}^2}p_{1[i}\hat{S}_{1(j)](l)}\hat{S}_{2(k)(l)}\hat{n}_{12}^{k}\bigg]+\mathcal{O}(c^{-8})\,.
\end{aligned}
\end{align}
The transformation of the Lorentz matrix is derived in the following section.

\section{Reduction via an action principle\label{sec:action}}
Spin in relativity can also be treated by an action principle
\cite{Goenner:Westpfahl:1967, Romer:Westpfahl:1969, Westpfahl:1969:2}, see also
\cite{Hanson:Regge:1974, Bailey:Israel:1975, Porto:2006,
Steinhoff:Schafer:2009:2, Steinhoff:2011} and appendix A of \cite{DeWitt:2011}. It is indeed possible to derive the
transformation to \emph{reduced} canonical variables using an action approach.
Let us start with the effective action for two interacting spinning compact objects in curved space
\begin{align}\label{effaction}
 S_{\text{eff}}=\int\dd t\, L_{\text{eff}}=\int\dd t\left(p_{1i}\dot z^{i}_{1}+p_{2i}\dot z^{i}_{2}-\frac{1}{2}S_{1ab}\Omega_{1}^{ab}-\frac{1}{2}S_{2ab}\Omega_{2}^{ab}-H_{\text{eff}}\left(\vct{z}_{I},\vct{p}_{J},S_{Kab},\Lambda^{Aa}_{L}\right)\right) \,.
\end{align}
%The variables in the Hamiltonian $H_{\text{eff}}\left(\vct{z}_{I},\vct{p}_{I},S_{Iab}\right)$ correspond to the covariant SSC $S_{ab}u^{b}=0$ which is meant to be fulfilled locally through a projection on to a local Lorentz basis by a Vierbein $e_{a\mu}$ which fulfills the condition
The variables in the Hamiltonian $H_{\text{eff}}\left(\vct{z}_{I},\vct{p}_{J},S_{Kab},\Lambda^{Aa}_{L}\right)$
are all independent from each other, no SSC is imposed yet on this stage. The variables in this action 
span therefore a too large phase space, because of the redundant $S_{I(0)(i)}$ and $\Lambda^{[0]a}$ degrees of freedom, which
makes the phase space still unphysical, but will give us some insight into the non-reduced Poisson brackets
between all the variables of the action. As usual the spin tensor $S_{ab}$ is defined locally through a projection onto a
local Lorentz basis by a vierbein $e_{a\mu}$ which fulfills the condition
\begin{align}
 e_{a}^{\;\;\mu}e_{b}^{\;\;\nu}g_{\mu\nu}=\eta_{ab}\,,\quad\eta_{ab}=\text{diag}(1,-1,-1,-1)
\,,\quad e_{\;\;\mu}^{a}e_{\;\;\nu}^{b}\eta_{ab}=g_{\mu\nu}
\end{align}
with the Lorentz indices $(a,b,c,d)$ and spacetime coordinate indices $(\mu,\nu,..)$
\begin{align}
 (a,b,c,d)\in\{(0),(i)\}\quad\text{and}\quad i\in\{1,2,3\}\,,\quad(\mu,\nu,\sigma,\ldots)\in\{0,1,2,3\}\,.
\end{align}
We need still another index label appearing in the definition of the angular velocity tensor $\Omega_{ab}$, whose generalized momentum is the spin tensor $S_{ab}$, which involves the Lorentz matrix $\Lambda_{A\mu}$ and therefore rotations and boosts with capital letters labeling body-fixed Lorentz indices in the sense that
\begin{align}\label{lambdaindex}
 \Lambda_{A\mu}\Lambda^{A}_{\;\;\;\nu}=g_{\mu\nu}\,,\quad\Lambda_{Aa}\Lambda^{A}_{\;\;\;b}=\eta_{ab}\quad\text{with}\quad(A,B,..)\in\{[0],[i]\}\,.
\end{align}
Notice that $\Lambda_{Ab}$ is time dependent. Now the definition is (the dot marking total time derivative)
\begin{equation}\label{omegadef}
 \Omega^{ab}\equiv\Lambda_{A}^{\;\;\;a}\dot\Lambda^{Ab}\quad\text{making}\quad\Omega^{ab}=-\Omega^{ba}=\Omega^{[ab]}.
\end{equation}
The minus sign in front of the spin kinematic term $\frac{1}{2}S_{Iab}\Omega_{I}^{ab}$ in the action (\ref{effaction}) is due to the sign convention for the antisymmetric angular velocity tensor $\Omega^{ab}$ and the used signature.
% \footnote{In fact, we should absorb this sign in the definition of the angular velocity; see also the next footnote.}.
Indeed, variation of the action gives (dropping boundary terms)
\begin{align}
\begin{aligned}
\delta S_{\text{eff}}&=\int\dd t \bigg[ \bigg( \dot z^{i}_{1} - \frac{\pa H_{\text{eff}}}{\pa p_{1i}} \bigg) \delta p_{1i}
	- \bigg( \dot p_{1i} + \frac{\pa H_{\text{eff}}}{\pa z^{i}_{1}} \bigg) \delta z^{i}_{1}\\
	&- \bigg( \frac{1}{2} \Omega_{1}^{ab} + \frac{\pa H_{\text{eff}}}{\pa S_{1ab}} \bigg) \delta S_{1ab}
	+ \bigg( \frac{1}{2} \dot S_{1ab} - S_{1ca} \Omega_{1b}{}^c+\frac{1}{2}\left(\frac{\partial H_{\text{eff}}}{\partial\Lambda^{Aa}_{1}}\Lambda^{A}_{1\;\,b}-\frac{\partial H_{\text{eff}}}{\partial\Lambda^{Ab}_{1}}\Lambda^{A}_{1\;\,a}\right) \bigg) \delta \theta_{1}^{ab}
\\&+ (1 \leftrightarrow 2) \bigg] \,,
\end{aligned}
\end{align}
where the variation of $\Lambda^{Ab}$ was written in terms of the antisymmetric symbol $\delta\theta^{ab}=\Lambda_A{}^a\delta\Lambda^{Ab}$ and we used the relation $\delta\Omega^{ab}=\frac{\dd}{\dd t}\delta\theta^{ab}+2\Omega_c{}^{[a}\delta\theta^{b]c}$. The variation of the Hamiltonian with respect to the independent degrees of freedom of the Lorentz matrices is also achieved by usage of $\delta\theta^{ab}$ according to
\begin{align}
\delta H_{\text{eff}}&=\frac{\partial H_{\text{eff}}}{\partial\Lambda^{Ab}}\delta\Lambda^{Ab}=\frac{\partial H_{\text{eff}}}{\partial\Lambda^{Ab}}\eta^{AB}\delta\Lambda_{B}^{\,\,\,\,\,b}=\frac{\partial H_{\text{eff}}}{\partial\Lambda^{Ab}}\Lambda^{A}_{\;\;\,c}\Lambda^{Bc}\delta\Lambda_{B}^{\,\,\,\,\,b}=\frac{\partial H_{\text{eff}}}{\partial\Lambda^{A[b}}\Lambda^{A}_{\;\;\,c]}\delta\theta^{cb}\\
        &=-\frac{1}{2}\left(\frac{\partial H_{\text{eff}}}{\partial\Lambda^{Aa}}\Lambda^{A}_{\;\;\,b}-\frac{\partial H_{\text{eff}}}{\partial\Lambda^{Ab}}\Lambda^{A}_{\;\;\,a}\right)\delta\theta^{ab}\,.
\end{align}
The equations of motion are therefore given by
\begin{equation}
\dot z^{i}_{I} = \frac{\pa H_{\text{eff}}}{\pa p_{Ii}} \,, \qquad
\dot p_{Ii} = - \frac{\pa H_{\text{eff}}}{\pa z^{i}_{I}} \,, \qquad
\dot S_{Iab} = 4 S_{Ic[a} \eta_{b]d} \frac{\pa H_{\text{eff}}}{\pa S_{Icd}}-2\eta_{c[a}\Lambda^{A}_{I\,b]}\frac{\partial H_{\text{eff}}}{\partial\Lambda_{I\,c}^{A}} \,,
\end{equation}
and from the definition 
\begin{equation*}
 \dot z^{i}_{I}=\{z^{i}_{I},H_{\text{eff}}\}\,,\qquad
 \dot p_{Ii}=\{p_{Ii},H_{\text{eff}}\}\,,\qquad
 \dot S_{Iab}=\{S_{Iab},H_{\text{eff}}\}
\end{equation*}
we can thus read off the Poisson brackets (see also (\ref{nrbrackets}))
\begin{align}
\{ z^{i}_{I}, p_{Jj} \} &= \delta_{ij}\delta_{IJ}\label{PBdim4a} \,,\\
\{ S_{Iab}, S_{Icd} \} &= S_{Ica} \eta_{bd} - S_{Ida} \eta_{bc} + S_{Idb} \eta_{ac} - S_{Icb} \eta_{ad}\label{PBdim4b} \,,\\
\{ S_{Iab},\Lambda^{A}_{I\,c}\}&=\eta_{bc}\Lambda^{A}_{I\,a}-\eta_{ac}\Lambda^{A}_{I\,b}\,,\label{PBdim4c}
\end{align}
%all other zero (the Poisson brackets for $\Lambda^{Ab}$ are not shown, but are given in, e.g., \cite{Hanson:Regge:1974}).
and all other brackets are zero. The goal is to transform (\ref{effaction}) into the canonical form
\begin{align}
\begin{aligned}\label{acan}
 \hat{S}_{\text{eff}}=\int\dd t\, \hat{L}_{\text{eff}}&=\int\dd t\bigg(\hat{p}_{1i}\dot{\hat{z}}^{i}_{1}+\hat{p}_{2i}\dot {\hat{z}}^{i}_{2}-\frac{1}{2}\hat{S}_{1(i)(j)}\hat{\Omega}_{1}^{(i)(j)}-\frac{1}{2}\hat{S}_{2(i)(j)}\hat{\Omega}_{2}^{(i)(j)}\\
                                                      &\quad\quad\qquad-H_{\text{can}}\left(\hat{\vct{z}}_{I},\hat{\vct{p}}_{J},\hat{\vct{S}}_{K},\hat{\Lambda}^{[i](j)}_{L}\right)\bigg)\,,
\end{aligned}
\end{align}
with $\hat{S}_{(i)(j)}=\epsilon_{ijk}\hat{S}_{(k)}$, $\epsilon_{ijk}=\frac{1}{2}(i-j)(j-k)(k-i)$, and $\hat{\Omega}^{(i)(j)}=\hat{\Lambda}_{[k]}^{\;\;\;(i)}\dot{\hat{\Lambda}}^{[k](j)}$.
Notice that with the used conventions the formula for the angular velocity
vector $\hat{\Omega}^{(i)} = - \hat{\Omega}_{(i)} = - \frac{1}{2} \epsilon_{ijk} \hat{\Omega}^{(j)(k)}$ involves a minus sign.
% \footnote{In Newtonian mechanics the angular velocity tensor is usually defined as $\hat{\Lambda}^{[k](i)}\dot{\hat{\Lambda}}^{[k](j)}=-\hat{\Omega}^{(i)(j)}$, i.e., with a different sign.}.
The hat labels functions depending on canonical position variables $\{\hat{z}_{I}^{i},\hat{\Lambda}^{[i](j)}_{J}\}$ with their
generalized momenta $\{\hat{p}_{Ii},\hat{S}_{J(i)(j)}\}$ in \emph{reduced} phase space, meaning an appropriate SSC is imposed
to get rid of $S_{I(0)(i)}$ and leading to canonical conjugate variables at
the same time. Variation of the action is completely analogous to above calculation.
Only the 4-indices $a, b, \dots$ have to be replaced by 3-indices $(i), (j), \dots$, so the Poisson brackets (\ref{PBdim4a})-(\ref{PBdim4c}) translate into
\begin{gather}
\{ \hat{z}^{i}_{I}, \hat{p}_{Jj} \} = \delta^i_j\delta_{IJ} \,,\\
\{ \hat{\Lambda}^{[i](j)}_{I}, \hat{S}^{(m)(n)}_{I} \} =
	- \delta_{jm} \hat{\Lambda}^{[i](n)}_{I} + \delta_{jn} \hat{\Lambda}^{[i](m)}_{I} \,, \\
\{ \hat{S}^{(i)(j)}_{I}, \hat{S}^{(k)(l)}_{I} \} =
	\delta_{jl} \hat{S}^{(i)(k)}_{I} - \delta_{jk} \hat{S}^{(i)(l)}_{I}
	+ \delta_{ik} \hat{S}^{(j)(l)}_{I} - \delta_{il} \hat{S}^{(j)(k)}_{I} \,,
\end{gather}
all other zero. We used $\eta_{(i)(j)} = - \delta_{ij}$ and the antisymmetry of the spin tensor.
So let us make the reduction in phase space explicit. We impose the covariant SSC, or to put it more exact the Mathisson-Pirani SSC with the 4-velocity
coupled to the 4-dimensional spin tensor. As already mentioned, the reason is that the considered potentials are only valid for this SSC.
% Therefor we have to start out from an already reduced action with some SSC imposed and a covariant one is the most appropriate.
This reduced action is then ready to be transformed to (\ref{acan}) while emerging with the Newton-Wigner SSC. First examine the term $\frac{1}{2}S_{ab}\Omega^{ab}$
(with suppressed particle label) and make a decomposition into time and space parts by using the supplementary conditions fixing the frame of reference
 (see also (\ref{DiracSSC}))
\begin{subequations} 
\begin{align}
 S_{ab}u^{b}&=0\quad\Leftrightarrow\quad S_{(i)b}u^{b}=0\quad\Leftrightarrow\quad S_{(i)(0)}+S_{(i)(j)}\frac{u^{(j)}}{u^{(0)}}=0\,,\label{cssc}\\
 \Lambda^{[i]a}u_{a}&=0\quad\Leftrightarrow\quad\Lambda^{[i](0)}+\Lambda^{[i](j)}\frac{u_{(j)}}{u_{(0)}}=0\,,\label{lambdassc}\\
 \Lambda_{[0]a}&=\frac{u_{a}}{\sqrt{u_{b}u^{b}}}\,.\label{lambda0ssc}
\end{align}
\end{subequations}
Here and in the following $u^a$ should be understood as given in terms of the canonical momentum $\vct{p}$ (not in terms of $\vct{v}$),
the pN approximate relations are given by Eqs.\ (\ref{u0red}) and (\ref{uired}).
Notice that we could have also chosen (\ref{supplcond}) as SSC with $p^{(0)}_{\text{D}}$ already
eliminated by the mass-shell constraint $p^{(0)}_{\text{D}}=\sqrt{m^2+p^{\text{D}}_{(i)}p^{\text{D}}_{(i)}}$.
But again due to the difference between Dixon's momentum and the canonical one in the Legendre transformation it is easier to work with the 4-velocity in the SSCs. As
the full derivation of the variable transformation formulae are quite cumbersome we have put the details in the Appendix \ref{sec:aaction} and shall present here only the key steps and
results. We start with the insertion of the constraints (\ref{cssc})-(\ref{lambda0ssc}) leading us the following `naive' reduced expression of the spin coupling term in the action
\begin{align}
\begin{aligned}
 \frac{1}{2}S_{ab}\Omega^{ab}&=-\frac{1}{2}S_{(j)(k)}\frac{u^{(k)}u_{(l)}}{u^{(0)}u_{(0)}}\tilde{\Omega}^{(l)(j)}-\frac{1}{2}S_{(j)(k)}\frac{u^{(k)}}{u^{(0)}}\Lambda_{[i]}^{\;\;\;(j)}\dot{\Lambda}^{[i](0)}+\frac{1}{2}S_{(k)(l)}\tilde{\Omega}^{(k)(l)}
\end{aligned}
\end{align}
with $\tilde{\Omega}^{(k)(l)}\equiv\Lambda_{[i]}^{\;\;\;(k)}\dot{\Lambda}^{[i](l)}$. Notice the formal difference to the definition of $\Omega^{(k)(l)}$
from $(\ref{omegadef})$, $\tilde{\Omega}^{(k)(l)}$ is therefore not necessarily antisymmetric, which is actually an unwanted feature.
After insertion of $\dot{\Lambda}^{[i](0)}$ by using (\ref{lambdassc}) and further algebraic manipulation we end up with the expression:
\begin{align}\label{somega}
\begin{aligned}
 \frac{1}{2}S_{ab}\Omega^{ab}&=\left(S_{(i)(j)}+S_{(i)(k)}\frac{u^{(k)}u_{(j)}}{u^{(0)}u_{(0)}}-S_{(j)(k)}\frac{u^{(k)}u_{(i)}}{u^{(0)}u_{(0)}}\right)\frac{\tilde{\Omega}^{(i)(j)}}{2}+\frac{1}{2}S_{(j)(k)}\frac{u^{(k)}\dot{u}^{(j)}}{u^{(0)}u_{(0)}}\,.
\end{aligned}
\end{align}
Next thing to do is to redefine variables so that the canonical structure of (\ref{acan}) is produced. Obviously one should start by shifting
$\tilde{\Omega}^{(i)(j)}$ to $\hat{\Omega}^{(i)(j)}$, which should be antisymmetric in order to be the correct velocity variable belonging to the spin tensor.
This is achieved by a redefinition of the Lorentz matrix according to
\begin{equation}\label{Lambdatrafo1}
 \Lambda^{[i](j)}=\hat{\Lambda}^{[i](k)}\left(\eta^{(j)}_{(k)}-\frac{u_{(k)}u^{(j)}}{u(u+u_{(0)})}\right)\,,\quad\text{with}\quad u_{(i)}u^{(i)}\equiv\vct{u}^2\quad\text{so that}\quad u_{a}u^{a}\equiv u^2=u_{(0)}^2+\vct{u}^2\,.
\end{equation}
After a further redefinition of the spin tensor to the canonical (hatted) one according to
\begin{equation}\label{spinredef}
 S_{(i)(j)}=\hat{S}_{(i)(j)}-\hat{S}_{(i)(k)}\frac{u_{(j)}u^{(k)}}{u(u+u_{(0)})}+\hat{S}_{(j)(k)}\frac{u_{(i)}u^{(k)}}{u(u+u_{(0)})}\,,
\end{equation}
we arrive at the following reduced expression for the spin coupling term with one term, the $\mathcal{Z}$-term, left to be cancelled by a proper
position variable shift, because this term includes a local accelaration
\begin{equation}\label{SOmegared}
 \frac{1}{2}S_{ab}\Omega^{ab}=\frac{1}{2}\hat{S}_{(i)(j)}\hat{\Omega}^{(i)(j)}-\mathcal{Z}\quad{\text{with}}\quad\mathcal{Z}\equiv\hat{S}_{(i)(j)}\frac{u^{(i)}\dot{u}^{(j)}}{u(u+u_{(0)})}\,.
\end{equation}
Notice that we could have also used the principle of general covariance to arrive at (\ref{SOmegared}),
because all the equations up to (\ref{SOmegared}) look the same in the special relativistic case for global Minkowski spacetime,
where the round brackets around the local indices of the corresponding variables are erased to yield coordinate indices.
The principle of general covariance in this case would be to rewrite the round brackets around the coordinate indices to arrive at
valid expression in curved spacetime of general relativity. Indeed we recover the result of Hanson and Regge \cite{Hanson:Regge:1974} for the transformation
to the Newton-Wigner spin variable $\hat{\vct{S}}$ and Lorentz matrix $\hat{\Lambda}^{[i](j)}$ in the special relativistic case, when the 4-velocity $u^{\mu}$
in our formulae is replaced by Dixon's momentum $p_{\mu}^{\text{D}}$ (\ref{Dixp}).
The same will be true for the transformation to the Newton-Wigner position variable in the special relativistic case.

To solve for the momenta in (\ref{SOmegared}) one has to insert the vierbein which is perturbatively calculated to the needed pN order, see Eqs.\ (\ref{e0c})-(\ref{eijc}). First we make an expansion of $\mathcal{Z}$ in powers of $\vct{u}^2$ (in the sense of a post-Newtonian approximation) 
\begin{align}\label{Zterm}
 \begin{aligned}
 \mathcal{Z}=\hat{S}_{(i)(j)}\frac{u^{(i)}\dot{u}^{(j)}}{u(u+u_{(0)})}=\hat{S}_{(i)(j)}u^{(i)}\dot{u}^{(j)}\left(\frac{1}{2u_{(0)}^2}-\frac{3\vct{u}^2}{8u_{(0)}^4}+\mathcal{O}\left(\vct{u}^4,c^{-10}\right)\right)\,.
 \end{aligned}
\end{align}
The goal is to find the shift of the position variable to its canonical one only approximately to linear order in $G$
and to leading order in spin-orbit, spin(1)-spin(2) and spin(1)-spin(1) interaction.
We insert (\ref{u0red}) into (\ref{Zterm}) and pN expand the result up to the order $c^{-8}$ leading to
\begin{align}
 \begin{aligned}
 \mathcal{Z}_{1}\simeq\frac{1}{2}\hat{S}_{1(i)(j)}u^{(i)}_{1}\dot{u}^{(j)}_{1}\left(1+2G\frac{m_{2}}{r_{12}}\right)+\mathcal{O}(c^{-10})\,.
 \end{aligned}
\end{align}
Next we insert (\ref{uired}) into this equation yielding the approximate expression
\begin{align}\label{Zcal}
 \begin{aligned}
 &\mathcal{Z}_{1}\simeq\frac{1}{2}\hat{S}_{1(i)(j)}u^{(i)}_{1}\dot{u}^{(j)}_{1}+G\frac{m_{2}}{r_{12}}\hat{S}_{1(i)(j)}u_{1}^{(i)}\frac{\dot{p}_{1j}}{m_{1}}+\mathcal{O}(c^{-10})\,\\
                &\simeq\frac{1}{2m_{1}}\hat{S}_{1(i)(j)}p_{1i}\dot{u}^{(j)}_{1}+\frac{1}{2}\hat{S}_{1(i)(j)}\Bigg[\frac{3Gp_{2i}}{2r_{12}}+\frac{Gn_{12}^{i}\piin}{2r_{12}}-\frac{Gm_{2}n_{12}^kS_{1}^{(i)(k)}}{m_{1}r_{12}^2}-\frac{Gn_{12}^kS_{2}^{(i)(k)}}{r_{12}^2}\Bigg]\frac{\dot{p}_{1j}}{m_{1}}+\mathcal{O}(c^{-10})\,.
\end{aligned}
\end{align}
We eliminate the time derivative of $u_{1}^{(j)}$ by shifting it on-shell (i.e. we neglect total time derivatives symbolized by $\approx$)
onto the momentum leaving us also with a time derivative of the canonical spin, which we will have to deal with later when we reconsider the spin redefinition. So
\begin{align}
\begin{aligned}
 \mathcal{Z}_{1}&\approx-\frac{1}{2m_{1}}\dot{\hat{S}}_{1(i)(j)}p_{1i}u^{(j)}_{1}-\frac{1}{2m_{1}}\hat{S}_{1(i)(j)}\dot{p}_{1i}u^{(j)}_{1}+\frac{1}{2}\hat{S}_{1(i)(j)}\Bigg[\frac{3Gp_{2i}}{2r_{12}}+\frac{Gn_{12}^{i}\piin}{2r_{12}}\\
                &\quad-\frac{Gm_{2}n_{12}^kS_{1}^{(i)(k)}}{m_{1}r_{12}^2}-\frac{Gn_{12}^kS_{2}^{(i)(k)}}{r_{12}^2}\Bigg]\frac{\dot{p}_{1j}}{m_{1}}+\mathcal{O}(c^{-10})\\
                &\simeq-\frac{1}{2m_{1}}\dot{\hat{S}}_{1(i)(j)}p_{1i}u^{(j)}_{1}+\frac{\hat{S}_{1(i)(j)}}{2m_{1}}\Bigg[\frac{p_{1i}}{m_{1}}-\frac{2Gm_{2}p_{1i}}{m_{1}r_{12}}+\frac{3Gp_{2i}}{r_{12}}+\frac{Gn_{12}^{i}\piin}{r_{12}}\\
                &\quad-\frac{2Gm_{2}n_{12}^kS_{1}^{(i)(k)}}{m_{1}r_{12}^2}-\frac{2Gn_{12}^kS_{2}^{(i)(k)}}{r_{12}^2}\Bigg]\dot{p}_{1j}+\mathcal{O}(c^{-10})\,.
 \end{aligned}
\end{align}
Now we are ready to return to the action (\ref{effaction}), wherein we insert Eqs.\ (\ref{SOmegared}) and (\ref{Zcal}) leading to the expression (for particle 1)
\begin{align}
 \begin{aligned}
  S_{\text{eff}}&=\int\dd t\left(p_{1i}\dot z^{i}_{1}-\frac{1}{2}S_{1ab}\Omega_{1}^{ab}-H_{\text{eff}}\right)\\
                &\approx\int\dd t\Bigg(-\dot{p}_{1j}z^{j}_{1}-\frac{1}{2}\hat{S}_{(i)(j)}\hat{\Omega}^{(i)(j)}-\frac{1}{2m_{1}}\dot{\hat{S}}_{1(i)(j)}p_{1i}u^{(j)}_{1}+\frac{\hat{S}_{1(i)(j)}}{2m_{1}}\Bigg[\frac{p_{1i}}{m_{1}}-\frac{2Gm_{2}p_{1i}}{m_{1}r_{12}}\\
                &+\frac{3Gp_{2i}}{r_{12}}+\frac{Gn_{12}^{i}\piin}{r_{12}}-\frac{2Gm_{2}n_{12}^kS_{1}^{(i)(k)}}{m_{1}r_{12}^2}-\frac{2Gn_{12}^kS_{2}^{(i)(k)}}{r_{12}^2}\Bigg]\dot{p}_{1j}-H_{\text{eff}}\Bigg)\,.
 \end{aligned}
\end{align}
This enables us to read off the position coordinate shift
\begin{align}\label{ztrafoA}
 \begin{aligned}
  z_{1}^{j}&=\hat{z}_{1}^{j}+\frac{\hat{S}_{1(i)(j)}}{2m_{1}}\Bigg[\frac{p_{1i}}{m_{1}}-\frac{2Gm_{2}p_{1i}}{m_{1}r_{12}}+\frac{3Gp_{2i}}{r_{12}}+\frac{Gn_{12}^{i}\piin}{r_{12}}-\frac{2Gm_{2}n_{12}^kS_{1}^{(i)(k)}}{m_{1}r_{12}^2}\\
                 &\qquad\qquad\qquad-\frac{2Gn_{12}^kS_{2}^{(i)(k)}}{r_{12}^2}\Bigg]+\mathcal{O}(c^{-6})\,.
 \end{aligned}
\end{align}
This formula coincides with Eq.\ (\ref{xNW}) when spin and position variables on the right hand side of Eq.\ (\ref{ztrafoA})
are provided with a hat in the highest pN terms (meaning all the linear in $G$ terms here), which is allowed when working in a perturbative scheme.
Again the Minkowski term here is shown for pedagogical reasons only, because the Minkowski case can be treated exactly and
in order to arrive at the pN order $c^{-4}$ one has to include one higher Minkowski term in (\ref{ztrafoA}), which changes the coefficient when transforming it perturbatively to the left of Eq.\ (\ref{ztrafoA}).
For the Minkowski case we state that we are always able to write
\begin{equation}\label{4g}
u^{(i)} = \frac{p_i}{\sqrt{m^2 + \vct{p}^2}}\,.
\end{equation}
It follows from Legendre transformation of the point particle Lagrangian ($u^{(i)}\equiv v^i$ in SRT)
\begin{equation}
 S_{\text{pp}}=\int \dd t\, L=m\int\dd\tau=m\int\dd t\,\sqrt{1-\vct{v}^2}\,,
\end{equation}
or from the 3-components of Dixon's momentum (\ref{Dixp}) which happens to be the same as the canonical one but only in the Minkowski case.
Then the addition to the action reads
\begin{equation}
\mathcal{Z} = \frac{\hat{S}_{(k)(j)} u^{(k)}}{u(u+u_{(0)})} \left[ \left( \delta_{ij} - \frac{p_i p_j}{m^2 + \vct{p}^2} \right) \frac{\dot{p}_i}{\sqrt{m^2 + \vct{p}^2}} \right] \,.
\end{equation}
This whole contribution can be absorbed by redefining the position variable as
\begin{align}\label{ztrafoM}
\begin{aligned}
z^{i}&= \hat{z}^{i} + \frac{1}{\sqrt{m^2 + \vct{p}^2}} \left( \delta_{ij} - \frac{p_i p_j}{m^2 + \vct{p}^2} \right) \frac{\hat{S}_{(k)(j)} u^{(k)}}{mu(u+u_{(0)})}\\ &=\hat{z}^{i} + \frac{1}{m^2 + \vct{p}^2} \left( \delta_{ij} - \frac{p_i p_j}{m^2 + \vct{p}^2} \right) \frac{\hat{S}_{(k)(j)} p_{k}}{mu(u+u_{(0)})}\,,\quad\text{notice}\quad\left(\hat{S}_{(k)(j)} p_{k}p_{j}\equiv 0\right)\\
&=\hat{z}^{i} +  \frac{\hat{S}_{(k)(i)} p_{k}}{m(m+\sqrt{m^2+\vct{p}^2})}\quad\text{with}\quad u \equiv \sqrt{ 1 - \vct{v}^2 }\,,\quad u_{(0)}\equiv 1\,.
\end{aligned} 
\end{align}
This is exactly the formula (B.11) from \cite{Hanson:Regge:1974}.
The expanded expression yields
\begin{align}
 z^i=\hat{z}^i-\frac{\hat{S}_{(i)(k)}p_{k}}{2m^2}\left(1-\frac{\vct{p}^2}{4m^2}\right)+\mathcal{O}(c^{-6})\,,
\end{align}
which thus matches the Minkowski terms in Eq.\ (\ref{xNW}).

As already mentioned the spin and the Lorentz matrix need another redefinition in order to cancel the term
$-\frac{1}{2m_{1}}\dot{\hat{S}}_{1(i)(j)}p_{1i}u^{(j)}_{1}$ from the action.
This is achieved by an infinitesimal rotation $\omega^{(i)(j)}=-\omega^{(j)(i)}$ of the local basis so that the canonical spin and Lorentz matrices are corotated according to
\begin{align}\label{spinrotation}
\begin{aligned}
- \frac{1}{2}\hat{S}_{(i)(j)}\hat{\Omega}^{(i)(j)}&\rightarrow-\frac{1}{2}\left[\hat{S}_{(i)(j)}+\omega_{(i)}{}^{(m)}\hat{S}_{(m)(j)}+\omega_{(j)}{}^{(m)}\hat{S}_{(i)(m)}\right]\hat{\Omega}_{\omega}^{(i)(j)}
\end{aligned}
\end{align}
As described in the Appendix you can read off
\begin{equation}
\omega_{(i)}{}^{(m)} = - \omega^{(i)(m)} = \frac{1}{2}\frac{p_{1i}u_{1}^{(m)}}{m_{1}}-\frac{1}{2}\frac{p_{1m}u_{1}^{(i)}}{m_{1}}
\end{equation}
and use Eq.\ (\ref{spinredef}) to determine the final spin redefinition to our approximation:
\begin{align}\label{SCorot}
 \begin{aligned}
  S_{1(i)(j)}&=\hat{S}_{1(i)(j)}-\hat{S}_{1(i)(k)}\frac{u_{1(j)}u_{1}^{(k)}}{2u_{1(0)}}+\hat{S}_{1(j)(k)}\frac{u_{1(i)}u_{1}^{(k)}}{2u_{1(0)}}+\frac{p_{1[i}u_{1}^{(m)]}}{m_{1}}\hat{S}_{(m)(j)}+\frac{p_{1[j}u_{1}^{(m)]}}{m_{1}}\hat{S}_{(i)(m)}\\
             &=\frac{1}{2}\hat{S}_{1(i)(j)}-\hat{S}_{1(i)(k)}\frac{u_{1(j)}u_{1}^{(k)}}{2u_{1(0)}}+\frac{p_{1[j}u_{1}^{(k)]}}{m_{1}}\hat{S}_{(i)(k)}-(i\leftrightarrow j)\,.
\end{aligned}
\end{align}
Next we insert Eq.\ (\ref{u0red}) for $u_{(0)}$ in the second term and Eq.\ (\ref{uired}) for $u^{(i)}$ the third term of (\ref{SCorot}) and make a pN expansion up to the order $c^{-7}$:
\begin{align}\label{Sbbb}
\begin{aligned}
  S_{1(i)(j)}&=\frac{1}{2}\hat{S}_{1(i)(j)}-\frac{1}{2}\hat{S}_{1(i)(k)}u_{1(j)}u_{1}^{(k)}\left(1+2G\frac{m_{2}}{r_{12}}\right)\\
             &\quad+\frac{p_{1[j}}{m_{1}}\hat{S}_{(i)(k)}\Bigg[\frac{p_{1k]}}{m_{1}}-\frac{2Gm_{2}p_{1k]}}{m_{1}r_{12}}+\frac{3Gp_{2k]}}{2r_{12}}+\frac{Gn_{12}^{k]}\piin}{2r_{12}}-\frac{Gm_{2}n_{12}^lS_{1}^{(k)](l)}}{m_{1}r_{12}^2}\\
             &\qquad\qquad\qquad-\frac{Gn_{12}^lS_{2}^{(k)](l)}}{r_{12}^2}\Bigg]-(i\leftrightarrow j)+\mathcal{O}(c^{-9})\,.
\end{aligned}
\end{align}
We are left with an insertion of the remaining 4-velocities. Again utilizing Eq.\ (\ref{uired}) a further expansion of (\ref{Sbbb}) yields
\begin{align}\label{Sccc}
\begin{aligned}
S_{1(i)(j)}&=\frac{1}{2}\hat{S}_{1(i)(j)}+\hat{S}_{1(i)(k)}\Bigg[\frac{1}{2}\frac{p_{1k}p_{1j}}{m_{1}^2}-\frac{Gm_{2}p_{1k}p_{1j}}{m_{1}^{2}r_{12}}+\frac{3}{2}\frac{Gp_{1j}p_{2k}}{m_{1}r_{12}}+\frac{Gp_{1j}n_{12}^{k}(\vct{p}_{2}\cdot\vct{n}_{12})}{2m_{1}r_{12}^2}\\
           &\qquad+\frac{Gm_{2}n_{12}^lS_{1}^{(l)(k)}p_{1j}}{m_{1}^2r_{12}^2}+\frac{Gn_{12}^lS_{2}^{(l)(k)}p_{1j}}{m_{1}r_{12}^2}\Bigg]-(i\leftrightarrow j)+\mathcal{O}(c^{-9})\,.
 \end{aligned}
\end{align}
The last step involves providing all spin and position variables with a hat in the highest pN terms of (\ref{Sccc}), so that we end up with the transformation formula
\begin{align}\label{Strafofinal}
 \begin{aligned}
  S_{1(i)(j)}&=\frac{1}{2}\hat{S}_{1(i)(j)}+\hat{S}_{1(i)(k)}\Bigg[\frac{1}{2}\frac{p_{1k}p_{1j}}{m_{1}^2}-\frac{Gm_{2}p_{1k}p_{1j}}{m_{1}^{2}\hat{r}_{12}}+\frac{3}{2}\frac{Gp_{1j}p_{2k}}{m_{1}\hat{r}_{12}}+\frac{Gp_{1j}\hat{n}_{12}^{k}(\vct{p}_{2}\cdot\hat{\vct{n}}_{12})}{2m_{1}\hat{r}_{12}^2}\\
             &\qquad+\frac{Gm_{2}\hat{n}_{12}^l\hat{S}_{1}^{(l)(k)}p_{1j}}{m_{1}^2\hat{r}_{12}^2}+\frac{G\hat{n}_{12}^l\hat{S}_{2}^{(l)(k)}p_{1j}}{m_{1}\hat{r}_{12}^2}\Bigg]-(i\leftrightarrow j)+\mathcal{O}(c^{-9})\,.
 \end{aligned}
\end{align}
Notice that Eq.\ (\ref{Strafofinal}) (or rather Eq.\ (\ref{SNW}), because one has to extend the Minkowski term to arrive at the order $c^{-7}$) is already high enough in pN order to be used for transforming effective NNLO Hamiltonians to
canonical ones, whereas for that purpose Eq.\ (\ref{ztrafoA}) needs to be extended to the order $c^{-10}$, which in turn means to calculate the vierbein components (\ref{e0c})-(\ref{eijc}) to higher pN orders. In order to get the Minkowski case right one starts with of
Eq.\ (\ref{spinredef}) and replaces the 4-velocities by Eq.\ (\ref{4g}).
The result can be expanded to match the Minkowski terms in Eq.\ (\ref{SNW}).

The most important relation derived in the present section is equation
(\ref{SOmegared}). To arrive at this expression the transformations of Lorentz
matrix and spin tensor shown in (\ref{Lambdatrafo1}) and (\ref{spinredef}) were
used. Notice that no approximation was used to arrive at (\ref{SOmegared}),
only the validity of the supplementary conditions (\ref{cssc}) and
(\ref{lambdassc}) is required. The calculation following (\ref{SOmegared}) is
devoted to the absorption of the term $\mathcal{Z}$ by various further variable 
transformations, most notably by a redefinition of the position. This is
necessary due to the presence of the acceleration $\dot{u}^{(i)}$ in
$\mathcal{Z}$. However, within a perturbative context such terms in the action
can be treated in a simpler way by the method developed in \cite{Schafer:1984,
Damour:Schafer:1991}, which in most cases amounts to an insertion of lower order
equations of motion. This actually corresponds to an \emph{implicit}
redefinition of variables, so a comparison with the Dirac bracket approach is
more difficult in this case and therefore we have not proceeded in this way here (e.g., it is
likely that the canonical momentum is implicitly redefined). However, for an
application at even higher pN orders one can take $- \mathcal{Z}$ as an addition
to the Hamiltonian, after $\dot{u}^{(i)}$ therein was eliminated using the
equations of motion. Also a mixed approach may be useful. One can always write
\begin{equation}
u^{(i)} = \frac{p_i}{m} + \mathcal{V}^i \,,
\end{equation}
where $\mathcal{V}^i$ includes all pN corrections to this relation, see
(\ref{uired}) for the NLO case. $\mathcal{Z}$ then reads
\begin{equation}
\mathcal{Z} = \frac{\hat{S}_{(i)(j)} u^{(i)}}{u(u+u_{(0)})}
	\left( \frac{\dot{p}_i}{m} + \dot{\mathcal{V}}^i \right) \,.
\end{equation}
The first term can now be absorbed by a transformation of the position of the
form
\begin{equation}\label{zredtrafo}
z^{i} = \hat{z}^{i} + \frac{\hat{S}_{(j)(i)} u^{(j)}}{mu(u+u_{(0)})} \,. 
\end{equation}
while the second term is considered as a contribution to the Hamiltonian. This
mixed approach thus consists of an explicit redefinition of the position
implementing the leading order flat space transformation to the Newton Wigner
position, followed by implicit variable redefinitions due to the insertion of
equations of motion.

The discussion in the last paragraph can even serve to modify the Feynman rules
of the EFT formalism to use reduced canonical variables from the very beginning.
First let us explain why such a modification of the Feynman rules provides an
improvement. It is most desirable to formulate the Feynman rules in terms of
reduced spin variables (either covariant or canonical). First, this
allows for intermediate simplifications and leads to more compact
expressions for the potentials. Second, further field variables would
appear at a later stage otherwise, after the potential modes where
supposed to be integrated out. This is due to the fact that the covariant SSC must be
handled at some stage, which introduces the velocity in the local frame and thus
further field variables, see (\ref{ulocal}); at lower orders, this issue is
just semantics, but at higher orders it is of practical relevance. To achieve this goal one
has to discuss the kinematic term $-\frac{1}{2} S_{ab} \Omega^{ab}$ in the
manifestly covariant Lagrangian, which is disregarded in the Routhian approach
as it contains no interactions with the field.\footnote{Notice that the Routhian
in \cite{Porto:Rothstein:2008:1} is therefore not manifestly covariant. By
adding $-\frac{1}{2} S_{ab} \Omega^{ab}$ to it one can get back to the
manifestly covariant Lagrangian contained in Eq.\ (1) of
\cite{Porto:Rothstein:2008:1}, see also Sect.\ III in \cite{Levi:2010} and Sect.\ 5.2.2 in \cite{Steinhoff:2011}.}
However, when the covariant SSC is eliminated at the level of the action this
kinematic term turns into (\ref{somega}), which corresponds to the complicated
kinematics described by the Dirac bracket for the covariant SSC and produces
further interactions with the gravitational field via the velocity in the local
frame. It should be stressed that due to the use of a Lagrangian all equations
of motion can be obtained by a variational principle at any stage, without the
need to resort to nonreduced Poisson brackets as in the Routhian approach. Still
we think one should transform the reduced covariant spin to a reduced canonical
one, as the kinematic terms (\ref{somega}) then simplify to (\ref{SOmegared}).
Incidentally this implies that the spin kinematics can be described by standard
reduced Poisson brackets, with some advantages already mentioned in the
Introduction. This procedure is essentially a straightforward adaption of the
approach in \cite{Steinhoff:Schafer:2009:2} to the EFT formalism, where both the reduction of the covariant spin and the transition to a
reduced canonical spin succeeded at the action level in a consistent and
transparent manner (see also \cite{Steinhoff:2011} for a more detailed
exposition).

Having this said, we propose the following steps to improve the Feynman rules.
First a manifestly covariant Lagrangian is obtained by adding
$-\frac{1}{2} S_{ab} \Omega^{ab}$ given by (\ref{SOmegared}) to the initial
Routhian defined by Eq.\ (7) in \cite{Porto:Rothstein:2008:1}.
Next the covariant SSC is inserted in the usual interaction terms of the
Routhian and the spin is transformed to the reduced canonical one using
(\ref{spinredef}). Finally one can use a mixed approach to eliminate the
acceleration in $\mathcal{Z}$ by writing
$\dot{u}^{(j)} = e^{(j)}{}_k \dot{u}^{k} + \dot{e}^{(j)}{}_k u^{k} + \dot{e}^{(j)}{}_0$
when $u^0 = 1$. Again the first term can be absorbed by a corresponding
redefinition of the position variable and the other terms simply provide further
interaction terms which must be included in the Feynman rules. It is left as a
future task to work this out in detail. The appearance of higher order time
derivatives poses no difficulties at this stage, as one is working with a
Lagrangian. After the Feynman rules were applied, one may even obtain a fully
reduced Hamiltonian in the usual way, if desired (i.e, by eliminating
accelerations and higher order time derivatives followed by a Legendre
transformation in the velocities).

\section{Final comparisons of potentials with Hamiltonians\label{sec:app}}

In this last section we make use of the transformation formulae we have found throughout the previous sections, which enable us to transform all the non-reduced effective Hamiltonians from \Sec{sec:Ltrafo} to reduced ones depending on standard canonical variables. We are especially interested in the effective NLO Hamiltonians, which shall be compared with their ADM counterparts that were calculated directly within the ADM approach, see \cite{Damour:Jaranowski:Schafer:2008:1,Steinhoff:Schafer:Hergt:2008,Hergt:Schafer:2008,Steinhoff:Hergt:Schafer:2008:1,Hergt:Steinhoff:Schafer:2010:1}. 
As the variable transformation of the covariant variables to Newton-Wigner ones has to be done in \textit{all} Hamiltonians up to 2pN order we will again get
NLO correction terms stemming from all the subleading order Hamiltonians starting with the Newtonian Hamiltonian from (\ref{HNewt}). Therein we replace the position variable $\vct{z}$ by $\hat{\vct{z}}$ utilizing Eq.\ (\ref{xNW}) or (\ref{ztrafoA}). The leading-order spin-orbit correction term emerging from this procedure is labeled as $H^{\text{N}}_{\text{LOSO}}$. The upper index refers to the Hamiltonian where the variable replacement is made and lower one is reference to the pN order and the specific Hamiltonian which is to be corrected by that. Likewise we label all the NLO correction terms and get
\begin{align}
 H^{\text{N}}_{\text{LOSO}}&=\frac{G}{\hat{r}_{12}^2}\Bigg[\frac{m_{1}}{2m_{2}}\piicSiinNW-\frac{m_{2}}{2m_{1}}\picSinNW\Bigg]\,,
\end{align}
\begin{align}
\begin{aligned}
 H^{\text{N}}_{\text{NLOSO}}&=\frac{G}{\hat{r}_{12}^2}\Bigg[\frac{m_{2}}{8m_{1}^3}\pipi\picSinNW-\frac{m_{1}}{8m_{2}^3}\piipii\piicSiinNW\Bigg]\\
     &\quad+\frac{G^2}{\hat{r}_{12}^3}\Bigg[\frac{m_{2}^2}{m_{1}}\picSinNW-\frac{3m_{2}}{2}\piicSinNW\\
     &\qquad\qquad+\frac{3m_{1}}{2}\picSiinNW-\frac{m_{1}^2}{m_{2}}\piicSiinNW\Bigg]\,,
\end{aligned}
\end{align}
\begin{align}
\begin{aligned}
H^{\text{N}}_{\text{NLO} S_{1}S_{2}}&=\frac{G}{m_{1}m_{2}\hat{r}_{12}^3}\Bigg[-\frac{3}{4}\pipii\SinNW\SiinNW+\frac{3}{4}\pinNW\SipiiNW\SiinNW\\
                                    &\qquad-\frac{1}{2}\SipiiNW\SiipiNW+\frac{3}{4}\piinNW\SinNW\SiipiNW\\
                                    &\qquad-\frac{3}{4}\pinNW\piinNW\SiSiiNW+\frac{1}{2}\pipii\SiSiiNW\Bigg]\\
    &\quad+\frac{G^2(m_{1}+m_{2})}{\hat{r}_{12}^4}\Bigg[\SinNW\SiinNW-\SiSiiNW\Bigg]\,,
\end{aligned}
\end{align}
\begin{align}
\begin{aligned}
H^{\text{N}}_{\text{NLO} S_{1}^2}&=\frac{G}{\hat{r}_{12}^3}\frac{m_{2}}{m_{1}^3}\Bigg[\frac{3}{8}\pipi\SinNW^2-\frac{3}{4}\pinNW\SinNW\SipiNW\\
                                 &\quad+\frac{1}{4}\SipiNW^2+\frac{3}{8}\pinNW^2\SiSiNW-\frac{1}{4}\pipi\SiSiNW\Bigg]+\frac{G^2m_{2}^2}{m_{1}\hat{r}_{12}^4}\left[\SinNW^2-\SiSiNW\right]\,.
\end{aligned} 
\end{align}
The same replacements of position variables must be performed in the EIH potential giving rise to the following correction terms to the NLO spin-orbit Hamiltonian:
\begin{align}
 \begin{aligned}
  H_{\text{NLOSO}}^{\text{EIH}}&=\frac{G}{\hat{r}_{12}^2}\Bigg(\left(-\frac{3m_{2}}{4m_{1}^3}\vct{p}_{1}^2+\frac{3}{4m_{1}^2}\pinNW\piinNW\right)\picSinNW\\
                                &\quad+\left(\frac{7}{4m_{1}^2}\pipii-\frac{3}{4m_{1}m_{2}}\vct{p}_{2}^2\right)\picSinNW\\
                                &\quad-\frac{1}{4m_{1}^2}\pinNW\picSipiiNW\Bigg)+\frac{G^2}{\hat{r}_{12}^3}\left(\frac{m_{2}}{2}+\frac{m_{2}^2}{2m_{1}}\right)\picSinNW\,.
 \end{aligned}
\end{align}
Next we replace variables in the LO spin-orbit Hamiltonian from Eq.\ (\ref{LOSO}), where we first have to insert the covariant SSC from Eq.\ (\ref{SSSC}) to arrive at a consistent expression. The spin replacement by Eq.\ (\ref{SNW}) yields the correction term
\begin{align}
\begin{aligned}
 H^{\text{LOSO}}_{\text{NLO SO}}&=\frac{G}{\hat{r}_{12}^2}\Bigg(\frac{m_{2}}{m_{1}^3}\vct{p}_{1}^2\vct{\hat{n}}_{12}\cdot\left(\vct{p}_{1}\times\vct{\hat{S}}_{1}\right)-\frac{1}{m_{1}^2}\left(\vct{p}_{1}\cdot\vct{p}_{2}\right)\vct{\hat{n}}_{12}\cdot\left(\vct{p}_{1}\times\vct{\hat{S}}_{1}\right)\\
                               &\quad\quad-\frac{1}{m_{1}^2}\left(\vct{\hat{n}}_{12}\cdot\vct{p}_{1}\right)\vct{p}_{1}\cdot\left(\vct{p}_{2}\times\vct{\hat{S}}_{1}\right)\Bigg)+(1\leftrightarrow 2)\,.
\end{aligned}
\end{align}
Equally replacing the position variables by Eq.\ (\ref{xNW}) in this Hamiltonian gives further correction terms for the NLO $S_{1}S_{2}$ and NLO $S_{1}^2$ Hamiltonians
\begin{align}
\begin{aligned}
H^{\text{LOSO}}_{\text{NLO} S_{1}S_{2}}&=\frac{G}{\hat{r}_{12}^3}\Bigg[\frac{1}{m_{1}^2}\Big(-3\pipi\SinNW\SiinNW+3\pinNW\SipiNW\SiinNW+2\pipi\SiSiiNW\\
            &+3\pinNW\SinNW\SiipiNW-2\SipiNW\SiipiNW-3\pinNW^2\SiSiiNW\Big)\\
            &+\frac{1}{m_{2}^2}\Big(-3\piipii\SiinNW\SinNW+3\piinNW\SiipiiNW\SinNW+2\piipii\SiSiiNW\\
            &-2\SiipiiNW\SipiiNW-3\piinNW^2\SiSiiNW+3\piinNW\SiinNW\SipiiNW\Big)\\
            &+\frac{1}{m_{1}m_{2}}\Big(6\pipii\SinNW\SiinNW-6\pinNW\SipiiNW\SiinNW\\
            &-6\piinNW\SinNW\SiipiNW+4\SipiiNW\SiipiNW\\
            &+6\pinNW\piinNW\SiSiiNW-4\pipii\SiSiiNW\Big)\Bigg]\,,
\end{aligned}
\end{align}
\begin{align}
\begin{aligned}
 H^{\text{LOSO}}_{\text{NLO}S_{1}^2}&=\frac{G}{\hat{r}_{12}^3}\Bigg[\frac{m_{2}}{m_{1}^3}\Big(-3\pipi\SinNW^2+6\pinNW\SinNW\SipiNW-2\SipiNW^2\\
            &\quad-3\pinNW^2\SiSiNW+2\pipi\SiSiNW\Big)+\frac{1}{m_{1}^2}\Big(3\pipii\SinNW^2\\
            &\quad-3\piinNW\SinNW\Sipi-3\pinNW\SinNW\SipiiNW\\
            &\quad+2\Sipi\Sipii+3\pinNW\piinNW\SiSiNW-2\pipii\SiSiNW\Big)\Bigg]\,.
\end{aligned}
\end{align}
Spin replacement in the LO $S_{1}S_{2}$ and LO $S_{1}^2$-Hamiltonian leads to correction terms
\begin{align}
\begin{aligned}
 H^{\text{LO}S_{1}S_{2}}_{\text{NLO}S_{1}S_{2}}&=\frac{G}{\hat{r}_{12}^3}\Bigg[\frac{1}{m_{1}^2}\Big(\frac{3}{2}\pipi\SinNW\SiinNW-\frac{3}{2}\pinNW\SipiNW\SiinNW\\
                                               &+\frac{1}{2}\SipiNW\SiipiNW-\frac{1}{2}\pipi\SiSiiNW\Big)+\frac{1}{m_{2}^2}\Big(\frac{3}{2}\piipii\SiinNW\SinNW\\
                                               &-\frac{3}{2}\piinNW\SiipiiNW\SinNW+\frac{1}{2}\SiipiiNW\SipiiNW-\frac{1}{2}\piipii\SiSiiNW\Big)\Bigg]\,,
\end{aligned}
\end{align}
\begin{equation}
H^{\text{LO}S_{1}^2}_{\text{NLO}S_{1}^2}=\frac{G\,C_{Q_{1}}\,m_{2}}{2m_{1}^3\hat{r}_{12}^3}\Bigg[3\pipi\SinNW-3\pinNW\SinNW\SipiNW+\SipiNW^2-\pipi\SiSiNW\Bigg]\,,
\end{equation}
whilst the replacement of the position variable in these two Hamiltonians leads to no new NLO correction terms.

\subsection{The NLO spin-orbit Hamiltonian of Levi}
We adopt Eq.\ (\ref{LevieffH2}) and make the transition to a canonical Hamiltonian, which depends solely on
the hatted variables $(\hat{\vct{x}}_{I},\hat{\vct{p}}_{J},\hat{\vct{S}}_{K})$. First thing to do is to eliminate
the $S^{(0)(i)}_{I}$ variables by imposing the SSC from Eq.\ (\ref{SSSC}).
The result is the effective Hamiltonian still depending on $(\vct{x}_{I},\vct{p}_{J},\vct{S}_{K})$:
\begin{align}\label{Hefflevi}
\begin{aligned}
H_{\text{SO(L),red}}^{\text{NLO(eff)}}&= 
\frac{Gm_2}{r_{12}^2}{\bf{S}}_1\cdot
\biggl[\frac{{\bf{p}}_1\times{\vct{n}_{12}}}{m_1}\left(\frac{p_1^2}{m_1^2}+\frac{{\bf{p}}_1\cdot{\bf{p}}_2}{m_1m_2}-\frac{p_2^2}{m_2^2}+\frac{3({\bf{p}}_1\cdot{\vct{n}_{12}})({\bf{p}}_2\cdot{\vct{n}_{12}})}{m_1m_2}\right)\\
&\quad+ 
\frac{{\bf{p}}_2\times{\vct{n}_{12}}}{m_2}\left(-\frac{{\bf{p}}_1\cdot{\bf{p}}_2}{m_1m_2}-\frac{3({\bf{p}}_1\cdot{\vct{n}_{12}})({\bf{p}}_2\cdot{\vct{n}_{12}})}{m_1m_2}\right) +
\frac{{\bf{p}}_1\times{\bf{p}}_2}{m_1m_2}\left(\frac{{\bf{p}}_1\cdot{\vct{n}_{12}}}{m_1}-\frac{{\bf{p}}_2\cdot{\vct{n}_{12}}}{m_2}\right)\biggr] \\ 
&\quad+\frac{G^2m_2}{r_{12}^3}{\bf{S}}_1\cdot\left[\frac{{\bf{p}}_1\times{\vct{n}_{12}}}{m_1}\left(6m_1+\frac{13}{2}m_2\right)+\frac{{\bf{p}}_2\times{\vct{n}_{12}}}{m_2}\left(-6m_1-\frac{17}{2}m_2\right) \right] + \left[ 1\leftrightarrow2 \right].
\end{aligned}
\end{align}
To make the transition of this Hamiltonian complete we put a hat on its variables and add up to it all the needed correction terms via
\begin{align}
 H_{\text{SO(L)}}^{\text{NLO(can)}}=H_{\text{SO(L),red}}^{\text{NLO(eff)}}+H^{\text{N}}_{\text{NLOSO}}+H_{\text{NLOSO}}^{\text{EIH}}+H^{\text{LOSO}}_{\text{NLOSO}}
\end{align}
which results in
\begin{align}
 \begin{aligned}
  H_{\text{SO(L)}}^{\text{NLO(can)}} &= \frac{Gm_2}{\hat{r}_{12}^2}{\bf{\hat{S}}}_1\cdot
\biggl[\frac{{\bf{p}}_1\times{\vct{\hat{n}}_{12}}}{m_1}\left(
\frac{5p_1^2}{8m_1^2}+\frac{{\bf{p}}_1\cdot{\bf{p}}_2}{4m_1m_2}-\frac{p_2^2}{4m_2^2}+\frac{9({\bf{p}}_1\cdot{\vct{\hat{n}}_{12}})({\bf{p}}_2\cdot{\vct{\hat{n}}_{12}})}{4m_1m_2}\right) \\
&+ \frac{{\bf{p}}_2\times{\vct{\hat{n}}_{12}}}{m_2}\left(
-\frac{{\bf{p}}_1\cdot{\bf{p}}_2}{m_1m_2}-3\frac{({\bf{p}}_1\cdot{\vct{\hat{n}}_{12}})({\bf{p}}_2\cdot{\vct{\hat{n}}_{12}})}{m_1m_2}\right)+\frac{{\bf{p}}_1\times{\bf{p}}_2}{m_1m_2}\left(\frac{{\bf{p}}_1\cdot{\vct{\hat{n}}_{12}}}{4m_1}-\frac{{\bf{p}}_2\cdot{\vct{\hat{n}}_{12}}}{m_2}\right)\biggr]
\\
&+\frac{G^2m_2}{\hat{r}_{12}^3}{\bf{\hat{S}}}_1\cdot
\left[\frac{{\bf{p}}_1\times{\vct{\hat{n}}_{12}}}{m_1}\left(\frac{11}{2}m_1+5m_2\right)+\frac{{\bf{p}}_2\times{\vct{\hat{n}}_{12}}}{m_2}\left(-6m_1-7m_2\right)\right]+ \left[ 1\leftrightarrow2 \right]\,,
 \end{aligned}
\end{align}
which is to be compared with the NLO spin-orbit ADM canonical Hamiltonian first derived by Damour, Jaranowski and Sch\"afer \cite{Damour:Jaranowski:Schafer:2008:1}, which reads
\begin{align}
\begin{aligned}\label{HDJS}
H_{\text{NLOSO}}^{\text{DJS}} &= \frac{Gm_2}{\hat{r}_{12}^2}{\bf{\hat{S}}}_1\cdot
\biggl[\frac{{\bf{p}}_1\times{\vct{\hat{n}}_{12}}}{m_1}\left(
\frac{5p_1^2}{8m_1^2}+\frac{3{\bf{p}}_1\cdot{\bf{p}}_2}{4m_1m_2}-\frac{3p_2^2}{4m_2^2}+\frac{3({\bf{p}}_1\cdot{\vct{\hat{n}}_{12}})({\bf{p}}_2\cdot{\vct{\hat{n}}_{12}})}{4m_1m_2}+\frac{3({\bf{p}}_2\cdot{\vct{\hat{n}}_{12}})^2}{2m_2^2}\right) \\
&\quad+ \frac{{\bf{p}}_2\times{\vct{\hat{n}}_{12}}}{m_2}\left(
-\frac{{\bf{p}}_1\cdot{\bf{p}}_2}{m_1m_2}-3\frac{({\bf{p}}_1\cdot{\vct{\hat{n}}_{12}})({\bf{p}}_2\cdot{\vct{\hat{n}}_{12}})}{m_1m_2}\right)+\frac{{\bf{p}}_1\times{\bf{p}}_2}{m_1m_2}\left(\frac{3{\bf{p}}_1\cdot{\vct{\hat{n}}_{12}}}{4m_1}-2\frac{{\bf{p}}_2\cdot{\vct{\hat{n}}_{12}}}{m_2}\right)\biggr]
\\
&\quad+\frac{G^2m_2}{\hat{r}_{12}^3}{\bf{\hat{S}}}_1\cdot
\left[\frac{{\bf{p}}_1\times{\vct{\hat{n}}_{12}}}{m_1}\left(\frac{11}{2}m_1+5m_2\right)+\frac{{\bf{p}}_2\times{\vct{\hat{n}}_{12}}}{m_2}\left(-6m_1-\frac{15}{2}m_2\right)\right] + \left[ 1\leftrightarrow2 \right]\,.
\end{aligned}
\end{align}
If both Hamiltonians are correct and are therefore to generate the same equations of motion, the difference between
these two should equal an infinitesimal canonical transformation, which in turn involves a generator function $g$
that is to be chosen appropriately as outlined in e.g., \cite{Levi:2010}. So it should hold
\begin{equation}\label{DiffNLO}
 \Delta H^{\text{can}}_{\text{NLOSO}}=H_{\text{NLOSO}}^{\text{DJS}}-H_{\text{SO(L)}}^{\text{NLO(can)}}\equiv\{H_{\text{N}},g\}=-\frac{\dd g}{\dd t}\,,
\end{equation}
where for the generator function one can make the following general ansatz to canonically transform a NLO spin-orbit Hamiltonian:
\begin{align}\label{gNLO}
\begin{aligned}
g&= \frac{Gm_2}{\hat{r}_{12}}{\bf{S}}_1 \cdot \biggl[\gamma_1\frac{{\bf{p}}_1\times{\bf{p}}_2}{m_1m_2}+\frac{{\bf{p}}_1\times{\vct{\hat{n}}_{12}}}{m_1}\left(\gamma_2\frac{{\bf{p}}_1\cdot{\vct{\hat{n}}_{12}}}{m_1}+\gamma_3\frac{{\bf{p}}_2\cdot{\vct{\hat{n}}_{12}}}{m_2}\right)
\\
&\qquad+
\frac{{\bf{p}}_2\times{\vct{\hat{n}}_{12}}}{m_2}\left(\gamma_4\frac{{\bf{p}}_1\cdot{\vct{\hat{n}}_{12}}}{m_1}+\gamma_5\frac{{\bf{p}}_2\cdot{\vct{\hat{n}}_{12}}}{m_2}\right)\biggr]\,.
\end{aligned}
\end{align}
It turns out Eq.\ (\ref{DiffNLO}) can be fulfilled by choosing the coefficients to be
\begin{equation}
 \gamma_1=\frac{1}{2},~~\gamma_2=0,~~\gamma_3=-\frac{1}{2},~~\gamma_4=0,~~\gamma_5=0\,.
\end{equation}
We kindly note that the same agreement of (\ref{Hefflevi}) with (\ref{HDJS}) was
already achieved by Levi \cite{Levi:2010}. However, the transformation to
canonical variables was found in \cite{Levi:2010} by comparing with the
Hamiltonian in \cite{Damour:Jaranowski:Schafer:2008:1}, whereas here we derived it
from general principles. For this reason our Eq.\ (\ref{xNW}) also contains
a term which is irrelevant for translation invariant quantities like the Hamiltonian (but would be
needed, e.g., for the center of mass), whereas such a term was omitted in
Eq.\ (121) of \cite{Levi:2010} as it is not needed there. Notice in \cite{Damour:Jaranowski:Schafer:2008:1} there was
also a variable transformation formula determined to achieve a comparison with a result which was obtained in
harmonic coordinates and with a spin whose lenght is non-conserved. Their transformation formula (6.11), which was
uniquely determined and comprises non-canonical and canonical transformations, contains a similar irrelevant term when comparing
translation invariant quantities. By our derivation of Eq. (\ref{xNW}) from general principles we also achieved direct justification of their transformation formula, that
was left as future work by the authors.

\subsection{The NLO spin-orbit Hamiltonian of Porto}

We also want to compare the result of Porto with the ADM NLO spin-orbit Hamiltonian from Eq.\ (\ref{HDJS}). We start out from the non-reduced effective Hamiltonian from Eq.\ (\ref{VNLOSOPO}) and insert the covariant SSC from Eq.\ (\ref{SSSC}), which results in the reduced effective NLO Hamiltonian in 'covariant' variables
\begin{align}
\begin{aligned}
H_{\text{SO(P),red}}^{\text{NLO(eff)}}&= 
\frac{Gm_2}{r_{12}^2}{\bf{S}}_1\cdot
\biggl[\frac{{\bf{p}}_1\times{{\bf n}_{12}}}{m_1}\left(\frac{p_1^2}{2m_1^2}+\frac{2{\bf{p}}_1\cdot{\bf{p}}_2}{m_1m_2}-\frac{2p_2^2}{m_2^2}+\frac{3({\bf{p}}_2\cdot{{\bf n}_{12}})^2}{m_2^2}\right)\\
&\quad+ 
\frac{{\bf{p}}_2\times{{\bf n}_{12}}}{m_2}\left(-\frac{3({\bf{p}}_2\cdot{{\bf n}_{12}})^2}{m_2^2}+\frac{{\bf{p}}_{2}^{2}}{m_2^2}-\frac{2({\bf{p}}_1\cdot{\bf{p}}_2)}{m_1 m_2}\right) \biggr] \\ 
&\quad+\frac{G^2m_2}{r_{12}^3}{\bf{S}}_1\cdot\left[\frac{{\bf{p}}_1\times{{\bf n}_{12}}}{m_1}\left(8m_1+\frac{13}{2}m_2\right)+\frac{{\bf{p}}_2\times{{\bf n}_{12}}}{m_2}\left(-7m_1-\frac{15}{2}m_2\right) \right] + \left[ 1\leftrightarrow2 \right].
 \end{aligned}
\end{align}
The transition to the canonical Hamiltonian is made as in the case of Levi's effective Hamiltonian by adding up the missing pieces and putting a hat on the variables 
of Porto's Hamiltonian:
\begin{align}
 H_{\text{SO(P)}}^{\text{NLO(can)}}=H_{\text{SO(P),red}}^{\text{NLO(eff)}}+H^{\text{N}}_{\text{NLOSO}}+H_{\text{NLOSO}}^{\text{EIH}}+H^{\text{LOSO}}_{\text{NLOSO}}
\end{align}
The result is
\begin{align}
 \begin{aligned}
  H_{\text{SO(P)}}^{\text{NLO(can)}} &= 
\frac{Gm_2}{\hat{r}_{12}^2}{\bf{\hat{S}}}_1\cdot
\biggl[\frac{{\bf{p}}_1\times{{\bf\hat{n}}_{12}}}{m_1}\biggl(\frac{5}{8}\frac{p_1^2}{m_1^2}+\frac{5{\bf{p}}_1\cdot{\bf{p}}_2}{4m_1m_2}-\frac{5p_2^2}{4m_2^2}+\frac{3({\bf{p}}_2\cdot{{\bf\hat{n}}_{12}})^2}{m_2^2}\\
&\qquad\qquad-\frac{3}{4m_{1}m_{2}}({\bf{p}}_1\cdot{{\bf\hat{n}}_{12}})({\bf{p}}_2\cdot{{\bf\hat{n}}_{12}})\biggr)-\frac{3}{4m_{1}^2m_{2}}({\bf{p}}_1\cdot{{\bf\hat{n}}_{12}})({\bf{p}}_1\times{\bf{p}}_{2})\\
&\quad+\frac{{\bf{p}}_2\times{{\bf\hat{n}}_{12}}}{m_2}\left(-\frac{3({\bf{p}}_2\cdot{{\bf\hat{n}}_{12}})^2}{m_2^2}+\frac{{\bf{p}}_{2}^{2}}{m_2^2}-\frac{2({\bf{p}}_1\cdot{\bf{p}}_2)}{m_1 m_2}\right) \biggr] \\ 
&\quad+\frac{G^2m_2}{\hat{r}_{12}^3}{\bf{\hat{S}}}_1\cdot\left[\frac{{\bf{p}}_1\times{{\bf\hat{n}}_{12}}}{m_1}\left(\frac{15}{2}m_1+5m_2\right)+\frac{{\bf{p}}_2\times{{\bf\hat{n}}_{12}}}{m_2}\left(-7m_1-6m_2\right) \right] + \left[ 1\leftrightarrow2 \right]\,.
 \end{aligned}
\end{align}
Again for the difference $\Delta H^{\text{can}}_{\text{NLOSO}}$ there should exist a generator function Eq.\ (\ref{gNLO}) with determined coefficients. The
coefficients for this case read
\begin{equation}
\gamma_1=\frac{3}{2},~~\gamma_2=0,~~\gamma_3=\frac{1}{2},~~\gamma_4=0,~~\gamma_5=-1 ,
\end{equation}
which means we have achieved `on-shell' agreement with the ADM Hamiltonian.

\subsection{The NLO spin(1)-spin(2) Hamiltonian of Porto and Rothstein}

For reason of completeness we also compare the NLO spin(1)-spin(2) Hamiltonian which results from the corresponding
effective potential that Porto and Rothstein have calculated, with the corresponding ADM Hamiltonian.
% although the comparison already succeeded in \cite{Steinhoff:Hergt:Schafer:2008:2} yet not in this formal way as outlined here.
We refer to Eq.\ (\ref{NLOS1S2}) and insert the SSC from Eq.\ (\ref{SSSC}) to arrive at the reduced effective Hamiltonian, which reads
\begin{align}
\begin{aligned}
 &H_{S_{1}S_{2},\text{red}}^{\text{NLO(eff)}}=~\frac{G}{r_{12}^3}\Bigg[\frac{1}{m_{1}^2}\Big(3\pipi\Sin\Siin-3\pin\Sipi\Siin\\
&\quad-3\pin\Sin\Siipi+2\Sipi\Siipi\\
&\quad+3\pin^2\SiSii-2\pipi\SiSii\Big)+\frac{1}{m_{2}^2}\Big(3\piipii\Sin\Siin\\
&\quad-3\piin\Sipii\Siin-3\piin\Sin\Siipii\\
&\quad+2\Sipii\Siipii+3\piin^2\SiSii-2\piipii\SiSii\Big)\\
&\quad+\frac{1}{m_{1}m_{2}}\Big(\frac{15}{2}\pin\piin\Sin\Siin\\
&\quad-\frac{21}{2}\pipii\Sin\Siin+\frac{9}{2}\piin\Sipi\Siin\\
&\quad+\frac{9}{2}\pin\Sipii\Siin-\frac{5}{2}\Sipi\Siipii\\
             &\quad+\frac{9}{2}\piin\Sin\Siipi-\frac{5}{2}\Sipii\Siipi\\
&\quad+\frac{9}{2}\pin\Sin\Siipii-\frac{9}{2}\pin\piin\SiSii\\
&\quad+\frac{11}{2}\pipii\SiSii\Big)\Bigg]+\frac{G^2(m_{1}+m_{2})}{r_{12}^4}\Bigg[5\SiSii-11\Sin\Siin\Bigg]\,.
\end{aligned}
\end{align}
We transform this `covariant' Hamiltonian to the canonical one by adding up to it the corresponding correction terms which follow from the variable transformation and replace the `covariant' variables with Newton-Wigner ones in the original Hamiltonian
\begin{align}
 H_{S_{1}S_{2},\text{red}}^{\text{NLO(can)}}=H_{S_{1}S_{2},\text{red}}^{\text{NLO(eff)}}+H^{\text{N}}_{\text{NLO}S_{1}S_{2}}+H^{\text{LOSO}}_{\text{NLO}S_{1}S_{2}}+H^{\text{LO}S_{1}S_{2}}_{\text{NLO}S_{1}S_{2}}\,,
\end{align}
the result being
\begin{align}
\begin{aligned}
 H_{S_{1}S_{2},\text{red}}^{\text{NLO(can)}}&=~\frac{G}{\hat{r}_{12}^3}\Bigg[\frac{1}{m_{1}^2}\Big(3\pipi\SinNW\SiinNW-3\pinNW\SipiNW\SiinNW\\
\quad&+\frac{1}{2}\SipiNW\SiipiNW-\frac{1}{2}\pipi\SiSiiNW\Big)+\frac{1}{m_{2}^2}\Big(3\piipii\SinNW\SiinNW\\
\quad&-3\piinNW\SinNW\SiipiiNW+\frac{1}{2}\SipiiNW\SiipiiNW-\frac{1}{2}\piipii\SiSiiNW\Big)\\
             \quad&+\frac{1}{m_{1}m_{2}}\Big(-\frac{15}{2}\pinNW\piinNW\SinNW\SiinNW\\
\quad&-\frac{21}{4}\pipii\SinNW\SiinNW+\frac{9}{2}\piinNW\SipiNW\SiinNW\\
\quad&-\frac{3}{4}\pinNW\SipiiNW\SiinNW-\frac{5}{2}\SipiNW\SiipiiNW\\
             \quad&-\frac{3}{4}\piinNW\SinNW\SiipiNW+\SipiiNW\SiipiNW\\
             \quad&+\frac{9}{2}\pinNW\SinNW\SiipiiNW+\frac{3}{4}\pinNW\piinNW\SiSiiNW\\
             \quad&+2\pipii\SiSiiNW\Big)\Bigg]+\frac{G^2(m_{1}+m_{2})}{\hat{r}_{12}^4}\Bigg[5\SiSiiNW-11\SinNW\SiinNW\Bigg]\,.
\end{aligned}
\end{align}
This Hamiltonian shall be compared with the corresponding ADM Hamiltonian which was calculated by us
in \cite{Steinhoff:Hergt:Schafer:2008:2,Steinhoff:Schafer:Hergt:2008}. It reads
\begin{align}
\begin{aligned}
 &H^{\text{ADM(can)}}_{\text{NLO}~S_{1}S_{2}}=~\frac{G}{\hat{r}_{12}^3}\Bigg[\frac{1}{m_{1}^2}\Big(3\pipi\SinNW\SiinNW-3\pinNW\SipiNW\SiinNW\\
\quad&-3\pinNW\SinNW\SiipiNW+\frac{3}{2}\SipiNW\SiipiNW-\frac{3}{2}\pipi\SiSiiNW\Big)\\
             \quad&+\frac{1}{m_{2}^2}\Big(3\piipii\SinNW\SiinNW-3\piinNW\SipiiNW\SiinNW\\
\quad&-3\piinNW\SinNW\SiipiiNW+\frac{3}{2}\SipiiNW\SiipiiNW-\frac{3}{2}\piipii\SiSiiNW\Big)\\
             \quad&+\frac{1}{m_{1}m_{2}}\Big(-\frac{15}{2}\pinNW\piinNW\SinNW\SiinNW\\
\quad&-\frac{21}{4}\pipii\SinNW\SiinNW+\frac{9}{2}\piinNW\SipiNW\SiinNW\\
\quad&+\frac{9}{4}\pinNW\SipiiNW\SiinNW-\frac{5}{2}\SipiNW\SiipiiNW\\
             \quad&+\frac{9}{4}\piinNW\SinNW\SiipiNW-\SipiiNW\SiipiNW\\
\quad&+\frac{9}{2}\pinNW\SinNW\SiipiiNW-\frac{21}{4}\pinNW\piinNW\SiSiiNW\\
\quad&+4\pipii\SiSiiNW\Big)\Bigg]+\frac{G^2(m_{1}+m_{2})}{\hat{r}_{12}^4}\Bigg[6\SiSiiNW-12\SinNW\SiinNW\Bigg]\,.
\end{aligned}
\end{align}
To generate the difference $\Delta H_{\text{NLO}~S_{1}S_{2}}^{\text{can}}=H^{\text{ADM(can)}}_{\text{NLO}~S_{1}S_{2}}-H_{S_{1}S_{2},\text{red}}^{\text{NLO(can)}}$ of the ADM Hamiltonian to the one resulting from an effective potential we make another ansatz for the generator function $g$:
\begin{align}
\begin{aligned}
 g^{\text{can}}_{\text{NLO}~S_{1}S_{2}}&=\frac{G}{\hat{r}_{12}^2}\Bigg[\frac{\gamma_{1}}{m_{1}}\SipiNW\SiinNW-\frac{\gamma_{2}}{m_{1}}\SinNW\SiipiNW+\frac{\gamma_{3}}{m_{1}}\pinNW\SiSiiNW\\
                                       &\quad\qquad+\frac{\gamma_{4}}{m_{1}}\pinNW\SinNW\SiinNW+(1 \leftrightarrow 2)\Bigg]\,.
\end{aligned}
\end{align}
The coefficients can be uniquely determined and yield
\begin{equation}
 \gamma_{1}=0\,,\quad \gamma_{2}=-1\,,\quad \gamma_{3}=-1\,,\quad \gamma_{4}=0\,.
\end{equation}
An alternative potential in the covariant SSC was derived by Levi
\cite{Levi:2008} and the procedure to obtain a fully reduced Hamiltonian
used in the present paper (and briefly presented before in
\cite{Hergt:Steinhoff:Schafer:2010:1}) was already applied therein. Therefore we
will not repeat it here. However, the transformation to canonical variables was
found in \cite{Levi:2008} by comparing with the Hamiltonian in
\cite{Steinhoff:Hergt:Schafer:2008:2}, whereas here we derived it from general
principles.

Notice that in \cite{Steinhoff:Hergt:Schafer:2008:2} it was found that the
result in \cite{Porto:Rothstein:2006} was incomplete, however, once completed it also fully agrees with the
corresponding NLO spin(1)-spin(2) ADM Hamiltonian \cite{Porto:Rothstein:2008:1}.
Though the result in \cite{Porto:Rothstein:2006} was derived within the EFT
approach, it was not based on the Routhian method but on a direct insertion of
a Newton-Wigner SSC.\footnote{However, in \cite{Porto:Rothstein:2006} this procedure is not
justified. For a proper way to directly implement a Newton-Wigner SSC
in an action see \cite{Barausse:Racine:Buonanno:2009}, in particular Eqs. (3.7) and (3.12) therein.} The
alternative NLO spin(1)-spin(2) potential derived by Levi even agrees exactly
with the corresponding ADM Hamiltonian if a suitable Newton-Wigner SSC is used
at the action level \cite{Levi:2008}. More work is needed to understand why such
a direct insertion of a Newton-Wigner SSC leads to correct fully reduced
Hamiltonians in some cases.

\subsection{The NLO spin(1)-spin(1) Hamiltonian of Porto and Rothstein}

The last Hamiltonian we want to verify is the effective NLO spin(1)-spin(1) Hamiltonian from Eq.\ (\ref{NLOS12}). From it we eliminate $S^{(0)(i)}_1$ by Eq.\ (\ref{SSSC}) resulting in
\begin{align}
 \begin{aligned}
  &H_{S_{1}^2,\text{red}}^{\text{NLO(eff)}}=~C_{Q_{1}}\frac{G}{r_{12}^3}\Bigg[\frac{m_{2}}{m_{1}^3}\bigg(\frac{3}{4}\pipi\Sin^2-3\pin\Sin\Sipi+\Sipi^2\\
              &+\frac{9}{2}\pin^2\SiSi-\frac{7}{4}\pipi\SiSi\bigg)+\frac{1}{m_{1}^2}\bigg(-\frac{15}{4}\pin\piin\Sin^2-\frac{21}{4}\pipii\Sin^2\\
              &+\frac{9}{2}\piin\Sin\Sipi-\frac{3}{2}\Sipi\Sipii+\frac{3}{2}\pin\Sin\Sipii-\frac{15}{4}\pin\piin\SiSi\\
              &+\frac{13}{4}\pipii\SiSi\bigg)+\frac{1}{m_{1}m_{2}}\bigg(\frac{9}{4}\piipii\Sin^2-\frac{3}{4}\piipii\SiSi\bigg)\Bigg]\\
              &+\frac{G^2m_{2}}{r_{12}^4}\Bigg[\left(2+\frac{1}{2}C_{Q_{1}}+\frac{m_{2}}{m_{1}}\Big(\frac{1}{2}+3C_{Q_{1}}\Big)\right)\SiSi+\left(-3-\frac{3}{2}C_{Q_{1}}+\frac{m_{2}}{m_{1}}\Big(\frac{1}{2}+6C_{Q_{1}}\Big)\right)\Sin^2\Bigg]\,.
 \end{aligned}
\end{align}
We proceed to the canonical Hamiltonian by collecting all matching correction terms and by replacing covariant variables with Newton-Wigner ones in the above Hamiltonian $H_{S_{1}^2,\text{red}}^{\text{NLO(eff)}}$
\begin{align}
 H_{S_{1}^2,\text{red}}^{\text{NLO(can)}}=H_{S_{1}^2,\text{red}}^{\text{NLO(eff)}}+H^{\text{N}}_{\text{NLO}S_{1}^2}+H^{\text{LOSO}}_{\text{NLO}S_{1}^2}+H^{\text{LO}S_{1}^2}_{\text{NLO}S_{1}^2}\,.
\end{align}
The result is
\begin{align}
\begin{aligned}
  &H_{S_{1}^2,\text{red}}^{\text{NLO(can)}}=~\frac{G}{\hat{r}_{12}^3}\Bigg[\frac{m_{2}}{m_{1}^3}\Bigg(\left(-\frac{21}{8}+\frac{9}{4}C_{Q_{1}}\right)\pipi\SinNW^2\\
&+\left(\frac{21}{4}-\frac{9}{2}C_{Q_{1}}\right)\pinNW\SinNW\SipiNW+\left(-\frac{7}{4}+\frac{3}{2}C_{Q_{1}}\right)\SipiNW^2\\
&+\left(-\frac{21}{8}+\frac{9}{2}C_{Q_{1}}\right)\pinNW^2\SiSiNW+\left(\frac{7}{4}-\frac{9}{4}C_{Q_{1}}\right)\pipi\SiSiNW\Bigg)\\
              &+\frac{1}{m_{1}^2}\Bigg(-\frac{15}{4}C_{Q_{1}}~\pinNW\piinNW\SinNW^2+\left(3-\frac{21}{4}C_{Q_{1}}\right)\pipii\SinNW^2\\
&+\left(-3+\frac{9}{2}C_{Q_{1}}\right)\piinNW\SinNW\SipiNW+\left(-3+\frac{3}{2}C_{Q_{1}}\right)\pinNW\SinNW\SipiiNW\\
              &+\left(2-\frac{3}{2}C_{Q_{1}}\right)\SipiNW\SipiiNW+\left(3-\frac{15}{4}C_{Q_{1}}\right)\pinNW\piinNW\SiSiNW\\
              &+\left(-2+\frac{13}{4}C_{Q_{1}}\right)\pipii\SiSiNW\Bigg)+\frac{C_{Q_{1}}}{m_{1}m_{2}}\Big(\frac{9}{4}\piipii\SinNW^2-\frac{3}{4}\piipii\SiSiNW\Big)\Bigg]\\
              &+\frac{G^2m_{2}}{\hat{r}_{12}^4}\Bigg[\left(2+\frac{1}{2}C_{Q_{1}}+\frac{m_{2}}{m_{1}}\Big(\frac{1}{2}+3C_{Q_{1}}\Big)\right)\SiSiNW+\left(-3-\frac{3}{2}C_{Q_{1}}-\frac{m_{2}}{m_{1}}\Big(\frac{1}{2}+6C_{Q_{1}}\Big)\right)\SinNW^2\Bigg]\,.
\end{aligned}
\end{align}
The comparison with the ADM NLO spin(1)-spin(1) Hamiltonian was already
briefly addressed \cite{Hergt:Steinhoff:Schafer:2010:1}, and we also refer to \cite{Hergt:Steinhoff:Schafer:2010:1} for details on the derivation of the ADM Hamiltonian.
Notice the two misprints in Eq.\ (32) of \cite{Hergt:Steinhoff:Schafer:2010:1} if one compares it with (\ref{SNW}).
The round bracket term should be corrected to read
\begin{equation}
 \left(1-\frac{3\vct{p}_{1}^2}{4m_{1}^2}\right)\quad\rightarrow\quad\left(1-\frac{\vct{p}_{1}^2}{4m_{1}^2}\right)\,,
\end{equation}
and the third term has a wrong particle label, rather it should read
\begin{equation}
 \frac{3G}{m_{1}\hat{r}_{12}}p_{1[i}\hat{S}_{1(j)(k)]}p_{1k}\quad\rightarrow\quad\frac{3G}{m_{1}\hat{r}_{12}}p_{1[i}\hat{S}_{1(j)(k)]}p_{2k}\,.
\end{equation}
It turned out that with the following general ansatz for the generator function
\begin{align}
 \begin{aligned}
  g^{\text{can}}_{\text{NLO}~S_{1}^2}&=\frac{G}{\hat{r}_{12}^2}\frac{m_{2}}{m_{1}^2}\Bigg[\gamma_{1}\pinNW\SiSiNW+\gamma_{2}\piinNW\SiSiNW+\gamma_{3}\SinNW\SipiNW\\
                                     &+\gamma_{4}\SinNW\SipiiNW+\gamma_{5}\pinNW\SinNW^2+\gamma_{6}\piinNW\SinNW^2\Bigg]
 \end{aligned}
\end{align}
 of an infinitesimal canonical transformation which belongs to this Hamiltonian, the difference to the ADM Hamiltonian
\begin{equation}
 \Delta H_{\text{NLO}~S_{1}^2}=\{H_{\text{N}},g^{\text{can}}_{\text{NLO}~S_{1}^2}\}
\end{equation}
 can be generated by choosing the coefficients to beat lower orders, this issue is
just semantics, but at higher orders it is of practical relevance
\begin{align}
 \gamma_{1}=-\frac{1}{2}+C_{Q_{1}},\,\gamma_{2}=0\,,\gamma_{3}=\frac{1}{2}\,,\gamma_{4}=0\,,\gamma_{5}=0\,,\gamma_{6}=0\,.
\end{align}
We have thus shown rather clearly how one can transform a non-reduced effective potential at NLO
to a canonical Hamiltonian in a very replicable and systematically way without
losing oneself in subtleties.

\section{Conclusions and Outlook\label{sec:conclusions}}

In this paper we have presented two different methods to transform non-reduced effective potentials,
 that can be calculated within the EFT approach, for spinning compact binary systems in general relativity to fully reduced canonical Hamiltonians,
 which then depend only on the physical degrees of freedom. The main subject we focused on were potentials of constrained systems with redundant spin
 degrees of freedom that had to be eliminated. The key difficulty was to achieve this scheme on curved spacetime at least perturbatively up to next-to-leading order.
 The first method involved working with Dirac brackets in \Sec{sec:db} to reduce the phase-space variables and the second method was to redefine variables
 in such a way that the action in \Sec{sec:action} becomes fully reduced. Both methods yielded the same transformation formulae for spin
 (\ref{SNW},\ref{Strafofinal}) and position variable (\ref{xNW},\ref{ztrafoA}) although the action approach is much more transparent and provides quite general
 transformation formulae that possess validity in full general relativity
given that the vierbein field of curved spacetime is known. Furthermore we could determine how the Lorentz matrices should transform in a gravitational field
 in order to arrive at a standard canonical representation, see Eq.\ (\ref{Lambdatrafo1}). Thus our formalism is also valid for effective potentials/Hamiltonians
 that depend on the Lorentz matrices, which would describe an asymmetric behavior of the physical system in consideration.

Interestingly the method with the Dirac brackets needs no transformation of the Lorentz matrices, which was expected,
 because the Hamiltonians are all independent of the Euler angles and the Dirac brackets operate on the level of the equations of motion,
 which are in turn generated by these Hamiltonians. In contrast the action approach needs the redefinition of the Lorentz matrices via
Eq.\ (\ref{Lambdatrafo1}) because the action depends on them while coupling to the spin.
Due to the knowledge of the vierbein field to next-to-leading pN order we could explicitly calculate the transformation formulae
 for transforming all next-to-leading order effective potentials known to date for two self-gravitating spinning compact objects to
 canonical Hamiltonians while eliminating the spin supplementary condition and performing a Legendre transformation.
 In the near future we plan to go to next-to-next-to-leading pN order in the transformation formulae in order to
 compare the non-reduced NNLO spin(1)-spin(2) effective potential by Levi \cite{Levi:2011} with the fully reduced
 NNLO spin(1)-spin(2) canonical ADM Hamiltonian calculated by Hartung and Steinhoff in \cite{Hartung:Steinhoff:2011:2}.
 Clearly for this purpose it is much easier to start from the generally valid formulae found by transforming the action
 directly instead of elaborating more on Dirac bracket calculations. The only piece missing is the vierbein field to
 next-to-next-to-leading pN order, which can be calculated, e.g., from the metric in harmonic coordinates \cite{Tagoshi:Ohashi:Owen:2001} 
and corrections \cite{Faye:Blanchet:Buonanno:2006}, and poses no conceptual problem. It is interesting to note that now after 37 years we
 finally succeeded with the proposal made by Hanson and Regge \cite{Hanson:Regge:1974} in their conclusion section to ``attempt to include gravitation in our formalism'' initiated in
\cite{Porto:Rothstein:2008:1,Barausse:Racine:Buonanno:2009,Yee:Bander:1993,Westpfahl:1969:2,Bailey:Israel:1975}.
Furthermore at the end of \Sec{sec:action} it was discussed how Eq.\ (\ref{SOmegared})
can be used to improve the Feynman rules of the EFT
formalism by formulating them in terms of reduced canonical spin variables.

One future task would be to extend this formalism for dipole (spin) interaction that we developed
 especially for weak deformation effects to higher multipoles that would account for stronger rotational
 deformation effects which will play an important role when the binary system is very close to the merger phase.
 Right before the merger there is also no more guarantee for the spin length to be conserved which would also be an attractive
 topic to deal with, because so far there has been no algebraic treatment of a change in the spin length, only analytically starting with Teukolsky's analysis \cite{Teukolsky:1974}. 
% document ends here\cite{Ibanez:Martin:1982}
%

\acknowledgments 
\ack

\appendix

\section{Dirac brackets\label{sec:DBA}}

Here we list all possible combinations of quantities to enter the Dirac bracket and their results except the ones containing the $\Lambda$-matrices.
The reason is that all Hamiltonians of interest in this work do not own any $\Lambda$-dependency, so there is no need to canonicalize the $\Lambda$-matrices
explicitly here with the Dirac brackets besides all the other variables $(z_{I}^i\,,p_{Ji},\,S_{K}^{ab})$, which really appear in the Hamiltonians.
This is different when eliminating the SSC and transforming to canonical variables on the action level, see \Sec{sec:action}.
Then the first thing one \emph{must} do is to transform the $\Lambda$-matrices to canonical ones being the hatted $\hat\Lambda$-matrices.
The results of the Dirac brackets read
\begin{subequations}
 \begin{align}
  \{z^{i}_{1},p_{1j}\}_{\text{D}}&\simeq\delta_{ij}+\frac{G}{r_{12}^2}\Bigg(-\frac{m_{2}}{m_{1}^2}n_{12}^{j}p_{1k}S_{1}^{(i)(k)}+\frac{3}{2m_{1}^2}n_{12}^{j}p_{2k}S_{1}^{(i)(k)}-\frac{1}{2m_{1}}n_{12}^{k}p_{2j}S_{1}^{(i)(k)}\nonumber\\
                       &\quad-\frac{1}{2m_{1}}\piin S_{1}^{(i)(j)}+\frac{3}{2m_{1}}n_{12}^{j}\piin n_{12}^{k}S_{1}^{(i)(k)}\Bigg)\nonumber\\
                       &\quad+\frac{G}{m_{1}r_{12}^3}\left(3n_{12}^{k}n_{12}^{j}S_{1}^{(i)(l)}S_{2(k)(l)}-S_{1}^{(i)(k)}S_{2(j)(k)}\right)\label{relzD}\\
                       &\quad+\frac{Gm_{2}}{m_{1}^2r_{12}^3}\left(3n_{12}^{k}n_{12}^{j}S_{1}^{(k)(l)}S_{1(i)(l)}-S_{1}^{(i)(k)}S_{1(j)(k)}\right)+\mathcal{O}\left(c^{-5}\right)\,,\nonumber\\
 \{z^{i}_{1},z_{1}^{j}\}_{\text{D}}&\simeq\frac{1}{m_1^2} \mathcal{P}_{1ik} \mathcal{P}_{1jl} S_1^{(k)(l)}-2G\frac{m_{2}S_{1}^{(i)(j)}}{m_{1}^2r_{12}}+\mathcal{O}\left(c^{-2}\right)\,,\\
 \{z^{i}_{1},S_{1}^{(j)(k)}\}_{\text{D}}&\simeq\frac{2p_{1[k}S_{1(j)](l)}\mathcal{P}_{1li}}{m_{1}^2}\nonumber\\
                         &\quad+\frac{G}{m_{1}r_{12}}\left(3S_{1(i)[(k)}p_{2j]}+\frac{4m_{2}}{m_{1}}S_{1(i)[(j)}p_{1k]}+\piin S_{1(i)[(k)}n_{12j]}\right)\\
                         &\quad+\frac{2GS_{1(i)[(j)}S_{2(k)](l)}n_{12l}}{m_{1}r_{12}^2}+\frac{2Gm_{2}S_{1(i)(l)}S_{1(l)[(j)}n_{12k]}}{m_{1}^2r_{12}^2}+\mathcal{O}\left(c^{-5}\right)\,,\nonumber
\end{align}
\begin{align}
 \{z^{i}_{1},z^{j}_{2}\}_{\text{D}}&\simeq\frac{G}{2m_{1}r_{12}}\left(3S_{1}^{(i)(j)}+S_{1}^{(i)(k)}n_{12k}n_{12}^{j}\right)\nonumber\\
                                    &\quad+\frac{G}{2m_{2}r_{12}}\left(3S_{2}^{(i)(j)}-S_{2}^{(j)(k)}n_{12k}n_{12}^{i}\right)+\mathcal{O}\left(c^{-2}\right)\,,\\
 \{z^{i}_{1},p_{2j}\}_{\text{D}}&\simeq\delta_{ij}-\{x^{i}_{1},p_{1j}\}_{\text{D}}\,,\\
 \{z^{i}_{1},S_{2}^{(j)(k)}\}_{\text{D}}&\simeq\frac{3GS_{2(i)[(k)}p_{2j]}}{m_{2}r_{12}}+\frac{GS_{2(l)[(k)}p_{2j]}n_{12}^i n_{12}^l}{m_{2}r_{12}}+\frac{2GS_{2(j)(l)}S_{1(i)[(k)}n_{12l]}}{m_{1}r_{12}^2}\nonumber\\
                                     &\quad+\frac{2GS_{2(k)(l)}S_{1(i)[(l)}n_{12j]}}{m_{1}r_{12}^2}+\mathcal{O}\left(c^{-5}\right)\,,
\end{align}
\begin{align}
 \{p_{1i},p_{1j}\}_{\text{D}}&=\{p_{1i},p_{2j}\}_{\text{D}}=\{p_{2i},p_{2j}\}_{\text{D}}=0\,,\\
 \{p_{1i},S_{1}^{(j)(k)}\}_{\text{D}}&\simeq\frac{G}{m_{1}r_{12}^2}\bigg[\piin S_{1(i)[(j)}p_{1k]}\nonumber\\
                      &\quad+p_{1[j}S_{1(k)](l)}\left(2\frac{m_{2}}{m_{1}}p_{1l}n_{12}^{i}-3p_{2l}n_{12}^{i}+p_{2i}n_{12}^l-3\piin n_{12}^l n_{12}^i\right)\bigg]\nonumber\\
                      &\quad+\frac{2Gp_{1[k}S_{1(j)](m)}}{m_{1}r_{12}^3}\left(3n_{12}^i n_{12}^l S_{2}^{(l)(m)}-S_{2(i)(m)}\right)\\
                      &\quad+\frac{2Gm_{2}p_{1[k}S_{1(j)](l)}}{m_{1}^2r_{12}^3}\left(3n_{12}^m n_{12}^i S_{1(m)(l)}-S_{1(i)(l)}\right)+\mathcal{O}\left(c^{-8}\right)\,,\nonumber\\
 \{p_{1i},S_{2}^{(j)(k)}\}_{\text{D}}&=-\{p_{1i},S_{1}^{(j)(k)}\}_{\text{D}}\quad (1\leftrightarrow 2)\,,\\
 \{S_{1}^{(i)(j)},S_{1}^{(k)(l)}\}_{\text{D}}&\simeq\mathcal{P}_{jl}S_{1(i)(k)}-\mathcal{P}_{jk}S_{1(i)(l)}+\mathcal{P}_{ik}S_{1(j)(l)}-\mathcal{P}_{il}S_{1(j)(k)}\nonumber\\
                          &+\frac{8Gm_{2}p_{1[i}S_{1)(j)][(k)}p_{1l]}}{m_{1}^2r_{12}}+\frac{6G}{m_{1}r_{12}}\left(p_{2[j}S_{1(i)][(k)}p_{1l]}+p_{1[j}S_{1(i)][(k)}p_{2l]}\right)\nonumber\\
                          &+\frac{2Gp_{1[j} S_{1(i)][(k)}n_{12l]}\piin}{m_{1}r_{12}}+\frac{2Gn_{12[j}S_{1(i)][(k)}p_{1l]}}{m_{1}r_{12}}\nonumber\\
                          &+\frac{4G}{m_{1}r_{12}^2}\left(p_{1[l}S_{1(k)][(i)}S_{2(j)](m)}n_{12}^m+p_{1[j}S_{1(i)][(l)}S_{2(k)](m)}n_{12}^m\right)\label{relSD}\\
                          &+\frac{4Gm_{2}}{m_{1}^2r_{12}^2}\left(n_{12[l}S_{1(k)](m)}p_{1[j}S_{1(i)](m)}+n_{12[j}S_{1(i)](m)}p_{1[k}S_{1(l)](m)}\right)\nonumber\\
                          &+\mathcal{O}\left(c^{-8}\right)\,,\nonumber\\
 \{S_{1}^{(i)(j)},S_{2}^{(k)(l)}\}_{\text{D}}&\simeq\frac{4G}{m_{2}r_{12}^2}\left(p_{2[k}S_{2(l)](m)}n_{12[i}S_{1(j)](m)}+n_{12}^m S_{1(m)[(j)}S_{2(i)][(k)}p_{2l]}\right)\nonumber\\
                          &\quad+\frac{4G}{m_{1}r_{12}^2}\left(p_{1[j}S_{1(i)](m)} n_{12[l}S_{2(k)](m)}+p_{1[i}S_{1(j)][(l)}S_{2(k)](m)}n_{12}^m\right)\nonumber\\
                          &\quad+\mathcal{O}\left(c^{-8}\right)\,.
 \end{align}
\end{subequations}
All remaining Dirac brackets follow via exchange of particle indices $1$ with $2$ in all terms.

\section{Full derivation of the transition to canonical variables via an action principle\label{sec:aaction}}

In this section we shall elaborate in great detail on the reduction of the action and the resulting transition formulae to canonical variables
presented in Sect. \ref{sec:action}. The first step involves the insertion 
of the constraints (\ref{cssc})-(\ref{lambda0ssc}) into the spin coupling term in the action, which yields
\begin{align}
\begin{aligned}
 \frac{1}{2}S_{ab}\Omega^{ab}&=\frac{1}{2}S_{ab}\Lambda_{A}^{\;\;\;a}\dot{\Lambda}^{Ab}=\frac{1}{2}S_{ab}\left(\Lambda_{[0]}^{\;\;\;a}\dot{\Lambda}^{[0]b}+\Lambda_{[i]}^{\;\;\;a}\dot{\Lambda}^{[i]b}\right)\stackrel{(\ref{lambda0ssc})+(\ref{cssc})}{=}\frac{1}{2}S_{ab}\Lambda_{[i]}^{\;\;\;a}\dot{\Lambda}^{[i]b}\\
                             &=\frac{1}{2}S_{(0)(j)}\Lambda_{[i]}^{\;\;\;(0)}\dot{\Lambda}^{[i](j)}+\frac{1}{2}S_{(j)(0)}\Lambda_{[i]}^{\;\;\;(j)}\dot{\Lambda}^{[i](0)}+\frac{1}{2}S_{(j)(k)}\Lambda_{[i]}^{\;\;\;(j)}\dot{\Lambda}^{[i](k)}\\
                           (\ref{cssc})\rightarrow  &=-\frac{1}{2}S_{(j)(k)}\frac{u^{(k)}u_{(l)}}{u^{(0)}u_{(0)}}\Lambda_{[i]}^{\;\;\;(l)}\dot{\Lambda}^{[i](j)}-\frac{1}{2}S_{(j)(k)}\frac{u^{(k)}}{u^{(0)}}\Lambda_{[i]}^{\;\;\;(j)}\dot{\Lambda}^{[i](0)}+\frac{1}{2}S_{(k)(l)}\Lambda_{[i]}^{\;\;\;(k)}\dot{\Lambda}^{[i](l)}\\
                           &=-\frac{1}{2}S_{(j)(k)}\frac{u^{(k)}u_{(l)}}{u^{(0)}u_{(0)}}\tilde{\Omega}^{(l)(j)}-\frac{1}{2}S_{(j)(k)}\frac{u^{(k)}}{u^{(0)}}\Lambda_{[i]}^{\;\;\;(j)}\dot{\Lambda}^{[i](0)}+\frac{1}{2}S_{(k)(l)}\tilde{\Omega}^{(k)(l)}
\end{aligned}
\end{align}
with $\tilde{\Omega}^{(k)(l)}\equiv\Lambda_{[i]}^{\;\;\;(k)}\dot{\Lambda}^{[i](l)}$. Notice the formal difference to the definition of $\Omega^{(k)(l)}$ from $(\ref{omegadef})$, $\tilde{\Omega}^{(k)(l)}$ is therefore not necessarily antisymmetric, which is actually an unwanted feature. To solve for $\dot{\Lambda}^{[i](0)}$ we use (\ref{lambdassc}):
\begin{align}
 \dot{\Lambda}^{[i](0)}=-\dot{\Lambda}^{[i](k)}\frac{u_{(k)}}{u_{(0)}}-\Lambda^{[i](k)}\frac{\dot{u}_{(k)}}{u_{(0)}}+\Lambda^{[i](k)}\frac{u_{(k)}}{u_{(0)}^2}\dot{u}_{(0)}\,.
\end{align}
So
\begin{align}\label{term1a}
 \begin{aligned}
  \frac{1}{2}S_{ab}\Omega^{ab}&=-\frac{1}{2}S_{(j)(k)}\frac{u^{(k)}u_{(l)}}{u^{(0)}u_{(0)}}\tilde{\Omega}^{(l)(j)}+\frac{1}{2}S_{(k)(l)}\tilde{\Omega}^{(k)(l)}\\
                              &\quad-\frac{1}{2}S_{(j)(k)}\frac{u^{(k)}}{u^{(0)}}\Lambda_{[i]}^{\;\;\;(j)}\left(-\dot{\Lambda}^{[i](l)}\frac{u_{(l)}}{u_{(0)}}-\Lambda^{[i](l)}\frac{\dot{u}_{(l)}}{u_{(0)}}+\Lambda^{[i](l)}\frac{u_{(l)}}{u_{(0)}^2}\dot{u}_{(0)}\right)\,.
 \end{aligned}
\end{align}
We make use of (\ref{lambdaindex}) to find
\begin{equation}
 \eta^{(i)(j)}=\Lambda_{A}^{\;\;\;(i)}\Lambda^{A(j)}=\Lambda_{[0]}^{\;\;\;(i)}\Lambda^{[0](j)}+\Lambda_{[k]}^{\;\;\;(i)}\Lambda^{[k](j)}\stackrel{(\ref{lambda0ssc})}{=}\frac{u^{(i)}u^{(j)}}{u^a u_{a}}+\Lambda_{[k]}^{\;\;\;(i)}\Lambda^{[k](j)}\,.
\end{equation}
Because of antisymmetry of the spin tensor the first term with the 4-velocities is irrelevant for (\ref{term1a}). We are left with
\begin{align}\label{somega1}
\begin{aligned}
 \frac{1}{2}S_{ab}\Omega^{ab}&=-\frac{1}{2}S_{(j)(k)}\frac{u^{(k)}u_{(l)}}{u^{(0)}u_{(0)}}\tilde{\Omega}^{(l)(j)}+\frac{1}{2}S_{(k)(l)}\tilde{\Omega}^{(k)(l)}+\frac{1}{2}S_{(j)(k)}\tilde{\Omega}^{(j)(l)}\frac{u^{(k)}u_{(l)}}{u^{(0)}u_{(0)}}\\
                             &\quad+\frac{1}{2}S_{(j)(k)}\frac{u^{(k)}\dot{u}_{(l)}}{u^{(0)}u_{(0)}}\eta^{(j)(l)}\\
                             &=\left(S_{(i)(j)}+S_{(i)(k)}\frac{u^{(k)}u_{(j)}}{u^{(0)}u_{(0)}}-S_{(j)(k)}\frac{u^{(k)}u_{(i)}}{u^{(0)}u_{(0)}}\right)\frac{\tilde{\Omega}^{(i)(j)}}{2}+\frac{1}{2}S_{(j)(k)}\frac{u^{(k)}\dot{u}^{(j)}}{u^{(0)}u_{(0)}}\,.
\end{aligned}
\end{align}
Next thing to do is to redefine variables so that the canonical structure of (\ref{acan}) is produced. Obviously one should start by shifting $\tilde{\Omega}^{(i)(j)}$ to $\hat{\Omega}^{(i)(j)}$, which should be antisymmetric in order to be the correct velocity variable belonging to the spin tensor. To find the correct redefinition it is useful to make an ansatz for the intrinsic redefinition of Lorentz matrices with an unknown function $\xi$ according to
\begin{equation}
 \Lambda^{[i](j)}=\hat{\Lambda}^{[i](k)}\left(\eta^{(j)}_{(k)}+\xi u_{(k)}u^{(j)}\right)\quad\text{so that} \quad\hat{\Lambda}^{[k](i)}\hat{\Lambda}^{[k](j)}=\delta_{ij}
\end{equation}
transforming $\hat{\Lambda}^{[i](j)}$ into a 3-dimensional rotation matrix yielding
\begin{equation}
\hat{\Omega}^{(i)(j)}=\hat{\Lambda}_{[k]}^{\;\;\;(i)}\dot{\hat{\Lambda}}^{[k](j)}=-\hat{\Omega}^{(j)(i)}\,.
\end{equation}
 It follows
\begin{align}
 \begin{aligned}
  \eta_{(i)}^{(j)}&=\Lambda_{A(i)}\Lambda^{A(j)}=\Lambda_{[0](i)}\Lambda^{[0](j)}+\Lambda_{[k](i)}\Lambda^{[k](j)}\\
                  &=\frac{u_{(i)}u^{(j)}}{u^a u_{a}}+\left(\hat{\Lambda}_{[k](i)}+\xi\hat{\Lambda}_{[k](l)}u^{(l)}u_{(i)}\right)\left(\hat{\Lambda}^{[k](j)}+\xi\hat{\Lambda}^{[k](p)}u_{(p)}u^{(j)}\right)\\
                  &=\frac{u_{(i)}u^{(j)}}{u^a u_{a}}+\eta_{(i)}^{(j)}+\xi\eta_{(i)}^{(p)}u_{(p)}u^{(j)}+\xi\eta_{(l)}^{(j)}u^{(l)}u_{(i)}+\xi^2\eta^{(p)}_{(l)}u^{(l)}u_{(p)}u_{(i)}u^{(j)}\\
                  &=\eta_{(i)}^{(j)}+\frac{u_{(i)}u^{(j)}}{u^a u_{a}}+2\xi u_{(i)}u^{(j)}+\xi^2\vct{u}^2u_{(i)}u^{(j)}\,.
 \end{aligned}
\end{align}
For convenience we define
\begin{equation}
 u_{(i)}u^{(i)}\equiv\vct{u}^2\quad\text{so that}\quad u_{a}u^{a}\equiv u^2=u_{(0)}^2+\vct{u}^2\,.
\end{equation}
To find and expression for $\xi$ we have to solve the equation
\begin{align}
 \begin{aligned}
  \frac{1}{u^2}+2\xi+\xi^2\vct{u}^2=0\,\Leftrightarrow\,\xi^2+\frac{2}{\vct{u}^2}\xi+\frac{1}{u^2\vct{u}^2}=0\,,
 \end{aligned}
\end{align}
so the solution is twofold
\begin{equation}
 \xi_{1,2}=-\frac{1}{u(u+su_{(0)})}\quad\text{with}\quad s\in\{1,-1\}\,.
\end{equation}
Taking the limit $\vct{u}^2=0$ the function $\xi$ should still be regular because this state corresponds to the rest frame limit which is therefore an adequate physical limit to take. This argument leaves us with the sole solution
\begin{equation}
\xi=-\frac{1}{u(u+u_{(0)})}\,. 
\end{equation}
So the Lorentz matrices undergo a redefinition
\begin{equation}\label{Lambdatrafo1a}
 \Lambda^{[i](j)}=\hat{\Lambda}^{[i](k)}\left(\eta^{(j)}_{(k)}-\frac{u_{(k)}u^{(j)}}{u(u+u_{(0)})}\right)\,.
\end{equation}
Next we take the time derivative of this expression
\begin{align}
 \begin{aligned}
  \dot{\Lambda}^{[i](j)}&=\left(\hat{\Lambda}^{[i](j)}-\hat{\Lambda}^{[i](k)}\frac{u_{(k)}u^{(j)}}{u(u+u_{(0)})}\right)^{.}\\
                        &=\dot{\hat{\Lambda}}^{[i](j)}-\dot{\hat{\Lambda}}^{[i](k)}\frac{u_{(k)}u^{(j)}}{u(u+u_{(0)})}-\hat{\Lambda}^{[i](k)}\Bigg(\frac{\dot{u}_{(k)}u^{(j)}}{u(u+u_{(0)})}+\frac{u_{(k)}\dot{u}^{(j)}}{u(u+u_{(0)})}\\
                        &\quad+u_{(k)}u^{(j)}\left(\frac{1}{u(u+u_{(0)})}\right)^{\cdotp}\Bigg)\,.
 \end{aligned}
\end{align}
It follows
\begin{align}
 \begin{aligned}
  \tilde{\Omega}^{(i)(j)}&=\Lambda_{[k]}^{\;\;\;(i)}\dot{\Lambda}^{[k](j)}\\
                         =&\left(\hat{\Lambda}^{\;\;\;(i)}_{[k]}-\hat{\Lambda}^{\;\;\;(l)}_{[k]}\frac{u_{(l)}u^{(i)}}{u(u+u_{(0)})}\right)\\
                         &\times\Bigg[\dot{\hat{\Lambda}}^{[k](j)}-\dot{\hat{\Lambda}}^{[k](p)}\frac{u_{(p)}u^{(j)}}{u(u+u_{(0)})}-\hat{\Lambda}^{[k](p)}\Bigg(\frac{\dot{u}_{(p)}u^{(j)}}{u(u+u_{(0)})}+\frac{u_{(p)}\dot{u}^{(j)}}{u(u+u_{(0)})}\\
                         &\quad+u_{(p)}u^{(j)}\left(\frac{1}{u(u+u_{(0)})}\right)^{\cdotp}\Bigg)\Bigg]\\
                         =&\,\hat{\Omega}^{(i)(j)}-\hat{\Omega}^{(i)(p)}\frac{u_{(p)}u^{(j)}}{u(u+u_{(0)})}-\eta^{(i)(p)}\Bigg(\frac{\dot{u}_{(p)}u^{(j)}}{u(u+u_{(0)})}+\frac{u_{(p)}\dot{u}^{(j)}}{u(u+u_{(0)})}\\
                         &+u_{(p)}u^{(j)}\left(\frac{1}{u(u+u_{(0)})}\right)^{\cdotp}\Bigg)-\hat{\Omega}^{(l)(j)}\frac{u_{(l)}u^{(i)}}{u(u+u_{(0)})}+\hat{\Omega}^{(l)(p)}\frac{u_{(l)}u^{(i)}u_{(p)}u^{(j)}}{u^2(u+u_{(0)})^2}\\
                          &+\eta^{(l)(p)}\frac{u_{(l)}u^{(i)}}{u(u+u_{(0)})}\left(\frac{\dot{u}_{(p)}u^{(j)}}{u(u+u_{(0)})}+\frac{u_{(p)}\dot{u}^{(j)}}{u(u+u_{(0)})}+u_{(p)}u^{(j)}\left(\frac{1}{u(u+u_{(0)})}\right)^{\cdotp}\right)\\
                         [=&]\,\hat{\Omega}^{(i)(j)}-\hat{\Omega}^{(i)(l)}\frac{u_{(l)}u^{(j)}}{u(u+u_{(0)})}-\hat{\Omega}^{(l)(j)}\frac{u_{(l)}u^{(i)}}{u(u+u_{(0)})}+\frac{\vct{u}^2u^{(i)}\dot{u}^{(j)}}{u^2(u+u_{(0)})^2}\,.
 \end{aligned}
\end{align}
The Symbol $[=]$ means we neglect all terms symmetric by interchanging $i\leftrightarrow j$ and keep those being
antisymmetric by this interchange, because only those will contribute when projected on to the spin tensor $S_{(i)(j)}$
or equally antisymmetric expressions. We insert above expression into Eq.\ (\ref{somega1}), which leads to a rather long expression. We define
\begin{align}
 \frac{1}{2}S_{ab}\Omega^{ab}-\frac{1}{2}S_{(j)(k)}\frac{u^{(k)}\dot{u}^{(j)}}{u^{(0)}u_{(0)}}\equiv\mathcal{W}
\end{align}
and expand it
\begin{align}
 \begin{aligned}
  2\mathcal{W}&=\left(S_{(i)(j)}+S_{(i)(k)}\frac{u^{(k)}u_{(j)}}{u^{(0)}u_{(0)}}-S_{(j)(k)}\frac{u^{(k)}u_{(i)}}{u^{(0)}u_{(0)}}\right)\tilde{\Omega}^{(i)(j)}\\
             &=\left(S_{(i)(j)}+S_{(i)(k)}\frac{u^{(k)}u_{(j)}}{u^{(0)}u_{(0)}}-S_{(j)(k)}\frac{u^{(k)}u_{(i)}}{u^{(0)}u_{(0)}}\right)\\
             &\quad\times\left(\hat{\Omega}^{(i)(j)}-\hat{\Omega}^{(i)(l)}\frac{u_{(l)}u^{(j)}}{u(u+u_{(0)})}-\hat{\Omega}^{(l)(j)}\frac{u_{(l)}u^{(i)}}{u(u+u_{(0)})}+\frac{\vct{u}^2u^{(i)}\dot{u}^{(j)}}{u^2(u+u_{(0)})^2}\right)
\end{aligned}
\end{align}
\begin{align}
\begin{aligned}
  2\mathcal{W}&=S_{(i)(j)}\hat{\Omega}^{(i)(j)}\\
             &\quad-S_{(i)(j)}\hat{\Omega}^{(i)(l)}\frac{u_{(l)}u^{(j)}}{u(u+u_{(0)})}-S_{(i)(j)}\hat{\Omega}^{(l)(j)}\frac{u_{(l)}u^{(i)}}{u(u+u_{(0)})}+S_{(i)(j)}\frac{\vct{u}^2u^{(i)}\dot{u}^{(j)}}{u^2(u+u_{(0)})^2}\\
             &\quad+S_{(i)(k)}\hat{\Omega}^{(i)(j)}\frac{u^{(k)}u_{(j)}}{u_{(0)}^2}-S_{(i)(k)}\hat{\Omega}^{(i)(l)}\frac{u^{(k)}u_{(l)}\vct{u}^2}{u(u+u_{(0)})u_{(0)}^2}-S_{(j)(k)}\hat{\Omega}^{(i)(j)}\frac{u^{(k)}u_{(i)}}{u_{(0)}^2}\\
             &\quad-S_{(j)(k)}\hat{\Omega}^{(j)(l)}\frac{u^{(k)}u_{(l)}\vct{u}^2}{u(u+u_{(0)})u_{(0)}^2}-S_{(j)(k)}\frac{\vct{u}^4u^{(k)}\dot{u}^{(j)}}{u^2u_{(0)}^2(u+u_{(0)})^2}\\
             &=S_{(i)(j)}\hat{\Omega}^{(i)(j)}+S_{(i)(j)}\hat{\Omega}^{(i)(l)}\left[\frac{-2u_{(l)}u^{(j)}}{u(u+u_{(0)})}+\frac{2u^{(j)}u_{(l)}}{u_{(0)}^2}-\frac{2u^{(j)}u_{(l)}\vct{u}^2}{u(u+u_{(0)})u_{(0)}^2}\right]\\
             &\quad+S_{(i)(j)}\frac{\vct{u}^2u^{(i)}\dot{u}^{(j)}}{u^2(u+u_{(0)})^2}\left(1+\frac{\vct{u}^2}{u_{(0)}^2}\right)\,.
 \end{aligned}
\end{align}
So the result is
\begin{align}
 2\mathcal{W}=S_{(i)(j)}\hat{\Omega}^{(i)(j)}+2S_{(i)(j)}\hat{\Omega}^{(i)(l)}\frac{u_{(l)}u^{(j)}}{(u+u_{(0)})u_{(0)}}+S_{(i)(j)}\frac{\vct{u}^2u^{(i)}\dot{u}^{(j)}}{u_{(0)}^2(u+u_{(0)})^2}\,.
\end{align}
Again we make an ansatz for the spin redefinition with an unknown function $\chi$ according to the rule
\begin{align}
 S_{(i)(j)}=\hat{S}_{(i)(j)}+\chi\hat{S}_{(i)(k)}u_{(j)}u^{(k)}-\chi\hat{S}_{(j)(k)}u_{(i)}u^{(k)}\,,\quad S_{(i)(j)}u^{(j)}=\hat{S}_{(i)(j)}u^{(j)}\left(1+\chi\vct{u}^2\right)
\end{align}
with $\hat{S}_{(i)(j)}$ as the supposed canonical spin which yields
\begin{align}
\begin{aligned}
 2\mathcal{W}&=\hat{S}_{(i)(j)}\hat{\Omega}^{(i)(j)}+\chi\hat{S}_{(i)(k)}\hat{\Omega}^{(i)(j)}u_{(j)}u^{(k)}-\chi\hat{S}_{(j)(k)}\hat{\Omega}^{(i)(j)}u_{(i)}u^{(k)}\\
             &\quad+2\hat{S}_{(i)(j)}\hat{\Omega}^{(i)(l)}\frac{u_{(l)}u^{(j)}}{(u+u_{(0)})u_{(0)}}(1+\chi\vct{u}^2)-\hat{S}_{(i)(j)}\frac{\vct{u}^2u^{(j)}\dot{u}^{(i)}}{u_{(0)}^2(u+u_{(0)})^2}(1+\chi\vct{u}^2)\\
             &=\hat{S}_{(i)(j)}\hat{\Omega}^{(i)(j)}+2\hat{S}_{(i)(j)}\hat{\Omega}^{(i)(l)}u_{(l)}u^{(j)}\left[\frac{1+\chi\vct{u}^2}{(u+u_{(0)})u_{(0)}}+\chi\right]\\
             &\quad+\hat{S}_{(i)(j)}\frac{\vct{u}^2u^{(i)}\dot{u}^{(j)}}{u_{(0)}^2(u+u_{(0)})^2}(1+\chi\vct{u}^2)\,.
\end{aligned}
\end{align}
We demand
\begin{equation}
 \frac{1+\chi\vct{u}^2}{(u+u_{(0)})u_{(0)}}+\chi=0\quad\Leftrightarrow\quad 1+\chi u(u+u_{(0)})=0\quad\curvearrowright\quad\chi=-\frac{1}{u(u+u_{(0)})}\,,
\end{equation}
from which follows the spin redefinition
\begin{equation}\label{spinredef1}
 S_{(i)(j)}=\hat{S}_{(i)(j)}-\hat{S}_{(i)(k)}\frac{u_{(j)}u^{(k)}}{u(u+u_{(0)})}+\hat{S}_{(j)(k)}\frac{u_{(i)}u^{(k)}}{u(u+u_{(0)})}\,,
\end{equation}
and
\begin{equation}
 2\mathcal{W}=\hat{S}_{(i)(j)}\hat{\Omega}^{(i)(j)}+\hat{S}_{(i)(j)}\frac{\vct{u}^2u^{(i)}\dot{u}^{(j)}}{u_{(0)}^2(u+u_{(0)})^2}\left(1-\frac{\vct{u}^2}{u(u+u_{(0)})}\right)\,.
\end{equation}
Hence the spin coupling term is reduced to the expression
\begin{align}
\begin{aligned}
 \frac{1}{2}S_{ab}\Omega^{ab}&=\frac{1}{2}\hat{S}_{(i)(j)}\hat{\Omega}^{(i)(j)}+\frac{1}{2}\hat{S}_{(i)(j)}\frac{\vct{u}^2u^{(i)}\dot{u}^{(j)}}{u_{(0)}^2(u+u_{(0)})^2}\left(1-\frac{\vct{u}^2}{u(u+u_{(0)})}\right)\\
                             &\quad+\frac{1}{2}\hat{S}_{(j)(i)}\frac{u^{(i)}\dot{u}^{(j)}}{u_{(0)}^2}\left(1-\frac{\vct{u}^2}{u(u+u_{(0)})}\right)\\
                             &=\frac{1}{2}\hat{S}_{(i)(j)}\hat{\Omega}^{(i)(j)}+\frac{1}{2}\hat{S}_{(i)(j)}\frac{\vct{u}^2u^{(i)}\dot{u}^{(j)}}{u_{(0)}^2(u+u_{(0)})^2}\frac{u_{0}}{u}-\frac{1}{2}\hat{S}_{(i)(j)}\frac{u^{(i)}\dot{u}^{(j)}}{u_{(0)}^2}\frac{u_{(0)}}{u}\\
                             &=\frac{1}{2}\hat{S}_{(i)(j)}\hat{\Omega}^{(i)(j)}-\frac{1}{2}\hat{S}_{(i)(j)}\frac{u^{(i)}\dot{u}^{(j)}}{uu_{(0)}}\left(-\frac{\vct{u^2}}{(u+u_{(0)})^2}+1\right)\\
                             &=\frac{1}{2}\hat{S}_{(i)(j)}\hat{\Omega}^{(i)(j)}-\hat{S}_{(i)(j)}\frac{u^{(i)}\dot{u}^{(j)}}{u(u+u_{(0)})}\,,
\end{aligned}
\end{align}
giving
\begin{equation}\label{SOmegared1}
 \frac{1}{2}S_{ab}\Omega^{ab}=\frac{1}{2}\hat{S}_{(i)(j)}\hat{\Omega}^{(i)(j)}-\mathcal{Z}\quad{\text{with}}\quad\mathcal{Z}\equiv\hat{S}_{(i)(j)}\frac{u^{(i)}\dot{u}^{(j)}}{u(u+u_{(0)})}\,.
\end{equation}

To solve for the momenta in (\ref{SOmegared1}) one has to insert the vierbein which is perturbatively calculated to the needed
pN order, see Eqs.\ (\ref{e0c})-(\ref{eijc}). First we make an expansion of $\mathcal{Z}$ in powers of $\vct{u}^2$ (in the sense of a post-Newtonian approximation) 
\begin{align}\label{Zterm1}
 \begin{aligned}
 \mathcal{Z}=\hat{S}_{(i)(j)}\frac{u^{(i)}\dot{u}^{(j)}}{u(u+u_{(0)})}=\hat{S}_{(i)(j)}u^{(i)}\dot{u}^{(j)}\left(\frac{1}{2u_{(0)}^2}-\frac{3\vct{u}^2}{8u_{(0)}^4}+\mathcal{O}\left(\vct{u}^4,c^{-10}\right)\right)\,.
 \end{aligned}
\end{align}
Next we insert (\ref{u0red}) into (\ref{Zterm1}) and pN expand the result up to the order $c^{-8}$ leading to
\begin{align}
 \begin{aligned}
 \mathcal{Z}_{1}\simeq\frac{1}{2}\hat{S}_{1(i)(j)}u^{(i)}_{1}\dot{u}^{(j)}_{1}\left(1+2G\frac{m_{2}}{r_{12}}\right)+\mathcal{O}(c^{-10})\,.
 \end{aligned}
\end{align}
After that we insert (\ref{uired}) into this equation yielding the approximate expression
\begin{align}\label{Zcal1}
 \begin{aligned}
 &\mathcal{Z}_{1}\simeq\frac{1}{2}\hat{S}_{1(i)(j)}u^{(i)}_{1}\dot{u}^{(j)}_{1}+G\frac{m_{2}}{r_{12}}\hat{S}_{1(i)(j)}u_{1}^{(i)}\frac{\dot{p}_{1j}}{m_{1}}+\mathcal{O}(c^{-10})\,\\
                &\simeq\frac{1}{2}\hat{S}_{1(i)(j)}\Bigg[\frac{p_{1i}}{m_{1}}-\frac{2Gm_{2}p_{1i}}{m_{1}r_{12}}+\frac{3Gp_{2i}}{2r_{12}}+\frac{Gn_{12}^{i}\piin}{2r_{12}}-\frac{Gm_{2}n_{12}^kS_{1}^{(i)(k)}}{m_{1}r_{12}^2}\\
                &\quad-\frac{Gn_{12}^kS_{2}^{(i)(k)}}{r_{12}^2}\Bigg]\dot{u}^{(j)}_{1}+G\frac{m_{2}}{r_{12}}\hat{S}_{1(i)(j)}\frac{p_{1i}\dot{p}_{1j}}{m_{1}^2}+\mathcal{O}(c^{-10})\\
                &\simeq\hat{S}_{1(i)(j)}p_{1i}\frac{1}{2m_{1}}\dot{u}^{(j)}_{1}+G\frac{m_{2}}{r_{12}}\hat{S}_{1(i)(j)}\frac{p_{1i}\dot{p}_{1j}}{m_{1}^2}+\frac{1}{2}\hat{S}_{1(i)(j)}\Bigg[-\frac{2Gm_{2}p_{1i}}{m_{1}r_{12}}+\frac{3Gp_{2i}}{2r_{12}}\\
                &\quad+\frac{Gn_{12}^{i}\piin}{2r_{12}}-\frac{Gm_{2}n_{12}^kS_{1}^{(i)(k)}}{m_{1}r_{12}^2}-\frac{Gn_{12}^kS_{2}^{(i)(k)}}{r_{12}^2}\Bigg]\frac{\dot{p}_{1j}}{m_{1}}\\
                &\simeq\frac{1}{2m_{1}}\hat{S}_{1(i)(j)}p_{1i}\dot{u}^{(j)}_{1}+\frac{1}{2}\hat{S}_{1(i)(j)}\Bigg[\frac{3Gp_{2i}}{2r_{12}}+\frac{Gn_{12}^{i}\piin}{2r_{12}}\\
                &\quad-\frac{Gm_{2}n_{12}^kS_{1}^{(i)(k)}}{m_{1}r_{12}^2}-\frac{Gn_{12}^kS_{2}^{(i)(k)}}{r_{12}^2}\Bigg]\frac{\dot{p}_{1j}}{m_{1}}+\mathcal{O}(c^{-10})\,.
\end{aligned}
\end{align}
We eliminate the time derivative of $u_{1}^{(j)}$ by shifting it on-shell (i.e. we neglect total time derivatives symbolized
by $\approx$) onto the momentum leaving us also with a time derivative of the canonical spin, which we will have to deal with later when we reconsider the spin redefinition.
\begin{align}
\begin{aligned}
 \mathcal{Z}_{1}&\approx-\frac{1}{2m_{1}}\dot{\hat{S}}_{1(i)(j)}p_{1i}u^{(j)}_{1}-\frac{1}{2m_{1}}\hat{S}_{1(i)(j)}\dot{p}_{1i}u^{(j)}_{1}+\frac{1}{2}\hat{S}_{1(i)(j)}\Bigg[\frac{3Gp_{2i}}{2r_{12}}+\frac{Gn_{12}^{i}\piin}{2r_{12}}\\
                &\quad-\frac{Gm_{2}n_{12}^kS_{1}^{(i)(k)}}{m_{1}r_{12}^2}-\frac{Gn_{12}^kS_{2}^{(i)(k)}}{r_{12}^2}\Bigg]\frac{\dot{p}_{1j}}{m_{1}}+\mathcal{O}(c^{-10})\\
                &\simeq-\frac{1}{2m_{1}}\dot{\hat{S}}_{1(i)(j)}p_{1i}u^{(j)}_{1}+\frac{1}{2}\hat{S}_{1(i)(j)}\bigg[\frac{p_{1i}}{m_{1}}-\frac{2Gm_{2}p_{1i}}{m_{1}r_{12}}+\frac{3Gp_{2i}}{2r_{12}}+\frac{Gn_{12}^{i}\piin}{2r_{12}}\\
                &\quad-\frac{Gm_{2}n_{12}^kS_{1}^{(i)(k)}}{m_{1}r_{12}^2}-\frac{Gn_{12}^kS_{2}^{(i)(k)}}{r_{12}^2}\bigg]\frac{\dot{p}_{1j}}{m_{1}}+\frac{1}{2}\hat{S}_{1(i)(j)}\bigg[\frac{3Gp_{2i}}{2r_{12}}+\frac{Gn_{12}^{i}\piin}{2r_{12}}\\
                &\quad-\frac{Gm_{2}n_{12}^kS_{1}^{(i)(k)}}{m_{1}r_{12}^2}-\frac{Gn_{12}^kS_{2}^{(i)(k)}}{r_{12}^2}\bigg]\frac{\dot{p}_{1j}}{m_{1}}+\mathcal{O}(c^{-10})\\
%                 &=-\frac{1}{2m_{1}}\dot{\hat{S}}_{1(i)(j)}p_{1i}u^{(j)}_{1}+\frac{1}{2m_{1}^2}\hat{S}_{1(i)(j)}p_{1i}\dot{p}_{1j}\\
%                 &\quad+\hat{S}_{1(i)(j)}\left[-\frac{Gm_{2}p_{1i}}{m_{1}r_{12}}+\frac{3Gp_{2i}}{2r_{12}}+\frac{Gn_{12}^{i}\piin}{2r_{12}}-\frac{Gm_{2}n_{12}^kS_{1}^{(i)(k)}}{m_{1}r_{12}^2}-\frac{Gn_{12}^kS_{2}^{(i)(k)}}{r_{12}^2}\right]\frac{\dot{p}_{1j}}{m_{1}}\\
                &=-\frac{1}{2m_{1}}\dot{\hat{S}}_{1(i)(j)}p_{1i}u^{(j)}_{1}+\frac{\hat{S}_{1(i)(j)}}{2m_{1}}\Bigg[\frac{p_{1i}}{m_{1}}-\frac{2Gm_{2}p_{1i}}{m_{1}r_{12}}+\frac{3Gp_{2i}}{r_{12}}+\frac{Gn_{12}^{i}\piin}{r_{12}}\\
                &\quad-\frac{2Gm_{2}n_{12}^kS_{1}^{(i)(k)}}{m_{1}r_{12}^2}-\frac{2Gn_{12}^kS_{2}^{(i)(k)}}{r_{12}^2}\Bigg]\dot{p}_{1j}+\mathcal{O}(c^{-10})\,.
 \end{aligned}
\end{align}
Now we are ready to return to the action (\ref{effaction}), wherein we insert Eqs.\ (\ref{SOmegared1}) and (\ref{Zcal1}) leading to the expression (for particle 1)
\begin{align}
 \begin{aligned}
  S_{\text{eff}}&=\int\dd t\left(p_{1i}\dot z^{i}_{1}-\frac{1}{2}S_{1ab}\Omega_{1}^{ab}-H_{\text{eff}}\right)\\
                &\approx\int\dd t\Bigg(-\dot{p}_{1j}z^{j}_{1}-\frac{1}{2}\hat{S}_{(i)(j)}\hat{\Omega}^{(i)(j)}-\frac{1}{2m_{1}}\dot{\hat{S}}_{1(i)(j)}p_{1i}u^{(j)}_{1}+\frac{\hat{S}_{1(i)(j)}}{2m_{1}}\Bigg[\frac{p_{1i}}{m_{1}}-\frac{2Gm_{2}p_{1i}}{m_{1}r_{12}}\\
                &+\frac{3Gp_{2i}}{r_{12}}+\frac{Gn_{12}^{i}\piin}{r_{12}}-\frac{2Gm_{2}n_{12}^kS_{1}^{(i)(k)}}{m_{1}r_{12}^2}-\frac{2Gn_{12}^kS_{2}^{(i)(k)}}{r_{12}^2}\Bigg]\dot{p}_{1j}-H_{\text{eff}}\Bigg)\,.
 \end{aligned}
\end{align}
This enables us to read off the position coordinate shift (for the Minkowski case only the leading order term is shown)
\begin{align}\label{ztrafoA1}
 \begin{aligned}
  z_{1}^{j}&=\hat{z}_{1}^{j}+\frac{\hat{S}_{1(i)(j)}}{2m_{1}}\Bigg[\frac{p_{1i}}{m_{1}}-\frac{2Gm_{2}p_{1i}}{m_{1}r_{12}}+\frac{3Gp_{2i}}{r_{12}}+\frac{Gn_{12}^{i}\piin}{r_{12}}-\frac{2Gm_{2}n_{12}^kS_{1}^{(i)(k)}}{m_{1}r_{12}^2}\\
                 &\qquad\qquad\qquad-\frac{2Gn_{12}^kS_{2}^{(i)(k)}}{r_{12}^2}\Bigg]+\mathcal{O}(c^{-6})\,.
 \end{aligned}
\end{align}
The spin and the Lorentz matrix need another redefinition in order to cancel the term
$-\frac{1}{2m_{1}}\dot{\hat{S}}_{1(i)(j)}p_{1i}u^{(j)}_{1}$ from the action. This is achieved by an infinitesimal
rotation $\omega^{(i)(j)}=-\omega^{(j)(i)}$ of the local basis so that the canonical spin and Lorentz matrices are corotated according to
\begin{align}\label{spinrotation1}
\begin{aligned}
- \frac{1}{2}\hat{S}_{(i)(j)}\hat{\Omega}^{(i)(j)}&\rightarrow-\frac{1}{2}\left[\hat{S}_{(i)(j)}+\omega_{(i)}{}^{(m)}\hat{S}_{(m)(j)}+\omega_{(j)}{}^{(m)}\hat{S}_{(i)(m)}\right]\hat{\Omega}_{\omega}^{(i)(j)}\\
                                                            &=-\frac{1}{2}\left[\hat{S}_{(i)(j)}+\omega_{(i)}{}^{(m)}\hat{S}_{(m)(j)}+\omega_{(j)}{}^{(m)}\hat{S}_{(i)(m)}\right]\hat{\Lambda}_{[k]\omega}^{\;\;\;(i)}\dot{\hat{\Lambda}}^{[k](j)}_{\omega}\\
&=-\frac{1}{2}\left[\hat{S}_{(i)(j)}+\omega_{(i)}{}^{(m)}\hat{S}_{(m)(j)}+\omega_{(j)}{}^{(m)}\hat{S}_{(i)(m)}\right]\left(\hat{\Lambda}_{[k]}^{\;\;\;(i)}+\omega^{(i)}{}_{(l)}\hat{\Lambda}_{[k]}^{\;\;\;(l)}\right)\\
                                                            &\quad\times\left(\dot{\hat{\Lambda}}^{[k](j)}+\omega^{(j)}{}_{(p)}\dot{\hat{\Lambda}}^{[k](p)}+\dot{\omega}^{(j)}{}_{(p)}\hat{\Lambda}^{[k](p)}\right)\\
                                                            &=-\frac{1}{2}\hat{S}_{(i)(j)}\hat{\Omega}^{(i)(j)}-\frac{1}{2}\hat{S}_{(i)(j)}\dot{\omega}^{(j)}{}_{(p)}\eta^{(i)(p)}\\
                                                            &=-\frac{1}{2}\hat{S}_{(i)(j)}\hat{\Omega}^{(i)(j)}-\frac{1}{2}\hat{S}_{(i)(j)}\dot{\omega}^{(j)(i)}\\
                                                            &\approx-\frac{1}{2}\hat{S}_{(i)(j)}\hat{\Omega}^{(i)(j)}-\frac{1}{2}\dot{\hat{S}}_{(i)(j)}\omega^{(i)(j)}\,.
\end{aligned}
\end{align}
We identify
\begin{equation}
\omega_{(i)}{}^{(m)} = - \omega^{(i)(m)} = \frac{1}{2}\frac{p_{1i}u_{1}^{(m)}}{m_{1}}-\frac{1}{2}\frac{p_{1m}u_{1}^{(i)}}{m_{1}}
\end{equation}
and use Eq.\ (\ref{spinredef1}) to determine the final spin redefinition to our approximation:
\begin{align}\label{SCorot1}
 \begin{aligned}
  S_{1(i)(j)}&=\hat{S}_{1(i)(j)}-\hat{S}_{1(i)(k)}\frac{u_{1(j)}u_{1}^{(k)}}{2u_{1(0)}}+\hat{S}_{1(j)(k)}\frac{u_{1(i)}u_{1}^{(k)}}{2u_{1(0)}}+\frac{p_{1[i}u_{1}^{(m)]}}{m_{1}}\hat{S}_{(m)(j)}+\frac{p_{1[j}u_{1}^{(m)]}}{m_{1}}\hat{S}_{(i)(m)}\\
             &=\frac{1}{2}\hat{S}_{1(i)(j)}-\hat{S}_{1(i)(k)}\frac{u_{1(j)}u_{1}^{(k)}}{2u_{1(0)}}+\frac{p_{1[j}u_{1}^{(k)]}}{m_{1}}\hat{S}_{(i)(k)}-(i\leftrightarrow j)\,.
\end{aligned}
\end{align}
Next we insert Eq.\ (\ref{u0red}) for $u_{(0)}$ in the second term and Eq.\ (\ref{uired}) for $u^{(i)}$ the third term of
(\ref{SCorot1}) and make a pN expansion up to the order $c^{-7}$:
\begin{align}\label{Sbbb1}
\begin{aligned}
  S_{1(i)(j)}&=\frac{1}{2}\hat{S}_{1(i)(j)}-\frac{1}{2}\hat{S}_{1(i)(k)}u_{1(j)}u_{1}^{(k)}\left(1+2G\frac{m_{2}}{r_{12}}\right)\\
             &\quad+\frac{p_{1[j}}{m_{1}}\hat{S}_{(i)(k)}\Bigg[\frac{p_{1k]}}{m_{1}}-\frac{2Gm_{2}p_{1k]}}{m_{1}r_{12}}+\frac{3Gp_{2k]}}{2r_{12}}+\frac{Gn_{12}^{k]}\piin}{2r_{12}}-\frac{Gm_{2}n_{12}^lS_{1}^{(k)](l)}}{m_{1}r_{12}^2}\\
             &\qquad\qquad\qquad-\frac{Gn_{12}^lS_{2}^{(k)](l)}}{r_{12}^2}\Bigg]-(i\leftrightarrow j)+\mathcal{O}(c^{-9})\,.
\end{aligned}
\end{align}
We are left with an insertion of the remaining 4-velocities. Again utilizing Eq.\ (\ref{uired}) a further expansion of (\ref{Sbbb1}) yields
\begin{align}\label{Sccc1}
\begin{aligned}
S_{1(i)(j)}&=\frac{1}{2}\hat{S}_{1(i)(j)}+\frac{1}{2}\hat{S}_{1(i)(k)}\frac{p_{1j}}{m_{1}}u_{1}^{(k)}+\frac{1}{2}\hat{S}_{1(i)(k)}\frac{p_{1k}}{m_{1}}u_{1}^{(j)}+\frac{Gm_{2}}{m_{1}^2r_{12}}\hat{S}_{1(i)(k)}p_{1j}p_{1k}\\
             &\quad+\frac{p_{1[j}}{m_{1}}\hat{S}_{(i)(k)}\Bigg[\frac{p_{1k]}}{m_{1}}-\frac{2Gm_{2}p_{1k]}}{m_{1}r_{12}}+\frac{3Gp_{2k]}}{2r_{12}}+\frac{Gn_{12}^{k]}\piin}{2r_{12}}-\frac{Gm_{2}n_{12}^lS_{1}^{(k)](l)}}{m_{1}r_{12}^2}\\
             &\quad-\frac{Gn_{12}^lS_{2}^{(k)](l)}}{r_{12}^2}\Bigg]-\frac{1}{2}\hat{S}_{1(i)(k)}\frac{p_{1j}p_{1k}}{m_{1}^2}-(i\leftrightarrow j)+\mathcal{O}(c^{-9})\\
%              &=\frac{1}{2}\hat{S}_{1(i)(j)}+\hat{S}_{1(i)(k)}\frac{p_{1}^{(j}u_{1}^{(k))}}{m_{1}}+\frac{Gm_{2}}{m_{1}^2r_{12}}\hat{S}_{1(i)(k)}p_{1j}p_{1k}-\frac{1}{2}\hat{S}_{1(i)(k)}\frac{p_{1j}p_{1k}}{m_{1}^2}\\
%              &\quad+\frac{p_{1[j}}{m_{1}}\hat{S}_{(i)(k)}\Bigg[\frac{3Gp_{2k]}}{2r_{12}}+\frac{Gn_{12}^{k]}\piin}{2r_{12}}-\frac{Gm_{2}n_{12}^lS_{1}^{(k)](l)}}{m_{1}r_{12}^2}-\frac{Gn_{12}^lS_{2}^{(k)](l)}}{r_{12}^2}\Bigg]\\
%              &\quad-(i\leftrightarrow j)+\mathcal{O}(c^{-9})\\
             &=\frac{1}{2}\hat{S}_{1(i)(j)}+\frac{Gm_{2}}{m_{1}^2r_{12}}\hat{S}_{1(i)(k)}p_{1j}p_{1k}+\hat{S}_{1(i)(k)}\frac{p_{1(j}}{m_{1}}\Bigg[\frac{p_{1k)}}{m_{1}}-\frac{2Gm_{2}p_{1k)}}{m_{1}r_{12}}+\frac{3Gp_{2k)}}{2r_{12}}\\
             &\quad+\frac{Gn_{12}^{k)}\piin}{2r_{12}}-\frac{Gm_{2}n_{12}^lS_{1}^{(k))(l)}}{m_{1}r_{12}^2}-\frac{Gn_{12}^lS_{2}^{(k))(l)}}{r_{12}^2}\Bigg]-\frac{1}{2}\hat{S}_{1(i)(k)}\frac{p_{1j}p_{1k}}{m_{1}^2}\\
             &\quad+\frac{p_{1[j}}{m_{1}}\hat{S}_{(i)(k)}\Bigg[\frac{3Gp_{2k]}}{2r_{12}}+\frac{Gn_{12}^{k]}\piin}{2r_{12}}-\frac{Gm_{2}n_{12}^lS_{1}^{(k)](l)}}{m_{1}r_{12}^2}-\frac{Gn_{12}^lS_{2}^{(k)](l)}}{r_{12}^2}\Bigg]\\
             &\quad-(i\leftrightarrow j)+\mathcal{O}(c^{-9})\\
             &=\frac{1}{2}\hat{S}_{1(i)(j)}+\hat{S}_{1(i)(k)}\Bigg[\frac{1}{2}\frac{p_{1k}p_{1j}}{m_{1}^2}-\frac{Gm_{2}p_{1k}p_{1j}}{m_{1}^{2}r_{12}}+\frac{3}{2}\frac{Gp_{1j}p_{2k}}{m_{1}r_{12}}+\frac{Gp_{1j}n_{12}^{k}(\vct{p}_{2}\cdot\vct{n}_{12})}{2m_{1}r_{12}^2}\\
             &\qquad+\frac{Gm_{2}n_{12}^lS_{1}^{(l)(k)}p_{1j}}{m_{1}^2r_{12}^2}+\frac{Gn_{12}^lS_{2}^{(l)(k)}p_{1j}}{m_{1}r_{12}^2}\Bigg]-(i\leftrightarrow j)+\mathcal{O}(c^{-9})\,.
 \end{aligned}
\end{align}
The last step involves providing all spin and position variables with a hat in the highest pN terms of (\ref{Sccc1}), so that we end up with the transformation formula
\begin{align}\label{Strafofinal1}
 \begin{aligned}
  S_{1(i)(j)}&=\frac{1}{2}\hat{S}_{1(i)(j)}+\hat{S}_{1(i)(k)}\Bigg[\frac{1}{2}\frac{p_{1k}p_{1j}}{m_{1}^2}-\frac{Gm_{2}p_{1k}p_{1j}}{m_{1}^{2}\hat{r}_{12}}+\frac{3}{2}\frac{Gp_{1j}p_{2k}}{m_{1}\hat{r}_{12}}+\frac{Gp_{1j}\hat{n}_{12}^{k}(\vct{p}_{2}\cdot\hat{\vct{n}}_{12})}{2m_{1}\hat{r}_{12}^2}\\
             &\qquad+\frac{Gm_{2}\hat{n}_{12}^l\hat{S}_{1}^{(l)(k)}p_{1j}}{m_{1}^2\hat{r}_{12}^2}+\frac{G\hat{n}_{12}^l\hat{S}_{2}^{(l)(k)}p_{1j}}{m_{1}\hat{r}_{12}^2}\Bigg]-(i\leftrightarrow j)+\mathcal{O}(c^{-9})\,.
 \end{aligned}
\end{align}

\providecommand{\jr}[1]{#1}
\providecommand{\etal}{~et~al.}

\providecommand{\href}[2]{#2}\begingroup\raggedright\endgroup


\begin{thebibliography}{10}

\bibitem{Goldberger:Rothstein:2006}
W.~D. Goldberger and I.~Z. Rothstein, ``An effective field theory of gravity
  for extended objects,''
  \href{http://dx.doi.org/10.1103/PhysRevD.73.104029}{{\em Phys. Rev. D} {\bf
  73} (2006)  104029},
\href{http://arxiv.org/abs/hep-th/0409156}{{\tt arXiv:hep-th/0409156}}.
%%CITATION = HEP-TH/0409156;%%.

\bibitem{Gilmore:Ross:2008}
J.~B. Gilmore and A.~Ross, ``{E}ffective field theory calculation of second
  post-{N}ewtonian binary dynamics,''
  \href{http://dx.doi.org/10.1103/PhysRevD.78.124021}{{\em Phys. Rev. D} {\bf
  78} (2008)  124021}, \href{http://arxiv.org/abs/0810.1328}{{\tt
  arXiv:0810.1328 [gr-qc]}}.

\bibitem{Chu:2009}
Y.-Z. Chu, ``n-body problem in general relativity up to the second
  post-{N}ewtonian order from perturbative field theory,''
  \href{http://dx.doi.org/10.1103/PhysRevD.79.044031}{{\em Phys. Rev. D} {\bf
  79} (2009)  044031}, \href{http://arxiv.org/abs/0812.0012}{{\tt
  arXiv:0812.0012 [gr-qc]}}.

\bibitem{Foffa:Sturani:2011}
S.~Foffa and R.~Sturani, ``{E}ffective field theory calculation of conservative
  binary dynamics at third post-{N}ewtonian order,''
  \href{http://dx.doi.org/10.1103/PhysRevD.84.044031}{{\em Phys. Rev. D} {\bf
  84} (2011)  044031},
\href{http://arxiv.org/abs/1104.1122}{{\tt arXiv:1104.1122 [gr-qc]}}.
%%CITATION = 1104.1122;%%.

\bibitem{Kol:Smolkin:2009}
B.~Kol and M.~Smolkin, ``Dressing the post-{N}ewtonian two-body problem and
  classical effective field theory,''
  \href{http://dx.doi.org/10.1103/PhysRevD.80.124044}{{\em Phys. Rev. D} {\bf
  80} (2009)  124044},
\href{http://arxiv.org/abs/0910.5222}{{\tt arXiv:0910.5222 [hep-th]}}.
%%CITATION = 0910.5222;%%.

\bibitem{Porto:2006}
R.~A. Porto, ``Post-{N}ewtonian corrections to the motion of spinning bodies in
  nonrelativistic general relativity,''
  \href{http://dx.doi.org/10.1103/PhysRevD.73.104031}{{\em Phys. Rev. D} {\bf
  73} (2006)  104031},
\href{http://arxiv.org/abs/gr-qc/0511061}{{\tt arXiv:gr-qc/0511061}}.
%%CITATION = GR-QC/0511061;%%.

\bibitem{Kol:Smolkin:2008}
B.~Kol and M.~Smolkin, ``{N}on-relativistic gravitation: from {N}ewton to
  {E}instein and back,''
  \href{http://dx.doi.org/10.1088/0264-9381/25/14/145011}{{\em Class. Quant.
  Grav.} {\bf 25} (2008)  145011}, \href{http://arxiv.org/abs/0712.4116}{{\tt
  arXiv:0712.4116 [hep-th]}}.

\bibitem{Porto:Rothstein:2008:1}
R.~A. Porto and I.~Z. Rothstein, ``Spin(1)spin(2) effects in the motion of
  inspiralling compact binaries at third order in the post-{N}ewtonian
  expansion,'' \href{http://dx.doi.org/10.1103/PhysRevD.78.044012}{{\em Phys.
  Rev. D} {\bf 78} (2008)  044012},
\href{http://arxiv.org/abs/0802.0720}{{\tt arXiv:0802.0720 [gr-qc]}}.
%%CITATION = 0802.0720;%%.

\bibitem{Levi:2008}
M.~Levi, ``Next-to-leading order gravitational spin1-spin2 coupling with
  {K}aluza-{K}lein reduction,''
  \href{http://dx.doi.org/10.1103/PhysRevD.82.064029}{{\em Phys. Rev. D} {\bf
  82} (2010)  064029},
\href{http://arxiv.org/abs/0802.1508}{{\tt arXiv:0802.1508 [gr-qc]}}.
%%CITATION = 0802.1508;%%.

\bibitem{Porto:Rothstein:2008:2}
R.~A. Porto and I.~Z. Rothstein, ``Next to leading order spin(1)spin(1) effects
  in the motion of inspiralling compact binaries,''
  \href{http://dx.doi.org/10.1103/PhysRevD.78.044013}{{\em Phys. Rev. D} {\bf
  78} (2008)  044013},
\href{http://arxiv.org/abs/0804.0260}{{\tt arXiv:0804.0260 [gr-qc]}}.
%%CITATION = 0804.0260;%%.

\bibitem{Porto:Rothstein:2008:2:err}
R.~A. Porto and I.~Z. Rothstein, ``Erratum: {N}ext to leading order
  spin(1)spin(1) effects in the motion of inspiralling compact binaries,''
  \href{http://dx.doi.org/10.1103/PhysRevD.81.029905}{{\em Phys. Rev. D} {\bf
  81} (2010)  029905(E)}.

\bibitem{Porto:2010}
R.~A. Porto, ``Next to leading order spin-orbit effects in the motion of
  inspiralling compact binaries,''
  \href{http://dx.doi.org/10.1088/0264-9381/27/20/205001}{{\em Class. Quant.
  Grav.} {\bf 27} (2010)  205001},
\href{http://arxiv.org/abs/1005.5730}{{\tt arXiv:1005.5730 [gr-qc]}}.
%%CITATION = 1005.5730;%%.

\bibitem{Levi:2010}
M.~Levi, ``Next-to-leading order gravitational spin-orbit coupling in an
  effective field theory approach,''
  \href{http://dx.doi.org/10.1103/PhysRevD.82.104004}{{\em Phys. Rev. D} {\bf
  82} (2010)  104004},
\href{http://arxiv.org/abs/1006.4139}{{\tt arXiv:1006.4139 [gr-qc]}}.
%%CITATION = 1006.4139;%%.

\bibitem{Perrodin:2010}
D.~L. Perrodin, ``Subleading spin-orbit correction to the {N}ewtonian potential
  in effective field theory formalism,'' in {\em Proceedings of the 12th Marcel
  Grossmann Meeting on General Relativity}.
\newblock World Scientific, Singapore, 2010.
\newblock \href{http://arxiv.org/abs/1005.0634}{{\tt arXiv:1005.0634 [gr-qc]}}.
\newblock
(to be published).
%%CITATION = 1005.0634;%%.

\bibitem{Levi:2011}
M.~Levi, ``{B}inary dynamics from spin1-spin2 coupling at fourth
  post-{N}ewtonian order,''
\href{http://arxiv.org/abs/1107.4322}{{\tt arXiv:1107.4322 [gr-qc]}}.
%%CITATION = 1107.4322;%%.

\bibitem{Fleming:1965}
G.~N. Fleming, ``Covariant position operators, spin, and locality,''
  \href{http://dx.doi.org/10.1103/PhysRev.137.B188}{{\em Phys. Rev.} {\bf 137}
  (1965)  B188--B197}.

\bibitem{Westpfahl:1967}
K.~Westpfahl, ``{R}elativistische {B}ewegungsprobleme. {I}. {D}as freie
  {S}pinteilchen,'' \href{http://dx.doi.org/10.1002/andp.19674750302}{{\em Ann.
  Phys. (Berlin)} {\bf 475} (1967)  113--135}.

\bibitem{Westpfahl:1969:1}
K.~Westpfahl, ``Relativistische {B}ewegungsprobleme. {V}. {Z}ur
  allgemein-relativistischen {D}ynamik klassischer {S}pinteilchen,''
  \href{http://dx.doi.org/10.1002/andp.19694770705}{{\em Ann. Phys. (Berlin)}
  {\bf 477} (1969)  345--360}.

\bibitem{Landau:Lifshitz:Vol1}
L.~D. Landau and E.~M. Lifshitz, {\em Mechanics}, vol.~1 of {\em Course of
  Theoretical Physics}.
\newblock Butterworth-Heinenann, Linacre House, Jordan Hill, Oxford OX2 8DP,
  3rd~ed., 1981.

\bibitem{Goldstein:Poole:Safko:2000}
H.~Goldstein, C.~Poole, and J.~Safko, {\em Classical Mechanics}.
\newblock Addison-Wesley, Cambridge, 3rd~ed., 2000.

\bibitem{Hanson:Regge:1974}
A.~J. Hanson and T.~Regge, ``{T}he relativistic spherical top,''
\href{http://dx.doi.org/10.1016/0003-4916(74)90046-3}{{\em Ann. Phys. (N.Y.)}
  {\bf 87} (1974)  498--566}.
%%CITATION = APNYA,87,498;%%.

\bibitem{Barausse:Racine:Buonanno:2009}
E.~Barausse, {\'E}.~Racine, and A.~Buonanno, ``{H}amiltonian of a spinning
  test-particle in curved spacetime,''
  \href{http://dx.doi.org/10.1103/PhysRevD.80.104025}{{\em Phys. Rev. D} {\bf
  80} (2009)  104025},
\href{http://arxiv.org/abs/0907.4745}{{\tt arXiv:0907.4745 [gr-qc]}}.
%%CITATION = 0907.4745;%%.

\bibitem{Steinhoff:Schafer:2009:2}
J.~Steinhoff and G.~Sch{\"a}fer, ``Canonical formulation of self-gravitating
  spinning-object systems,''
  \href{http://dx.doi.org/10.1209/0295-5075/87/50004}{{\em Europhys. Lett.}
  {\bf 87} (2009)  50004},
\href{http://arxiv.org/abs/0907.1967}{{\tt arXiv:0907.1967 [gr-qc]}}.
%%CITATION = 0907.1967;%%.

\bibitem{Steinhoff:2011}
J.~Steinhoff, ``Canonical formulation of spin in general relativity,''
  \href{http://dx.doi.org/10.1002/andp.201000178}{{\em Ann. Phys. (Berlin)}
  {\bf 523} (2011)  296--353},
\href{http://arxiv.org/abs/1106.4203}{{\tt arXiv:1106.4203 [gr-qc]}}.
%%CITATION = 1106.4203;%%.

\bibitem{Steinhoff:Schafer:Hergt:2008}
J.~Steinhoff, G.~Sch{\"a}fer, and S.~Hergt, ``{ADM} canonical formalism for
  gravitating spinning objects,''
  \href{http://dx.doi.org/10.1103/PhysRevD.77.104018}{{\em Phys. Rev. D} {\bf
  77} (2008)  104018},
\href{http://arxiv.org/abs/0805.3136}{{\tt arXiv:0805.3136 [gr-qc]}}.
%%CITATION = 0805.3136;%%.

\bibitem{Steinhoff:Wang:2009}
J.~Steinhoff and H.~Wang, ``Canonical formulation of gravitating spinning
  objects at 3.5 post-{N}ewtonian order,''
  \href{http://dx.doi.org/10.1103/PhysRevD.81.024022}{{\em Phys. Rev. D} {\bf
  81} (2010)  024022},
\href{http://arxiv.org/abs/0910.1008}{{\tt arXiv:0910.1008 [gr-qc]}}.
%%CITATION = 0910.1008;%%.

\bibitem{Arnowitt:Deser:Misner:1962}
R.~L. Arnowitt, S.~Deser, and C.~W. Misner, ``The dynamics of general
  relativity,'' in {\em Gravitation: An Introduction to Current Research},
  L.~Witten, ed., pp.~227--265.
\newblock John Wiley, New York,
1962.
\newblock
%%CITATION = GR-QC/0405109;%%.

\bibitem{Arnowitt:Deser:Misner:2008}
R.~L. Arnowitt, S.~Deser, and C.~W. Misner, ``Republication of: {T}he dynamics
  of general relativity,''
  \href{http://dx.doi.org/10.1007/s10714-008-0661-1}{{\em Gen. Relativ.
  Gravit.} {\bf 40} (2008)  1997--2027},
\href{http://arxiv.org/abs/gr-qc/0405109}{{\tt arXiv:gr-qc/0405109}}.
%%CITATION = GR-QC/0405109;%%.

\bibitem{Schwinger:1963:1}
J.~S. Schwinger, ``Quantized gravitational field,''
\href{http://dx.doi.org/10.1103/PhysRev.130.1253}{{\em Phys. Rev.} {\bf 130}
  (1963)  1253--1258}.
%%CITATION = PHRVA,130,1253;%%.

\bibitem{Yee:Bander:1993}
K.~Yee and M.~Bander, ``Equations of motion for spinning particles in external
  electromagnetic and gravitational fields,''
  \href{http://dx.doi.org/10.1103/PhysRevD.48.2797}{{\em Phys. Rev. D} {\bf 48}
  (1993)  2797--2799},
\href{http://arxiv.org/abs/hep-th/9302117}{{\tt arXiv:hep-th/9302117}}.
%%CITATION = HEP-TH/9302117;%%.

\bibitem{Damour:Jaranowski:Schafer:2008:3}
T.~Damour, P.~Jaranowski, and G.~Sch{\"a}fer, ``{E}ffective one body approach
  to the dynamics of two spinning black holes with next-to-leading order
  spin-orbit coupling,''
  \href{http://dx.doi.org/10.1103/PhysRevD.78.024009}{{\em Phys. Rev. D} {\bf
  78} (2008)  024009}, \href{http://arxiv.org/abs/0803.0915}{{\tt
  arXiv:0803.0915 [gr-qc]}}.

\bibitem{Barausse:Buonanno:2009}
E.~Barausse and A.~Buonanno, ``Improved effective-one-body {H}amiltonian for
  spinning black-hole binaries,''
  \href{http://dx.doi.org/10.1103/PhysRevD.81.084024}{{\em Phys. Rev. D} {\bf
  81} (2010)  084024},
\href{http://arxiv.org/abs/0912.3517}{{\tt arXiv:0912.3517 [gr-qc]}}.
%%CITATION = 0912.3517;%%.

\bibitem{Nagar:2011}
A.~Nagar, ``{E}ffective one-body {H}amiltonian of two spinning black-holes with
  next-to-next-to-leading order spin-orbit coupling,''
\href{http://link.aps.org/doi/10.1103/PhysRevD.84.084028}{{\em Phys. Rev. D} {\bf
  81} (2011)  084028},
\href{http://arxiv.org/abs/1106.4349}{{\tt arXiv:1106.4349 [gr-qc]}}.
%%CITATION = 1106.4349;%%.

\bibitem{Porto:Ross:Rothstein:2010}
R.~A. Porto, A.~Ross, and I.~Z. Rothstein, ``{S}pin induced multipole moments
  for the gravitational wave flux from binary inspirals to third
  post-{N}ewtonian order,''
  \href{http://dx.doi.org/10.1088/1475-7516/2011/03/009}{{\em JCAP} (2011)
  no.~3, 009}, \href{http://arxiv.org/abs/1007.1312}{{\tt arXiv:1007.1312
  [gr-qc]}}.

\bibitem{Kidder:1995}
L.~E. Kidder, ``Coalescing binary systems of compact objects to
  (post)$^{5/2}$-{N}ewtonian order. {V}. {S}pin effects,''
  \href{http://dx.doi.org/10.1103/PhysRevD.52.821}{{\em Phys. Rev. D} {\bf 52}
  (1995)  821--847},
\href{http://arxiv.org/abs/gr-qc/9506022}{{\tt arXiv:gr-qc/9506022}}.
%%CITATION = GR-QC/9506022;%%.

\bibitem{Poisson:1998}
E.~Poisson, ``Gravitational waves from inspiraling compact binaries: {T}he
  quadrupole-moment term,''
  \href{http://dx.doi.org/10.1103/PhysRevD.57.5287}{{\em Phys. Rev. D} {\bf 57}
  (1998)  5287--5290},
\href{http://arxiv.org/abs/gr-qc/9709032}{{\tt arXiv:gr-qc/9709032}}.
%%CITATION = GR-QC/9709032;%%.

\bibitem{Tagoshi:Ohashi:Owen:2001}
H.~Tagoshi, A.~Ohashi, and B.~J. Owen, ``Gravitational field and equations of
  motion of spinning compact binaries to 2.5 post-{N}ewtonian order,''
  \href{http://dx.doi.org/10.1103/PhysRevD.63.044006}{{\em Phys. Rev. D} {\bf
  63} (2001)  044006},
\href{http://arxiv.org/abs/gr-qc/0010014}{{\tt arXiv:gr-qc/0010014}}.
%%CITATION = GR-QC/0010014;%%.

\bibitem{Faye:Blanchet:Buonanno:2006}
G.~Faye, L.~Blanchet, and A.~Buonanno, ``{H}igher-order spin effects in the
  dynamics of compact binaries. {I}. {E}quations of motion,''
  \href{http://dx.doi.org/10.1103/PhysRevD.74.104033}{{\em Phys. Rev. D} {\bf
  74} (2006)  104033},
\href{http://arxiv.org/abs/gr-qc/0605139}{{\tt arXiv:gr-qc/0605139}}.
%%CITATION = GR-QC/0605139;%%.

\bibitem{Damour:Jaranowski:Schafer:2008:1}
T.~Damour, P.~Jaranowski, and G.~Sch{\"a}fer, ``{H}amiltonian of two spinning
  compact bodies with next-to-leading order gravitational spin-orbit
  coupling,'' \href{http://dx.doi.org/10.1103/PhysRevD.77.064032}{{\em Phys.
  Rev. D} {\bf 77} (2008)  064032},
\href{http://arxiv.org/abs/0711.1048}{{\tt arXiv:0711.1048 [gr-qc]}}.
%%CITATION = 0711.1048;%%.

\bibitem{Steinhoff:Hergt:Schafer:2008:2}
J.~Steinhoff, S.~Hergt, and G.~Sch{\"a}fer, ``Next-to-leading order
  gravitational spin(1)-spin(2) dynamics in {H}amiltonian form,''
  \href{http://dx.doi.org/10.1103/PhysRevD.77.081501}{{\em Phys. Rev. D} {\bf
  77} (2008)  081501(R)},
\href{http://arxiv.org/abs/0712.1716}{{\tt arXiv:0712.1716 [gr-qc]}}.
%%CITATION = 0712.1716;%%.

\bibitem{Hartung:Steinhoff:2010}
J.~Hartung and J.~Steinhoff, ``Next-to-leading order spin-orbit and
  spin(a)-spin(b) {H}amiltonians for $n$ gravitating spinning compact
  objects,'' \href{http://dx.doi.org/10.1103/PhysRevD.83.044008}{{\em Phys.
  Rev. D} {\bf 83} (2011)  044008},
\href{http://arxiv.org/abs/1011.1179}{{\tt arXiv:1011.1179 [gr-qc]}}.
%%CITATION = 1011.1179;%%.

\bibitem{Steinhoff:Hergt:Schafer:2008:1}
J.~Steinhoff, S.~Hergt, and G.~Sch{\"a}fer, ``{S}pin-squared {H}amiltonian of
  next-to-leading order gravitational interaction,''
  \href{http://dx.doi.org/10.1103/PhysRevD.78.101503}{{\em Phys. Rev. D} {\bf
  78} (2008)  101503(R)},
\href{http://arxiv.org/abs/0809.2200}{{\tt arXiv:0809.2200 [gr-qc]}}.
%%CITATION = 0809.2200;%%.

\bibitem{Hergt:Schafer:2008}
S.~Hergt and G.~Sch{\"a}fer, ``Higher-order-in-spin interaction {H}amiltonians
  for binary black holes from {P}oincar{\'e} invariance,''
  \href{http://dx.doi.org/10.1103/PhysRevD.78.124004}{{\em Phys. Rev. D} {\bf
  78} (2008)  124004},
\href{http://arxiv.org/abs/0809.2208}{{\tt arXiv:0809.2208 [gr-qc]}}.
%%CITATION = 0809.2208;%%.

\bibitem{Hergt:Steinhoff:Schafer:2010:1}
S.~Hergt, J.~Steinhoff, and G.~Sch{\"a}fer, ``The reduced {H}amiltonian for
  next-to-leading-order spin-squared dynamics of general compact binaries,''
  \href{http://dx.doi.org/10.1088/0264-9381/27/13/135007}{{\em Class. Quant.
  Grav.} {\bf 27} (2010)  135007},
\href{http://arxiv.org/abs/1002.2093}{{\tt arXiv:1002.2093 [gr-qc]}}.
%%CITATION = 1002.2093;%%.

\bibitem{Hergt:Schafer:2008:2}
S.~Hergt and G.~Sch{\"a}fer, ``Higher-order-in-spin interaction {H}amiltonians
  for binary black holes from source terms of {K}err geometry in approximate
  {ADM} coordinates,'' \href{http://dx.doi.org/10.1103/PhysRevD.77.104001}{{\em
  Phys. Rev. D} {\bf 77} (2008)  104001},
\href{http://arxiv.org/abs/0712.1515}{{\tt arXiv:0712.1515 [gr-qc]}}.
%%CITATION = 0712.1515;%%.

\bibitem{Wang:Steinhoff:Zeng:Schafer:2011}
H.~Wang, J.~Steinhoff, J.~Zeng, and G.~Sch{\"a}fer, ``Leading-order spin-orbit
  and spin(1)-spin(2) radiation-reaction {H}amiltonians,''\href{http://dx.doi.org/10.1103/PhysRevD.84.124005}{{\em
  Phys. Rev. D} {\bf 84} (2011)  124005},
  \href{http://arxiv.org/abs/1109.1182}{{\tt arXiv:1109.1182 [gr-qc]}}.

\bibitem{Hartung:Steinhoff:2011:1}
J.~Hartung and J.~Steinhoff, ``Next-to-next-to-leading order post-{N}ewtonian
  spin-orbit {H}amiltonian for self-gravitating binaries,''
  \href{http://dx.doi.org/10.1002/andp.201100094}{{\em Ann. Phys. (Berlin)}
  {\bf 523} (2011)  783--790},
\href{http://arxiv.org/abs/1104.3079}{{\tt arXiv:1104.3079 [gr-qc]}}.
%%CITATION = 1104.3079;%%.

\bibitem{Hartung:Steinhoff:2011:2}
J.~Hartung and J.~Steinhoff, ``{N}ext-to-next-to-leading order post-{N}ewtonian
  spin(1)-spin(2) {H}amiltonian for self-gravitating binaries,''
  \href{http://dx.doi.org/10.1002/andp.201100163}{{\em Ann. Phys. (Berlin)}
  {\bf 523} (2011)  919--924},
\href{http://arxiv.org/abs/1107.4294}{{\tt arXiv:1107.4294 [gr-qc]}}.
%%CITATION = 1107.4294;%%.

\bibitem{Goenner:Gralewski:Westpfahl:1967}
H.~Goenner, U.~Gralewski, and K.~Westpfahl, ``{G}ravitative {S}elbstkr{\"a}fte
  und {S}trahlungsverluste klassischer {S}pinteilchen (erste {N}{\"a}herung),''
  \href{http://dx.doi.org/10.1007/BF01326226}{{\em Z. Phys.} {\bf 207} (1967)
  186--208}.

\bibitem{Barker:OConnell:1975}
B.~M. Barker and R.~F. O'Connell, ``Gravitational two-body problem with
  arbitrary masses, spins, and quadrupole moments,''
\href{http://dx.doi.org/10.1103/PhysRevD.12.329}{{\em Phys. Rev. D} {\bf 12}
  (1975)  329--335}.
%%CITATION = PHRVA,D12,329;%%.

\bibitem{DEath:1975}
P.~D. D'Eath, ``Interaction of two black holes in the slow-motion limit,''
  \href{http://dx.doi.org/10.1103/PhysRevD.12.2183}{{\em Phys. Rev. D} {\bf 12}
  (1975)  2183--2199}.

\bibitem{Bennewitz:Westpfahl:1971}
F.~Bennewitz and K.~Westpfahl, ``{S}elbstwechselwirkung von
  {G}ravitationsfeldern schnell bewegter {P}ol-{D}ipolquellen,''
  \href{http://dx.doi.org/10.1007/BF01893619}{{\em Commun. math. Phys.} {\bf
  23} (1971)  296--318}.

\bibitem{Ibanez:Martin:Ruiz:1984}
J.~Ib{\'a}{\~n}ez, J.~Martin, and E.~Ruiz, ``Gravitational interaction of two
  spinning particles in general relativity. {II},''
  \href{http://dx.doi.org/10.1007/BF00762538}{{\em Gen. Rel. Grav.} {\bf 16}
  (1984)  225--242}.

\bibitem{Bona:Fustero:Verdaguer:1983}
C.~Bona, X.~Fustero, and E.~Verdaguer, ``{C}ross sections for relativistic
  spinning particles,'' \href{http://dx.doi.org/10.1103/PhysRevD.28.317}{{\em
  Phys. Rev. D} {\bf 28} (1983)  317--324}.

\bibitem{Barker:OConnell:1979}
B.~M. Barker and R.~F. O'Connell, ``The gravitational interaction: Spin,
  rotation, and quantum effects---a review,''
  \href{http://dx.doi.org/10.1007/BF00756587}{{\em Gen. Relativ. Gravit.} {\bf
  11} (1979)  149--175}.

\bibitem{Thorne:Hartle:1985}
K.~S. Thorne and J.~B. Hartle, ``Laws of motion and precession for black holes
  and other bodies,''
\href{http://dx.doi.org/10.1103/PhysRevD.31.1815}{{\em Phys. Rev. D} {\bf 31}
  (1985)  1815--1837}.
%%CITATION = PHRVA,D31,1815;%%.

\bibitem{Hergt:2011}
S.~Hergt, {\em {H}amiltonsche {F}ormulierung und {B}ehandlung nichtlinearer
  {E}igendrehimpulsbeitr{\"a}ge in {B}in{\"a}rsystemen der {A}llgemeinen
  {R}elativit{\"a}tstheorie}.
\newblock PhD thesis, Friedrich-Schiller-Universit{\"a}t Jena, 2011.

\bibitem{Damour:Schafer:1991}
T.~Damour and G.~Sch{\"a}fer, ``Redefinition of position variables and the
  reduction of higher order {L}agrangians,''
\href{http://dx.doi.org/10.1063/1.529135}{{\em J. Math. Phys.} {\bf 32} (1991)
  127--134}.
%%CITATION = JMAPA,32,127;%%.

\bibitem{Newton:Wigner:1949}
T.~D. Newton and E.~P. Wigner, ``Localized states for elementary systems,''
\href{http://dx.doi.org/10.1103/RevModPhys.21.400}{{\em Rev. Mod. Phys.} {\bf
  21} (1949)  400--406}.
%%CITATION = RMPHA,21,400;%%.

\bibitem{Steinhoff:Schafer:2009:1}
J.~Steinhoff and G.~Sch{\"a}fer, ``Comment on two recent papers regarding
  next-to-leading order spin-spin effects in gravitational interaction,''
  \href{http://dx.doi.org/10.1103/PhysRevD.80.088501}{{\em Phys. Rev. D} {\bf
  80} (2009)  088501},
\href{http://arxiv.org/abs/0903.4772}{{\tt arXiv:0903.4772 [gr-qc]}}.
%%CITATION = 0903.4772;%%.

\bibitem{Goenner:Westpfahl:1967}
H.~Goenner and K.~Westpfahl, ``Relativistische {B}ewegungsprobleme. {II}. {D}er
  starre {R}otator,'' \href{http://dx.doi.org/10.1002/andp.19674750505}{{\em
  Ann. Phys. (Berlin)} {\bf 475} (1967)  230--240}.

\bibitem{Romer:Westpfahl:1969}
H.~R{\"o}mer and K.~Westpfahl, ``{R}elativistische {B}ewegungsprobleme. {IV}.
  {R}otator-{S}pinteilchen in schwachen {G}ravitationsfeldern,''
  \href{http://dx.doi.org/10.1002/andp.19694770506}{{\em Ann. Phys. (Berlin)}
  {\bf 477} (1969)  264--276}.

\bibitem{Westpfahl:1969:2}
K.~Westpfahl, ``Relativistische {B}ewegungsprobleme. {VI}.
  {R}otator-{S}pinteilchen und allgemeine {R}elativit{\"a}tstheorie,''
  \href{http://dx.doi.org/10.1002/andp.19694770706}{{\em Ann. Phys. (Berlin)}
  {\bf 477} (1969)  361--371}.

\bibitem{Bailey:Israel:1975}
I.~Bailey and W.~Israel, ``Lagrangian dynamics of spinning particles and
  polarized media in general relativity,''
  \href{http://dx.doi.org/10.1007/BF01609434}{{\em Commun. math. Phys.} {\bf
  42} (1975)  65--82}.

\bibitem{DeWitt:2011}
B.~S. DeWitt, \href{http://dx.doi.org/10.1007/978-3-540-36911-0}{{\em {B}ryce
  {D}e{W}itt's Lectures on Gravitation}}, vol.~826 of {\em Lecture Notes in
  Physics}.
\newblock Springer, Berlin, 1st~ed., 2011.

\bibitem{Schafer:1984}
G.~Sch{\"a}fer, ``Acceleration-dependent {L}agrangians in general relativity,''
  \href{http://dx.doi.org/10.1016/0375-9601(84)90947-2}{{\em Phys. Lett. A}
  {\bf 100} (1984)  128--129}.

\bibitem{Landau:Lifshitz:Vol2:2}
L.~D. Landau and E.~M. Lifshitz, {\em Klassische Feldtheorie}, vol.~2 of {\em
  Lehrbuch der theoretischen Physik}.
\newblock Harri Deutsch, Thun und Frankfurt am Main, 12th~ed., 1997.

\bibitem{Dirac:1950}
P.~A.~M. Dirac, ``Generalized {H}amiltonian dynamics,''
\href{http://dx.doi.org/10.4153/CJM-1950-012-1}{{\em Canad. J. Math.} {\bf 2}
  (1950)  129--148}.
%%CITATION = CJMAA,2,129;%%.

\bibitem{Dirac:1951}
P.~A.~M. Dirac, ``The {H}amiltonian form of field dynamics,''
\href{http://dx.doi.org/10.4153/CJM-1951-001-2}{{\em Canad. J. Math.} {\bf 3}
  (1951)  1--23}.
%%CITATION = CJMAA,3,1;%%.

\bibitem{Dirac:1958:1}
P.~A.~M. Dirac, ``Generalized {H}amiltonian dynamics,''
\href{http://dx.doi.org/10.1098/rspa.1958.0141}{{\em Proc. R. Soc. A} {\bf 246}
  (1958)  326--332}.
%%CITATION = PRSLA,A246,326;%%.

\bibitem{Dirac:1964}
P.~A.~M. Dirac, {\em Lectures on Quantum Mechanics}.
\newblock Yeshiva University Press, New York, 1964.

\bibitem{Hanson:Regge:Teitelboim:1976}
A.~J. Hanson, T.~Regge, and C.~Teitelboim, {\em Constrained Hamiltonian
  Systems}.
\newblock Academia Nazionale dei Lincei, Roma, 1976.
\newblock \url{http://hdl.handle.net/2022/3108}.

\bibitem{Henneaux:Teitelboim:1992}
M.~Henneaux and C.~Teitelboim, {\em Quantization of Gauge Systems}.
\newblock Princeton University Press, Princeton, 1992.

\bibitem{Pons:2005}
J.~M. Pons, ``On {D}irac's incomplete analysis of gauge transformations,''
  \href{http://dx.doi.org/10.1016/j.shpsb.2005.04.004}{{\em Stud. Hist. Philos.
  Mod. Phys.} {\bf 36} (2005)  491--518},
\href{http://arxiv.org/abs/physics/0409076}{{\tt arXiv:physics/0409076}}.
%%CITATION = PHYSICS/0409076;%%.

\bibitem{Anderson:Bergmann:1951}
J.~L. Anderson and P.~G. Bergmann, ``Constraints in covariant field theories,''
\href{http://dx.doi.org/10.1103/PhysRev.83.1018}{{\em Phys. Rev.} {\bf 83}
  (1951)  1018--1025}.
%%CITATION = PHRVA,83,1018;%%.

\bibitem{Rosenfeld:1930}
L.~Rosenfeld, ``Zur {Q}uantelung der {W}ellenfelder,''
  \href{http://dx.doi.org/10.1002/andp.19303970107}{{\em Ann. Phys. (Berlin)}
  {\bf 397} (1930)  113--152}.

\bibitem{Salisbury:2007}
D.~C. Salisbury, ``{R}osenfeld, {B}ergmann, {D}irac and the invention of
  constrained {H}amiltonian dynamics,'' in {\em Proceedings of the 11th Marcel
  Grossmann Meeting on General Relativity}, H.~Kleinert, R.~T. Jantzen, and
  R.~Ruffini, eds., p.~2467.
\newblock World Scientific, Singapore, 2008.
\newblock
\href{http://arxiv.org/abs/physics/0701299}{{\tt arXiv:physics/0701299}}.
\newblock
%%CITATION = PHYSICS/0701299;%%.

\bibitem{Porto:Rothstein:2008:1:err}
R.~A. Porto and I.~Z. Rothstein, ``Erratum: {S}pin(1)spin(2) effects in the
  motion of inspiralling compact binaries at third order in the
  post-{N}ewtonian expansion,''
  \href{http://dx.doi.org/10.1103/PhysRevD.81.029904}{{\em Phys. Rev. D} {\bf
  81} (2010)  029904(E)}.

\bibitem{Porto:Rothstein:2006}
R.~A. Porto and I.~Z. Rothstein, ``Calculation of the first nonlinear
  contribution to the general-relativistic spin-spin interaction for binary
  systems,'' \href{http://dx.doi.org/10.1103/PhysRevLett.97.021101}{{\em Phys.
  Rev. Lett.} {\bf 97} (2006)  021101},
\href{http://arxiv.org/abs/gr-qc/0604099}{{\tt arXiv:gr-qc/0604099}}.
%%CITATION = GR-QC/0604099;%%.

\bibitem{Teukolsky:1974}
S.~A. Teukolsky, {\em {P}erturbations of a rotating black hole}.
\newblock PhD thesis, California Institute of Technology, Pasadena, California,
  1974.
\newblock \url{http://resolver.caltech.edu/CaltechETD:etd-08022006-094950}.

\end{thebibliography}
\end{document}